\documentclass[11pt,a4paper]{article}
\usepackage[ margin=1 in]{geometry}
\usepackage[utf8]{inputenc}
\usepackage{amsmath}\usepackage{amssymb}
\usepackage{xcolor}\usepackage{feynmp-auto}
\usepackage{caption}
  \usepackage{subcaption}
 \usepackage{float}
\usepackage{cancel}
\usepackage{nicematrix}

\usepackage{graphicx}\usepackage{array}  \usepackage[normalem]{ulem}
 
\usepackage{feynmp} 
\usepackage[compat=1.1.0]{tikz-feynhand}
\usepackage[compat=1.1.0]{tikz-feynman}\tikzfeynmanset{ Spblob/.style={ /tikz/shape=circle,
/tikz/draw=black,   /tikz/minimum size=4mm,
/tikz/fill=white, /tikz/pattern=crosshatch,
}
}

\usepackage{tikz}\usepackage{jheppub}
\usepackage{slashed}\usepackage{dsfont}
\usepackage[boxsize=2em,aligntableaux=center]{ytableau}
 
\usepackage{soul}

\newcommand{\yss}{\ytableausetup{boxsize=0.5em}}
\newcommand{\ysn}{\ytableausetup{boxsize=2em}}
\newcommand{\fundamental}{\yss\ydiagram{1}\ysn}
\newcommand{\vev}[1]{\langle #1 \rangle}
\newcommand{\LSp}{\Lambda_{\rm Sp}}
\newcommand{\CSp}{{{\widehat\Lambda}_{\rm Sp}^3}}
 \usepackage{orcidlink}
\usepackage{multirow}
\begin{document}

\title{Solving the strong CP problem with massless grand-color quarks
}
\preprint{ UMN-TH-4328/24}
\author[a]{Ravneet Bedi\orcidlink{0000-0002-7104-1753}}
\affiliation[a]{School of Physics and Astronomy, University of Minnesota, Minneapolis, Minnesota 55455, USA}

\author[a]{Tony Gherghetta\orcidlink{0000-0002-8489-1116}}

\author[b,c,d]{Keisuke Harigaya\orcidlink{0000-0001-6516-3386}}
\affiliation[b]{ Department of Physics, The University of Chicago, Chicago, Illinois, 60637, USA} 
 
\affiliation[c]{Enrico Fermi Institute and Kavli Institute for Cosmological Physics, The University of Chicago, Chicago, Illinois 60637, USA}
\affiliation[d]{Kavli Institute for the Physics and Mathematics of the Universe (WPI), The University of Tokyo Institutes for Advanced Study,
The University of Tokyo, Kashiwa, Chiba 277-8583, Japan}

\emailAdd{bedi0019@umn.edu}
\emailAdd{tgher@umn.edu}
\emailAdd{kharigaya@uchicago.edu}

\abstract{
We propose a solution to the strong CP problem that specifically relies on massless quarks and has no light axion. The QCD color group $SU(3)_c$ is embedded into a larger, simple gauge group (grand-color) where one of the massless, colored fermions enjoys an anomalous chiral symmetry, rendering the strong CP phase unphysical. The grand-color gauge group $G_{\rm GC}$ is Higgsed down to $SU(3)_c\times G_{c'}$, after which $G_{c'}$ eventually confines at a lower scale, spontaneously breaking the chiral symmetry and generating a real, positive mass to the massless, colored fermion. Since the chiral symmetry has a $G_{c'}$ anomaly, there is no corresponding light Nambu-Goldstone boson. The anomalous chiral symmetry can be an accidental symmetry that arises from an exact discrete symmetry without introducing a domain wall problem. Potential experimental signals of our mechanism include vector-like quarks near the TeV scale, pseudo Nambu-Goldstone bosons below the 10 GeV scale,  
light dark matter decay, and primordial gravitational waves from the new strong dynamics.
}

\maketitle
\section{Introduction}

A popular solution to the strong CP problem in the Standard Model (SM) is the Peccei-Quinn (PQ) mechanism~\cite{Peccei:1977hh,Peccei:1977ur} in which an anomalous $U(1)$ PQ symmetry is spontaneously broken giving rise to a Nambu-Goldstone boson, the axion~\cite{Weinberg:1977ma,Wilczek:1977pj}. Importantly, the global PQ symmetry is also explicitly broken by nonperturbative QCD dynamics, generating an axion potential with a minimum that cancels the strong CP phase, $\bar\theta$, thereby solving the strong CP problem. The axion mass can be precisely predicted using chiral perturbation theory, and even though the original electroweak scale axion~\cite{Weinberg:1977ma, Wilczek:1977pj} has been ruled out, extra model building to make the axion lighter (and thus invisible~\cite{Kim:1979if,Shifman:1979if,Zhitnitsky:1980tq,Dine:1981rt}) or heavier than the QCD prediction (due to UV modifications of QCD~\cite{Dimopoulos:1979pp,Tye:1981zy,Holdom:1982ex,Holdom:1985vx,Rubakov:1997vp,Berezhiani:2000gh,Hook:2014cda,Fukuda:2015ana,Gherghetta:2016fhp,Agrawal:2017ksf,Gaillard:2018xgk,Hook:2019qoh,Csaki:2019vte,Gherghetta:2020keg,Gherghetta:2020ofz,Valenti:2022tsc,Dunsky:2023ucb}) has motivated a huge experimental effort to search for axions (see for example,~\cite{Adams:2022pbo}). 

Despite the simplicity of the axion and the PQ mechanism, there is an even simpler solution to the strong CP problem, namely, assuming that the up quark is massless~\cite{Georgi:1981be,Choi:1988sy}. A massless up quark implies that there is an (anomalous) $U(1)$ chiral symmetry which can be used to rotate away the strong CP phase without requiring a light axion. At the QCD scale, $\Lambda_{\rm QCD}\sim 250$ MeV, non-perturbative dynamics explicitly breaks this symmetry, generating an effective up-quark mass~\cite{Georgi:1981be,Choi:1988sy,Kaplan:1986ru,Banks:1994yg,Dine:2014dga}. Interestingly, the strong CP problem is still solved because the QCD dynamics generates a complex mass with a phase 
that again cancels
the strong CP phase or alternatively, a vacuum expectation value (VEV) for the QCD $\eta'$ cancels the strong CP phase. 
Due to the difficulty of nonperturbative QCD calculations, the significance of this nonperturbative QCD contribution to the up quark mass had remained unresolved for a long time (see for example~\cite{Creutz:2003xc,Srednicki:2005wc,Dine:2017swf,Frison:2017mod}).
Recent lattice QCD calculations of the meson spectrum near the physical masses obtained an up quark ($\overline{\rm MS}$) mass $m_u(2~{\rm GeV})=2.16_{-0.26}^{+0.49}$MeV~\cite{FlavourLatticeAveragingGroupFLAG:2021npn,ParticleDataGroup:2022pth} implying that the up quark is unlikely to be massless above the QCD scale. As a complementary check, lattice QCD calculations in~\cite{Alexandrou:2020bkd} computed the dependence of the pion mass on the dynamical strange quark mass and also showed that the nonperturbative QCD contribution to the up quark mass can only be a fraction of the up quark mass.

The fact that there is strong lattice evidence for a perturbative contribution to the up quark mass above the QCD scale suggests that to  obtain a massless up-quark type solution solution there 
should be new strong dynamics at UV scales which generates this ``perturbative mass" (assuming the up Yukawa coupling is zero). 
Indeed, this possibility was explored in Ref.~\cite{Agrawal:2017evu} where the QCD gauge group, $SU(3)_c$, was embedded into an $SU(3)\times SU(3)\times SU(3)$ product group with each generation of quarks
separately charged under only one of the $SU(3)$ groups. A nonzero value of the up quark Yukawa coupling was then generated from small instantons at a scale $\gg \Lambda_{\rm QCD}$, 
which explicitly breaks the chiral symmetry. The small instanton contribution is enhanced (relative to QCD) due to the larger gauge coupling of each individual $SU(3)$ factor at UV scales. Furthermore, the nontrivial SM flavor structure was generated by dimension five operators that arise from the symmetry breaking $SU(3)^3\rightarrow SU(3)$. Similarly, in Ref.~\cite{Gupta:2020vxb}, the enhanced effects of small instantons in the UV completion of a composite Higgs model was used to explicitly break the anomalous $U(1)$ symmetry and generate the up quark Yukawa coupling.
More recently, Ref.~\cite{Cordova:2024ypu} used the small instantons of a gauged flavor symmetry (with a
non-invertible symmetry structure below the gauged flavor symmetry breaking scale) to generate a down-type quark mass.

\begin{figure}[t!]
    \centering
  {
\begin{tikzpicture}[scale=0.6]
  \begin{feynman}  
\shade[top color=black!20,bottom color=black!60]  (3,6.5) rectangle (1,5);
   \draw [ very thick](1,5) -- (3,5);\node at (4,5){${M}_{P}$}; \draw [very thick] (2,-9) -- (2,5); 
 \draw (1.5,.25) -- (2.5,.25);\node at (4,.25){$M_{\rm GC }$
    };
    \node (MP2) at (2,4){};
    \node (MP1) at (2,-1){};
   \draw [ very thick](-6,3) circle (1.1cm);\node (gc) at (-6,3){$G_{\rm GC}$};
   \draw [ very thick](-9,-3) circle (1.1cm);
   \node (gcp) at (-9,-3){$G_{c'}$}; 
   \draw [ very thick](-3,-3) circle (1.1cm);
   \node (qcd) at (-3,-3){$SU(3)_{c}$};
    \draw (1.5,-6) -- (2.5,-6); \node at (4,-6){${\Lambda}_{c'}$};
    \node  (GcpConf) at (-9,-6){\textit{Confines at }$\Lambda_{c'}$
    };
    \node (GcAux0) at (-6,0.5){\textit{Spontaneously }}; 
    \node (GcAux1) at (-6,0){\textit{broken at $M_{\rm GC}$}};
     \node  at (-6,-3.45){$\cancel{U(1)}$ 
      };
  \draw [  thick] (gc) -- (GcAux0);
  \draw [->, thick] (GcAux1) -- (qcd);\draw [->, thick] (GcAux1) -- (gcp);
   \draw [->, thick] (gcp) -- (GcpConf);
     \draw [very thick](1,-9) -- (3,-9); 
     \node  at (4,-9){$\quad{\Lambda}_{\rm QCD}$}; \node (qcdConf) at (-3,-9.2){\textit{Confines at} $\Lambda_{\rm QCD}$}; \draw [->, thick]   (qcd) -- (qcdConf);
     \shade[top color=black!40,bottom color=black!10]  (3,-9) rectangle (1,-10) ;
   \diagram* { (gcp) -- [photon]
 (qcd),
};
  
\end{feynman}
\end{tikzpicture}}

    \caption{
    A schematic diagram of the scales associated with our mechanism. Below the Planck scale $M_P$, the grand-color group, $G_{\rm GC}$, is Higgsed at the scale $\sim M_{\rm GC}$ to the QCD group $SU(3)_c$ and a residual group, $  G_{c'}$. When $G_{c'}$ confines at the scale ${\Lambda}_{c'}$, the anomalous $U(1)$ chiral symmetry is spontaneously broken. This breaking is then communicated to QCD via the massive gauge bosons left over from the Higgsing of $G_{\rm GC}$ or alternatively, additional Higgs fields. 
    }
    \label{fig:scales}
\end{figure}
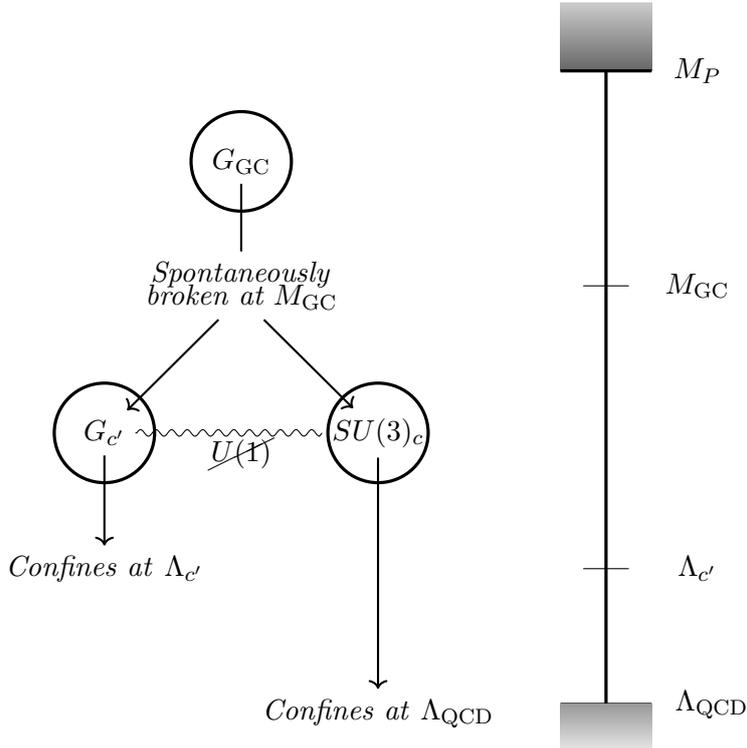

In this paper, we also propose a massless up-quark type solution by introducing new strong dynamics at a UV scale. However, instead of only explicitly breaking the chiral symmetry associated with a massless quark, we generate a quark mass by {\it spontaneously} breaking the chiral symmetry. This is achieved by embedding the QCD group, $SU(3)_c$, into a larger, simple group, $G_{\rm GC}=SU(2N+3)$~\cite{Gherghetta:2016fhp}, referred to as the grand-color group~\cite{Valenti:2022tsc}. 
In particular, the SM quarks are combined with grand-color partner fermions into the fundamental representation of the grand-color group. The grand-color group is then Higgsed to $SU(3)_c\times G_{c'}$ which gives rise to the following two features. 
First, the Higgsing generates, via heavy gauge boson exchange, dimension-six, four-fermion terms  that contain the massless quark and the grand color partner fermions. Second, when $G_{c'}$ confines, the grand-color fermion bilinear condensate spontaneously breaks the chiral symmetry, generating a nonzero quark mass via the dimension six terms. A schematic diagram of the scales associated with our mechanism is shown in figure~\ref{fig:scales}. 
Importantly, since the chiral symmetry is spontaneously broken, the generated quark mass is not suppressed by the product of SM Yukawa couplings and given that the chiral symmetry is anomalous 
with respect to $G_{c'}$, there is also no light axion. Alternatively, the symmetry breaking can be communicated from $G_{c'}$ to $SU(3)_c$ via scalar bosons with Yukawa couplings to quarks such as extra Higgses, or via fermions charged under both $SU(3)_c$ and $G_{c'}$.

Our mechanism is first applied to a model containing one fermion generation with $G_{c'}=Sp(2N)$ which ensures that the SM electroweak symmetry is not broken. We assume the right-handed up quark is massless and therefore has an anomalous chiral symmetry. 
When this symmetry is spontaneously broken by the $Sp(2N)$ dynamics, a positive up quark Yukawa coupling is generated and is proportional to the down quark Yukawa coupling. If $Sp(2N)$ confines at the scale $\LSp$, then a sufficiently large up Yukawa coupling can be generated in this minimal setup provided $11\lesssim N\lesssim 35$ for $10^3\, {\rm GeV} \lesssim \LSp \lesssim 10^{12}\, \rm GeV$. Smaller values of $N$ $(\gtrsim 5)$ can also explain the observed up Yukawa by introducing an extra pair of vector-like quarks which obtain a Dirac mass or mix with the up quark via the $Sp(2N)$ dynamics. Furthermore, with the vector-like quarks, we show how the anomalous chiral symmetry can accidentally arise from an exact $\mathbb{Z}_2$ symmetry. The model has a domain wall problem, which can be solved by imposing a $\mathbb{Z}_3$ symmetry instead. Thus, our solution to the strong CP problem is based on a simple exact symmetry, without imposing an anomalous chiral symmetry that is actually {\it not} a symmetry.

We then analyze models with more fermion generations, beginning with two generations.
We compute the Yukawa couplings arising from the $Sp(2N)$ dynamics  by carefully minimizing the potential of the $Sp(2N)$ pions and find that
a minimal two-generation extension of the one-generation model is ruled out because it leads to an unacceptably large strange quark mass. However, this problem can easily be avoided by again adding an extra pair of vector-like quarks. It is then shown that a three-generation model has a Higgs-pion mixing problem which makes it difficult to obtain
an order one top quark Yukawa coupling. There are several ways this problem can be avoided, but we focus on
the solution that charges the third generation of fermions under an extra $SU(2)$ or $SU(3)$ gauge group. The SM flavor structure arises from the extra gauge interactions.
Interestingly,
we show how the chiral symmetry can again be accidental in the low-energy theory from a $\mathbb{Z}_2$ or $\mathbb{Z}_3$ symmetry. We also show that  threshold corrections to the strong CP phase are sufficiently suppressed in the third generation model with extra gauge interactions. 

Finally, instead of having only fermions charged under the anomalous chiral symmetry, there is also the possibility of introducing chirally-charged extra Higgs fields.  We present a three-generation model with two-(or more) Higgs doublets, that has some advantages over the fermion models, except for a hierarchy problem which may be addressed with supersymmetry. In this class of models, the chiral symmetry breaking is mediated from $G_{c'}$ to $SU(3)_c$ via the extra Higgses,
unlike the minimal chiral fermion model where the mediation occurs via heavy gauge bosons. We also comment on the possibility of mediating the chiral symmetry breaking via fermions in higher representations of $SU(2N+3)$.

The massless up quark solution has inspired other UV modifications of QCD that also invoke an anomalous $U(1)$ symmetry with no light axion. Early work in Refs.~\cite{Dimopoulos:1979qi,Pagels:1979uu,Eichten:1980du,Lane:1980je,Cohen:1981fa} considered massless quarks in technicolor scenarios which incorporated QCD color. Since the technicolor strong dynamics breaks electroweak symmetry, there is no Higgs field and all Yukawa couplings originate from gauge interactions.
A more phenomenologically
viable scenario was studied in Ref.~\cite{Hook:2014cda} which
considered a mirror QCD model with a $\mathbb{Z}_2$ symmetry where the anomalous $U(1)$ symmetry is associated
with extra massless exotic quarks. In the UV, the anomalous $U(1)$ rotations of the massless quarks can then be used to set all theta angles to zero. When the mirror QCD confines, it spontaneously breaks the anomalous $U(1)$ symmetry, generating a mass for the QCD-charged exotic quarks. In the IR, the solution to the strong CP problem can be interpreted as either the mirror QCD $\eta'$ VEV cancelling the strong CP phase or the phase in the complex mass aligning with the strong CP phase.
This model is similar in spirit to our approach, except that our mechanism relies on grand color rather than mirror-QCD dynamics and furthermore, we directly generate the up quark mass from the new strong dynamics.

Our model has several phenomenological implications. A Nambu-Goldstone boson (NGB) corresponding to the spontaneous breaking of an approximate baryon symmetry can be the dark matter that decays into SM particles via the weak anomaly. Moreover, some of the NGBs that couple to gluons and photons are much lighter than the $G_{c'}$ confinement scale. Phenomenologically, these are similar to ``heavy QCD axions" and may be below the 10 GeV scale.
Successful models require new vector-like quarks which may be near the TeV scale and therefore accessible at colliders. Finally, the confinement of $G_c'$ could involve a first-order phase transition that produces primordial gravitational waves.

The outline of our paper is as follows. In section~\ref{sec:toy} we present a one-flavor toy model to illustrate the basic features our mechanism. A more realistic model is then constructed in section~\ref{sec:grandcolor} where we first present details of a one fermion generation model in section~\ref{subsec:one_gen_model}, followed by a two-fermion generation model in section~\ref{sec:Two_gen}. The full three-generation case is then discussed in section~\ref{sec:3gen}. A potential issue for generating the top quark mass arising from the Higgs-pion mixing is discussed in section~\ref{sec:difficulty}. This can be resolved with extra $SU(2)$ or $SU(3)$ gauge interactions as detailed in section~\ref{sec:extragaugeint}. Arguments for why corrections to the strong CP phase are sufficiently suppressed are presented in section~\ref{sec:CPcorrections} with  phenomenological consequences of our scenario given in section~\ref{sec:pheno} and the accidental chiral symmetry in our model is discussed in section~\ref{sec:twogen_accidental}. A class of models with extra Higgses or higher fermion representations is discussed in section~\ref{sec:Higgs} and the possibility of explaining the dark matter with the lightest pion is studied in section~\ref{sec:darkmatter}. A summary and concluding remarks are given in section~\ref{sec:conclusion}. The appendices contain further aspects of the computation including a discussion on the stability of the grand-color symmetry-breaking vacuum in appendix~\ref{sec:GC_breaking}, a proof of $\bar{\theta}=0$ in appendix~\ref{sec:proof}, a derivation of the four-fermion operators due to gauge boson exchange in appendix~\ref{app:4fermiops}, details of the pion potentials and vacuum alignment for the one and two-generation models in appendix~\ref{app:CPT} and a computation of the flavor invariants for the three-generation models is given in appendix~\ref{app:flavor_inv}.

\section{Generating fermion mass by grand color: one-flavor toy model}

\label{sec:toy}
To understand the essence of our proposed mechanism, we first present a toy model with grand-color gauge group $G_{\rm GC}=SU(N+3)$ and one flavor of massless Weyl fermions $\Psi_U$,$\Psi_{\bar{U}}$ at the UV scale, transforming in the fundamental, anti-fundamental representation, respectively. The $\theta$ term can then be simply removed by a fermion chiral rotation $\Psi_{U,\bar U}\rightarrow e^{i \alpha_{U,\bar U}} \Psi_{U,\bar U}$ and the model preserves CP symmetry.
At a lower energy scale, we assume that $SU(N+3)$ is spontaneously broken to $SU(N)\times SU(3)$ by appropriate Higgs fields. The grand-color fermion $\Psi_U$ then decomposes into $\psi_U (\Box,{\bf 1})$ and $U({\bf 1},\Box)$, where the $SU(N)\times SU(3)$ charges are shown in the parentheses. Similarly, $\Psi_{\bar{U}}$ decomposes into $\psi_{\bar{U}} (\bar{\Box},{\bf 1})$ and $\bar{U}({\bf 1},\bar{\Box})$.
The $SU(3)$ fermions $U$ and $\bar{U}$ are the toy version of the up quark and $SU(3)$ is the toy version of QCD $SU(3)$. As we will see, $U$ and $\bar{U}$ can obtain a mass by $SU(N)$ dynamics.

For $N >3$, $SU(N)$ confines at $\Lambda_{SU(N)}$, above the $SU(3)$ confinement scale. The chiral symmetry of $\psi_U$ and $\psi_{\bar{U}}$ is explicitly broken by the $SU(N)$ anomaly and spontaneously broken by the fermion condensate $\langle \psi_U \psi_{\bar{U}} \rangle \neq 0$.
Since $U$ and $\bar{U}$ are in the same $SU(N+3)$ multiplets as $\psi_U$ and $\psi_{\bar{U}}$, respectively, the broken chiral symmetry of $\psi_U$,$\psi_{\bar{U}}$ also induces a broken $U$,$\bar{U}$ chiral symmetry. 
This indeed occurs via the exchange of heavy gauge bosons, which generates the dimension-six term
\begin{align}
\mathcal{L}\supset \frac{g_{\rm GC}^2}{2M_{\rm GC}^2} \psi_{U}^\dag \bar{\sigma}^\mu U {\psi}_{\bar{U}}^\dag \bar{\sigma}_\mu \bar{U} + {\rm h.c.} = \frac{g_{\rm GC}^2}{M_{\rm GC}^2} \psi_{U}^\dag {\psi}_{\bar{U}}^\dag U \bar{U} + {\rm h.c.},
\label{eq:4fermionop}
\end{align}
where $\bar{\sigma}^\mu=(1,-\vec\sigma)$ with $\sigma_{1,2,3}$ the Pauli matrices, $g_{\rm GC}$ is the $SU(N+3)$ gauge 
coupling and $M_{\rm GC}$ is the mass scale of the heavy gauge bosons.
The four-fermion operator in \eqref{eq:4fermionop} now connects the chiral symmetries of $\psi_U,\psi_{\bar{U}}$ and $U, {\bar U}$. 
When the condensation $\langle \psi_U \psi_{\bar{U}} \rangle \neq 0$ breaks the chiral symmetry, it generates a non-zero up quark mass $\sim \Lambda_{SU(N)}^3/M_{\rm GC}^2$. For $\theta=0$, the sign of the condensate is negative, $\langle \psi_U \psi_{\bar{U}} \rangle<0$ (see appendix~\ref{sec:proof}), 
and the up quark mass term is positive, thereby solving the strong CP problem. 
The absence of the strong CP phase can also be understood from the parity conservation theorem in Ref.~\cite{Vafa:1984xg}. Although, unlike the assumptions in \cite{Vafa:1984xg}, some of the gauge bosons obtain a mass, but that does not affect the positivity of the path-integral measure and parity should not be spontaneously broken.

Note that the effect of the anomaly
vanishes  in the large $N$ limit while the condensation does not, so the chiral symmetry breaking can be considered as dominantly spontaneous rather than explicit, with a light NGB. However, for finite $N$, there is no light NGB and the model is distinct from the axion solution to the strong CP problem. In fact, below the confinement scale, an $SU(N)$ $\eta'$ meson plays the role of a heavy axion to preserve the CP symmetry.

The implication of the spontaneous nature of the symmetry breaking can be further illuminated by adding vector-like quarks $\Psi_D, \Psi_{\bar{D}}$ with mass $m_D\ll \Lambda_{SU(N)}$, much below the confinement scale $\Lambda_{SU(N)}$. The quarks $U, \bar{U}$ still obtain a mass $\sim\Lambda_{SU(N)}^3/M_{\rm GC}^2$ from the condensate of $\psi_U$, $\psi_{\bar{U}}$, but there is no dependence on $m_D$.
This is in contrast with the case where the $SU(N)$ is Higgsed above the confinement scale and the mass of $U$ and $\bar{U}$ is generated only by $SU(N)$ instantons, which now give a mass suppressed by $m_D$. As we will see, the spontaneous nature of the chiral symmetry breaking is crucial for generating a sufficiently large quark mass.

Furthermore, the spontaneous nature allows the chiral symmetry to be an accidental symmetry that arises from another exact symmetry.%
\footnote{The model with bi-fundamental fermions $\psi_B,\bar{\psi}_B$ in Ref.~\cite{Hook:2014cda} also has this feature.
We can impose a $\mathbb{Z}_3$ symmetry under which $\psi_B$ has charge $+1$.
The $\mathbb{Z}_3$ symmetry does not have a color or mirror-color anomaly and forbids the mass term of the bi-fundamental fermions. The model, however, has a domain wall problem arising from the $\mathbb{Z}_3$ symmetry.}
In the model with $\Psi_U,\Psi_{\bar{U}},\Psi_D$ and $\Psi_{\bar{D}}$, this can be seen by imposing a $\mathbb{Z}_2$ symmetry under which $\Psi_{U}$ and $\Psi_{D}$ are odd. 
The $\mathbb{Z}_2$ symmetry does not have an $SU(N+3)$ anomaly and therefore can be an exact gauge symmetry. At the renormalizable level the quark mass terms are forbidden and instead the $U,D$ quarks obtain masses from the $SU(N)$ dynamics.
Note that it is crucial to generate the masses by spontaneous breaking; explicit breaking by $SU(N)$ instantons would only generate an $\mathbb{Z}_2$ preserving effective interaction $U\bar{U} D \bar{D}$ (implying  $m_U,m_D$ still remain zero).
The $\mathbb{Z}_2$ symmetry is spontaneously broken by the $SU(N)$ dynamics, but a linear combination of a continuous flavor symmetry subgroup ($SU(2)_A$ described below) and the $\mathbb{Z}_2$ symmetry remains unbroken, preventing the formation of stable domain walls. There are three massless NGBs, arising from the flavor symmetry breaking $SU(2)_V\times SU(2)_A\rightarrow SU(2)_V$, but their shift symmetry does not have a color anomaly and therefore they are not QCD axions. We can also impose $U(1)$ gauge symmetries on the theory to explicitly break the shift symmetry and give a mass to the NGBs, or have some of the NGBs eaten by gauge bosons. 
This $\mathbb{Z}_2$ extension can be easily incorporated into the models with extra vector-like quarks, to be discussed later. 
However, the accidental chiral symmetry is violated by dimension six, four-fermion operators $\Psi_U \Psi_{\bar{U}} \Psi_U \Psi_{\bar{U}}$,
$\Psi_D \Psi_{\bar{D}} \Psi_D \Psi_{\bar{D}}$,
$\Psi_U \Psi_{\bar{U}} \Psi_D \Psi_{\bar{D}}$,
suppressed by a UV cut-off scale, $M_{\rm UV}$. The condensation of $SU(N)$-charged fermions then generates $SU(3)_c$-charged fermion masses $\sim \Lambda_{SU(N)}^3/M^2_{\rm UV}$ that may be complex. To avoid too large a strong CP phase, $M_{\rm GC}< 10^{-5} M_{\rm UV}$ is required, which constrains the parameter space. As will be shown in section~\ref{sec:twogen_accidental}, for
realistic models with NGBs that have masses much below the chiral symmetry breaking scale and couple to gluons, this bound will be modified.

The higher dimensional operators also introduce a domain wall problem, since they explicitly break the $SU(2)_A$ symmetry. The mass of the field that comprises domain walls is $\Lambda_{SU(N)}^2/M_{\rm UV}$ and the energy density inside the domain walls is $\Lambda_{SU(N)}^6/M_{\rm UV}^2$. The resultant domain wall tension $\sigma$ is $\Lambda_{SU(N)}^4/M_{\rm UV}$. Domain walls dominate the universe at a temperature
\begin{equation}
T\sim \left(\frac{\sigma}{M_{\rm Pl}}\right)^{1/2} \simeq 10~{\rm eV} \left(\frac{M_{\rm Pl}}{M_{\rm UV}}\right)^{1/2} \left(\frac{\Lambda_{SU(N)}}{10^5~{\rm GeV}}\right)^2,
\end{equation}
so the universe becomes domain wall
dominated much before the cosmological epoch today.

To improve the quality of the accidental symmetry and avoid the domain wall problem, we can add more fermions and impose a higher-order symmetry. For example, with two vector-like fermions in addition to the up quark, we may impose a $\mathbb{Z}_3$ symmetry. Higher dimensional operators that violate CP and introduce the domain wall problem now have dimension nine. In order that domain walls never dominate the energy density of the universe before the present dark energy epoch (with $\rho_{\rm DE}\sim {\rm meV}^4$),  now requires $\Lambda_{SU(N)}\lesssim 10^7$ GeV $(M_{\rm UV}/M_{\rm Pl})^{5/11}$.

In the next section, we construct a realistic model using the chiral symmetry breaking by grand color and further require that $SU(N)$ is broken down to an $Sp$ or $SO$ group (similar to the grand color axion model~\cite{Valenti:2022tsc}), so that electroweak symmetry is not spontaneously broken.
A variety of models may be constructed, but the essence of the mechanism is universal: we impose a chiral symmetry on massless fermions to remove the strong CP phase and the chiral symmetry is spontaneously and explicitly broken by the strong dynamics of the grand-color partner gauge group of $SU(3)$.

\section{A quark mass by grand color}
\label{sec:grandcolor}

In this section, we construct a class of models where the strong CP problem is solved without any light axion by embedding the SM quarks into the grand color group. We discuss models with one and two generations of fermions. These models will then be extended to realistic models with three generations in section~\ref{sec:3gen}.

\subsection{Grand color and confinement of \texorpdfstring{$Sp(2N)$}{}}

We employ the gauge dynamics proposed in~\cite{Valenti:2022tsc} which
involves a UV modification of QCD where $SU(3)_c$ is embedded into an enlarged color gauge group, $G_{\text{\rm GC}}$ with the same flavor structure as QCD.  The grand color group, $G_{\text{\rm GC}}$, is then broken down to $SU(3)_c\times G_{c^\prime}$, and $G_{c^\prime}$ eventually confines.
If $G_{c^\prime}=SU(N)$ (for $N>2$), the electroweak symmetry is spontaneously broken by the $G_{c^\prime}$ dynamics, essentially reproducing the dynamics of technicolor models that are inconsistent with electroweak precision data and the light Higgs boson. 
Instead, if $G_{c^\prime}=Sp(2N)$, the confinement does not break the electroweak symmetry.
Furthermore, as we will show, the $G_{c^\prime}$ dynamics will play a role in generating a quark mass via chiral symmetry breaking.

{\renewcommand{\arraystretch}{0.8}
\begin{table}[]
    \centering
    \setlength\extrarowheight{3pt}

\begin{tabular}{c|c|c|c|c|c|c}
&$SU(2N+3)$ & $U(1)_{Y'}$ &$Sp(2N)$&$SU(3)_c$&$SU(2)_L$&$U(1)_{Y}$\\ \hline 
\multirow{2}{*}{$\Psi_q\equiv \left(\begin{array}{c}
     q\\\psi_q
\end{array}\right)$}  
& \multirow{2}{*}{\fundamental} & \multirow{2}{*}{$\frac{1}{4N+6}$} & 
$\textbf{1}$&\fundamental& \multirow{2}{*}{\fundamental} &$1/6$\\
 & & &\fundamental  &$\textbf{1}$  &   &$0$ \\ 
 [3pt]\hline
\multirow{2}{*}{$\Psi_{\bar u}\equiv \begin{pmatrix} \bar u\\\psi_{\bar u}\end{pmatrix}$}  &\multirow{2}{*}{$\overline{\fundamental}$}& \multirow{2}{*}{$-\frac{1}{2}-\frac{1}{4N+6}$} &  $\textbf{1}$&$\bar{\fundamental}$&  \multirow{2}{*}{$\textbf{1}$ } & $-2/3$\\
 &  &&  \fundamental  &$\textbf{1}$& &$-1/2$ \\ [3pt] \hline
\multirow{2}{*}{$\Psi_{\bar d}\equiv \begin{pmatrix} \bar d\\\psi_{\bar d}\end{pmatrix}$}  &  \multirow{2}{*}{$\overline{\fundamental}$}&  \multirow{2}{*}{$\frac{1}{2}-\frac{1}{4N+6}$}&$\textbf{1}$ &$\bar{\fundamental}$&\multirow{2}{*}{$\textbf{1}$ } &$1/3$ \\
& && \fundamental  &  $\textbf{1}$ & &$1/2$ \\
\end{tabular}
\caption{Representations of the SM quarks ($q$,$\bar u$,$\bar d$) and their grand-color partners ($\psi_q$,$\psi_{\bar u}$,$\psi_{\bar d}$) under the various gauge groups. The grand color group is $SU(2N+3)$ and $U(1)_{Y'}$ is an additional abelian group as written in \eqref{eq:GC_breaking_pattern}.
The intermediate confining group is $Sp(2N)$ and  $SU(3)_c,SU(2)_L,U(1)_Y$ are the SM color, weak isospin and hypercharge group, respectively.}   
    \label{tab:GC_fermion_charges}
\end{table}}

Assuming a grand-color group $G_{\rm GC}=SU(2N+3)$, the gauge symmetry-breaking chain is 
\begin{equation}
SU(2N+3)\times U(1)_{Y'}\xrightarrow{M_{\rm GC}} SU(2N)\times SU(3)_c\times  U(1)\times U(1)_{Y'}\xrightarrow{M_{\rm Sp}} Sp(2N)\times SU(3)_c\times  U(1)_Y,
    \label{eq:GC_breaking_pattern}
\end{equation}
where $G_{\rm GC}$ is first broken at the scale $M_{\rm GC}$, followed by a second breaking to $Sp(2N)$ at the scale $M_{\rm Sp}$.
The first breaking can be achieved by the VEV of an adjoint scalar while the second breaking can be achieved by a rank-2 anti-symmetric scalar. This two-step breaking is required to stabilize the vacuum. It is not possible to have a one-step breaking by a single $SU(2N+3)$ anti-symmetric representation (see  appendix~\ref{sec:GC_breaking}).
The SM quarks ($q$,$\bar u$,$\bar d$) combine with exotic fermions ($\psi_q$,$\psi_{\bar u}$,$\psi_{\bar d}$) to form grand-color fermion multiplets ($\Psi_q$,$\Psi_{\bar u}$,$\Psi_{\bar d}$), where all fermions are left-handed Weyl spinors. The gauge charges of the SM quarks and their grand-color partners are given in Table~\ref{tab:GC_fermion_charges}.  Note also that $\psi_q=(\psi_{q_u},\psi_{q_d})$ is a SM electroweak doublet.
Above the grand-color scale $M_{\rm GC}$, the Yukawa interactions between the fermions and the SM $SU(2)_L$ doublet Higgs field $H=(\varphi^+,\varphi^0)^T$ with hypercharge $+1/2$ are given by
\begin{align}
{\cal L} = -y_u \Psi_q \Psi_{\bar u}H- y_d \Psi_q \Psi_{\bar d}{\widetilde H} +{\rm h.c.} = -y_u ( q {\bar u} + \psi_q \psi_{\bar u})H-y_d  (  q {\bar d} + \psi_q \psi_{\bar d}){\widetilde H} +{\rm h.c.}\,,
\label{eq:HiggsfermionL}
\end{align} 
where ${\widetilde H}=i\sigma_2H^\dagger$ and $y_{u,d}$ are the Yukawa coupling matrices. Note that for $SU(2)_L$ doublets, such as $\Psi_q$ and $H$, we use the convention that $\varepsilon^{12}=+1$ in $\mathcal{L}\supset -\Psi_q\Psi_{\bar{u}}H\equiv-\varepsilon^{ij}\Psi_{q_i}H_j\Psi_{\bar{u}}$. The $Sp(2N)$ sign convention is given in appendix~\ref{sec:proof}. The exchange of heavy gauge bosons with a mass $M_{\rm GC}$ and gauge coupling $g_{\rm GC}$, associated with the breaking $SU(2N+3) \to SU(3)_c\times SU(2N)\times U(1)$, generates the following four-fermion interactions%
\footnote{Note that as in Ref.~\cite{Valenti:2022tsc}, we assume that the grand color-breaking scalars do not couple to the fermions.}
\begin{align}
\mathcal{L}&= -\frac{g_{\rm GC}^2}{2M_{\rm GC}^2} \left|\sum_{i}\psi_{\chi_i}^\dag \bar{\sigma}^\mu \chi_i -\sum_{i}\bar{\chi}_i^\dag \bar{\sigma}^\mu \psi_{\bar{\chi}_i}\right|^2 + {\rm h.c.} \nonumber \\
&\supset \frac{g_{\rm GC}^2}{2M_{\rm GC}^2} \psi_{q_i}^\dag \bar{\sigma}^\mu q_i\left( {\psi}_{\bar{u}_j}^\dag \bar{\sigma}_\mu \bar{u}_j +{\psi}_{\bar{d}_j}^\dag \bar{\sigma}_\mu \bar{d}_j  \right)+ {\rm h.c.},
    \label{eq:GC_int}
\end{align}
where in the second line, we have assumed $\chi_i=q_i$ and $\bar{\chi}_i=\bar{u}_i,\bar{d}_i$ in the fundamental and  antifundamental representation of $SU(3)_c$, respectively, with $\psi_{\chi_i}$, $\psi_{\bar\chi_i}$ the corresponding grand-color partners as given in Table~\ref{tab:GC_fermion_charges}.

After the second symmetry breaking stage in \eqref{eq:GC_breaking_pattern} at $M_{\rm Sp}$, the $Sp(2N)$ group is assumed to confine at a scale $\LSp$, below $M_{\rm Sp}$ but above the electroweak scale. With $2F$ Weyl fermions in the fundamental representation of $Sp(2N)$, the strong dynamics breaks the (approximate) global symmetry $SU(2F) \rightarrow Sp(2F)$, provided $ {2F}\lesssim (5-8)N$~\cite{Sannino:2009aw}, which is easily satisfied in our model. Correspondingly, the quark bilinears
form condensates, which in the large $N$ limit are estimated to be
\begin{align}
\label{eq:quark condensate}
     \langle \psi_I\psi_J\rangle  \simeq \frac{N}{16\pi^2} \LSp^3 \Sigma_0^{IJ} \equiv \CSp \Sigma_0^{IJ}  \,,
\end{align}
where $I,J,\dots$ are $SU(2F)$ flavor indices, $\Sigma_0$ denotes the vacuum configuration and $\psi_I = (\psi_{q_u}, \psi_{q_d}, \psi_{\bar{u}}, \psi_{\bar{d}}, \cdots)^T$ (with $T$ denoting the transpose).

In summary, the essence of the model is as follows: We impose a classical $U(1)_\xi$ chiral symmetry
$\Psi_{\xi} \rightarrow e^{i\alpha} \Psi_{\xi}$ on one of the quarks, $\Psi_{\xi}=(\xi,\psi_\xi)^T$. 
Performing a $U(1)_\xi$ chiral rotation removes the strong CP phase. The $U(1)_\xi$ symmetry of $\psi_{\xi}$ is then {\it spontaneously} broken by the fermion bilinear condensate in Eq.~\eqref{eq:quark condensate}. Since $\xi$ is in the same $SU(2N+3)$ multiplet as $\psi_{\xi}$, the quark $\xi$ should feel the same symmetry breaking and obtain a mass. This can indeed occur via the heavy gauge boson exchange that generates the four-fermion operator in Eq.~\eqref{eq:GC_int}. In the minimal model, $\xi$ is identified as one of the right-handed SM quarks. Alternatively, we may add a vector-like fermion and also generate the $\xi$ quark mass by $Sp(2N)$ dynamics.

\subsection{One-generation model}\label{subsec:one_gen_model}

To illustrate the basic mechanism, we first present a model with a single fermion generation. Although this is a toy model, the analysis will subsequently be extended to realistic models with multiple generations.

\subsubsection{The minimal setup}
\label{sec:onegenmin}

We consider a single SM quark generation $\{\Psi_{q_u},\Psi_{q_d}, \Psi_{\bar u},\Psi_{\bar d}\}$ charged under the grand-color group and impose a chiral symmetry on $\Psi_{\bar{u}}$ which forbids $\Psi_{\bar{u}}$ Yukawa couplings. This then allows the $\theta$ term to be zero and all Yukawa couplings to be real and positive.
The fermions charged under the $Sp(2N)$ group have an approximate global $SU(4)$ symmetry associated with $SU(4)$ transformations of $\psi = \left(\psi_{q_u},\psi_{q_d},\psi_{\bar{u}},\psi_{\bar{d}}
\right)^T$. This symmetry is broken down by the $Sp(2N)$ dynamics to $Sp(4)$, which gives $15-10=5$ NGBs, or ``pions", in the coset space $SU(4)/Sp(4)$. The dynamics of the pions $\Pi^\alpha(x)$ can be studied in the non-linear sigma model with a non-linear sigma field
\begin{equation}
\Sigma(x)=\exp\left(i\Pi^\alpha(x)\widetilde{T}^\alpha/f\right) \Sigma_0,\label{eq:pion_exp}
\end{equation}
where $\Sigma_0$ denotes the vacuum of the theory, as defined in \eqref{eq:quark condensate}, $f$ is the decay constant associated with the global symmetry breaking $SU(4)\rightarrow Sp(4)$, and $\widetilde{T}^\alpha$ are the generators of $SU(4)$ not in $Sp(4)$. 
To preserve electroweak symmetry, the vacuum is chosen with $\Sigma_{0, \,\rm EW}\equiv {\rm diag}(i\sigma_2, -i\sigma_2)$ which is equivalent to the fermion condensates
\begin{align}
    \langle \psi_{q_u}\psi_{q_d}\rangle =  {\langle \psi_{\bar{d}}\psi_{\bar{u}}\rangle} =\CSp\,.
    \label{eq:2condensate}
\end{align}
These condensates  
spontaneously break the chiral symmetry of $\Psi_{\bar{u}}$ and 
generate the Yukawa coupling of $\bar{u}$, as will be shown below. Here we assumed $\vev{\psi_{q_u} \psi_{q_d}}>0$ by a baryon number rotation, and fixed the sign of $\vev{\psi_{\bar{d}} \psi_{\bar{u}}}=-\vev{\psi_{\bar{u}} \psi_{\bar{d}}}>0$. As discussed in appendix~\ref{sec:proof}, this is done by choosing $\theta=0$ which implies $\langle\psi_{q_u}\psi_{\bar{u}}\rangle=\langle\psi_{q_d}\psi_{\bar{d}}\rangle < 0$ and then relating the signs of these condensates to those in the vacuum
\eqref{eq:2condensate} using a non-anomalous flavor transformation in $SU(4)/Sp(4)$.

The $SU(4)/Sp(4)$ coset space can be more conveniently understood in terms of the $SO(6)/SO(5)$ coset space, which has the same structure. The pions are then in the $\bf 5$ of $SO(5)$. To identify the quantum numbers, we can decompose the pions as $\bf 1+\bf 4$ of $SO(4)\subset SO(5)$. The $\bf 4$ representation  is equivalent to the $\bf (2,2)$ representation of $SU(2)\times SU(2)\cong SO(4)$. Thus, the NGBs in $SU(4)/Sp(4)$ transform as an electroweak singlet and a Higgs-like doublet. The Higgs-like NGBs, $\Pi_H$, obtain a mass-squared due to electroweak loops 
given by
\begin{equation}
    m_{\Pi_H}^2\simeq
    \frac{g_2^2}{16\pi^2}\LSp^2\,,\label{eq:Pion_mass}
\end{equation}
where $g_2$ is the $SU(2)_L$ gauge coupling (ignoring hypercharge contributions), and the overall constant in \eqref{eq:Pion_mass} is assumed to be order one. The gauge contributions are assumed to dominate the corrections to the pion potential and thus the positive mass squared in \eqref{eq:Pion_mass} implies that electroweak symmetry remains unbroken after $Sp(2N)$ confinement.
This is indeed the case unless some of the Yukawa couplings are ${\cal O}(1)$ (as will be discussed in section~\ref{sec:3gen}).
The flavor singlet, however, remains massless. It is associated with the spontaneous breaking of the $U(1)$ baryon symmetry due to condensation.%
\footnote{The spontaneous breaking of baryon number does not cause proton decay since the SM quarks possess an independent baryon symmetry, under which $Sp(2N)$ quarks are not charged, due to the   $SU(3)_c$ gauge symmetry, and also lepton number is not violated. This independent baryon symmetry also guarantees the absence of neutron-antineutron oscillations.}

\begin{figure}[h]
    \centering
    \begin{tikzpicture}[scale=0.4] 
  \begin{feynman}  
\vertex   (b) ; 
\vertex [above=1.5em of b] (b1);
\vertex [below=1.5em of b] (b2);
\vertex [below left=of b2](a1) ;
\vertex [above left=of b1](a2) ;
\vertex [right=4.5em of b] (e1);
\vertex [ dot,right  =4em of e1] (e)  {};
\vertex [ right  =5em of e] (e2)  {};
\vertex [Spblob,above   =3em of e1] (f)  {};
\vertex [Spblob,below   =3em of e1] (c)  {};
\vertex [  left=1 em of e](d ) {$y_d$} ;

\vertex [  left=.1em of b1](x ) {\footnotesize\( g_{\rm GC}\)} ;

\vertex [  left=.1em of b2](y ) {\footnotesize\(g_{\rm GC}\)} ;
\vertex [above=1.5em of c] {\small\( \vev{\psi_{q_u}\psi_{q_d}}\)};
\vertex [below=1.5em of f] { \small\(   \vev{\psi_{\bar{u}}\psi_{\bar{d}}}\)};
\vertex [left=0.5em of a2] {\( \overline{u}\)};
\vertex [left=0.5em of a1] {\( q\)};
\diagram* {
(a1)--[fermion](b2)--[photon](b1),(a2)--[fermion](b1),
  (b1) -- [fermion, bend left ,edge label=\(  \psi_{\bar{u}}  \)] (f), 

(b2) -- [fermion, bend right , edge label'=\( \psi_{q}  \)] (c), 
(e) -- [fermion,  bend right , edge label'=\( \psi_{\bar{d}} \) ] (f),(e)-- [fermion,  bend left , edge label=\( \psi_q \) ] (c) ,(e) -- [scalar, edge label'=\( H \) ] (e2) ,
};  
\vertex [ right  =2em of e2]   {$\equiv$}; 
\vertex [right=3cm of e2](ax) ; 
\vertex [crossed dot,right=of ax] (bx) {};
\vertex [right=of bx] (e1x);
\vertex [above=1em of bx] {\(y_d\)}; 
    \vertex [below left=of ax](a1x){\( {q}\)};\vertex [above left=of ax](a2x){\({\overline{u}  }\)};
\diagram* {
(a1x)--[fermion](ax),(a2x)--[fermion](ax),
(ax) -- [scalar,   very thick,  edge label'=\(
\Pi^{ H}\)] (bx) -- [scalar , edge label'=\({H }\)] (e1x),
}; \end{feynman} 
\end{tikzpicture}
      \caption{
      Feynman diagrams depicting how the up-quark Yukawa coupling is generated. The left diagram shows the $Sp(2N)$ quark interactions that connect the SM quarks to the Higgs field while the right diagram shows the equivalent pion picture where the effect is seen as Higgs-pion mixing induced by the down Yukawa coupling $y_d$.}
     \label{fig:barBaruQu_term}
\end{figure}
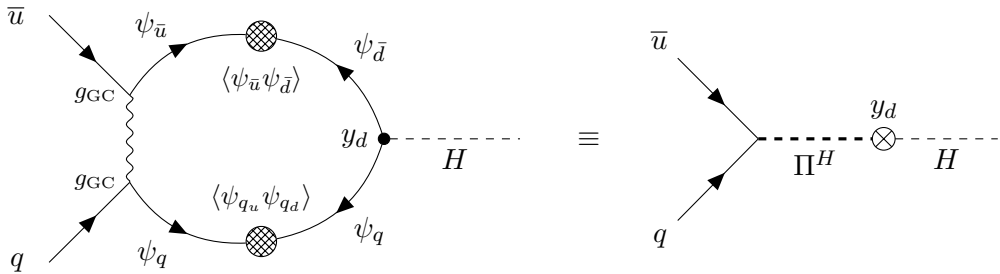
To see how the condensates \eqref{eq:2condensate} generate a mass for the massless up quark we next consider corrections to the NGB Yukawa interactions due to the grand-color gauge bosons as well as corrections to the NGB potential from interactions with the SM Higgs.
When the Higgs field breaks electroweak symmetry these corrections will 
generate an up quark mass.
The relevant terms in the Lagrangian are given by
\begin{equation}
    \mathcal{L}\supset \frac{g_{\rm GC}^2}{M_{GC}^2}q\Bar{u}\psi^\dag_q\psi^\dag_{\bar{u}}
    -y_d\psi_q{\psi}_{\bar d}\widetilde{H} +{\rm h.c.}\,,\label{eq:one_flavor_model}
\end{equation} 
where the first term, due to heavy gauge boson exchange, is computed in appendix~\ref{app:4fermiops} and the second term follows from \eqref{eq:HiggsfermionL}. As shown in the left diagram of figure~\ref{fig:barBaruQu_term}, the $Sp(2N)$ quark interactions in \eqref{eq:one_flavor_model} together with the fermion condensates in \eqref{eq:2condensate} are responsible for generating the up quark Yukawa coupling.

In the pion picture, a non-zero up Yukawa coupling arises from the Yukawa interaction in \eqref{eq:one_flavor_model} and can be understood as a result of the pion-Higgs mixing between $H$ and the Higgs-like NGBs $\Pi^H$ (see right diagram of figure~\ref{fig:barBaruQu_term}), or equivalently, a non-zero pion VEV for the real, neutral component of $\Pi^H$, induced by electroweak symmetry breaking.
To compute the induced VEV effect, we consider the fluctuations around the condensate in \eqref{eq:quark condensate}  
\begin{equation}
  \psi^I {\psi}^J \simeq \CSp \left(\Sigma_0^{IJ}+\frac{i}{f}(\widetilde{T}^\alpha\Pi^\alpha)^{IK}\Sigma_0^{KJ}+\dots \right),
    \label{eq:fermion_condensate}
\end{equation}
where $I,J,K= q_u,q_d, \bar{u}, \bar{d}$ and there is a sum over $\alpha=1,\dots,5$.
A non-zero condensate of $\psi_{q_d}\psi_{\bar{d}}$ corresponds to the displacement from the electroweak-symmetric vacuum 
by rotations of $(\psi_{q_d}, \psi_{\bar{u}})$ and $(\psi_{q_u}, \psi_{\bar{d}})$.  In the basis $(\psi_{q_u}, \psi_{q_d}, \psi_{\bar{u}}, \psi_{\bar{d}})$, the  generator of the corresponding NGB direction $\Pi^h$, where $\Pi^h/\sqrt{2}$ is the real part of the electromagnetic neutral component of $\Pi^H$, is
\begin{align}
\widetilde{T}^h=\frac{1}{2\sqrt{2}}
\begin{pmatrix}
 0 & 0 & 0 & -i \\
 0 & 0 & i & 0 \\
 0 & -i & 0 & 0 \\
 i & 0 & 0 & 0 
\end{pmatrix},
\end{align}
so that the quark bilinears are mapped into
\begin{align}
\psi_{q_d} \psi_{\bar{d}} = \psi_{q_u} \psi_{\bar{u}} \simeq  \frac{\CSp }{2\sqrt{2}f}  \Pi^h 
\,,
\end{align} 
where we have omitted ${\rm Im}\, \Pi^H$ since it does not obtain a VEV. The Lagrangian \eqref{eq:one_flavor_model} can be rewritten in terms of pions
\begin{equation}
    \mathcal{L}\supset \frac{g_{\rm GC}^2\CSp}{2\sqrt{2}fM_{GC}^2} q_u\Bar{u} \, \Pi^h
    -y_d\frac{\CSp }{\sqrt{2}f}  \Pi^h\,h-\frac{1}{2}m^2_{\Pi^H}\left(\Pi^{h}\right)^2\,,\label{eq:one_flavor_model_PiLag}
\end{equation} 
where $h/\sqrt{2}$ is the real part of $\varphi^0$
and we have also included the mass term from \eqref{eq:Pion_mass}.
The $\Pi^h\,h$ term generates a mixing between the Higgs $H$ and the Higgs-like NGBs $\Pi^H$. Equivalently, it can be thought of as a tadpole term below the electroweak symmetry breaking scale, $v$, i.e. when 
$\vev{h}=v$ it induces a non-zero VEV
\begin{align}
    \vev{\Pi^h} \simeq -y_d v \frac{N \LSp}{ \sqrt{2} g_2^2 f}.
\end{align}
This also corresponds to a non-zero condensate of $\psi_{q_u}\psi_{\bar{u}} <0$. The four-fermion operator in \eqref{eq:one_flavor_model} then generates a non-zero up quark mass proportional to the down Yukawa coupling
\begin{equation}
   y_u =  \frac{m_u}{v} \simeq \frac{N}{4}\frac{g_{GC}^2}{g_2^2}\frac{\LSp^2}{M_{\rm GC}^2}y_d \equiv \epsilon y_d\,,    
   \label{eq:upYukawa_woVLQ}
\end{equation} 
where we have used $f\simeq \sqrt{N}\LSp/4\pi.$
Note that the generated up quark mass is positive, so the strong CP phase $\bar{\theta}$ 
is indeed zero rather than $\pi$. An alternate proof for $\bar{\theta}=0$ without using the pion picture is given in appendix~\ref{sec:proof}.

The fact that the up Yukawa coupling is proportional to $y_d$ can also be understood using a symmetry argument. When $y_d$ = 0, there are two chiral symmetries associated with $\bar{u}$ and $\bar{d}$. The condensation of $\psi_{\bar{u}} \psi_{\bar{d}}$ breaks these symmetries to a diagonal subgroup which does not have an $Sp(2N)$ anomaly. This unbroken chiral symmetry forces the up and down quarks to be massless.

The observed magnitude of the up Yukawa coupling can be obtained from \eqref{eq:upYukawa_woVLQ} by relating the mass scales $\LSp$ and $M_{\rm GC}$ via the running of the gauge couplings. In particular, since $SU(2N)$ and $SU(3)_c$ unify into a single group at $M_{\rm GC}$, and $SU(2N)$ breaks down to $Sp(2N)$ at the scale $M_{\rm Sp}$, we obtain
\begin{eqnarray}
     \frac{2\pi}{\alpha_{\rm Sp}(\LSp)}+b_{\rm Sp}\log\left(\frac{M_{\rm Sp}}{\LSp}\right) +b_{\rm 2N}\log\left(\frac{M_{\rm GC}}{M_{\rm Sp}}\right) 
     =\frac{2\pi}{\alpha_{c}(m_Z)}+b_3\log\left(\frac{M_{\rm GC}}{m_Z}\right)\,,
     \label{eq:composite_running}
\end{eqnarray}
where $m_Z$ is the $Z$-boson mass, $\alpha_{c}$ and $\alpha_{\rm Sp}$ are the fine-structure constants of $SU(3)_c$ and $Sp(2N)$, respectively, and $b_3$, $b_{\rm Sp}$ and $b_{\rm 2N}$ is $\beta$-function coefficients of QCD, $Sp(2N)$ and $SU(2N)$, respectively. The $\beta$-function coefficients, $b_N$ and $b_{\rm Sp}$ for $SU(N)$ and $Sp(2N)$, respectively are
\begin{align}
\label{eq:b}
b_{\rm Sp}= &   \frac{11}{3}(N+1) - \frac{2}{3}F_{\rm Sp}\,, \nonumber \\
  b_{\rm N}= & \frac{11}{3} N - \frac{2}{3}F_{\rm N} \,,
\end{align}
where $F_{\rm Sp}$ and $F_{\rm N}$ are the number of fermion flavors charged under $Sp(2N)$ and $SU(N)$, respectively.
Since $\LSp <M_{\rm Sp} < M_{\rm GC}$, it suffices to consider two limiting cases, $M_{\rm Sp}=M_{\rm GC}$ and $M_{\rm Sp}=\LSp$, and find the range of up Yukawa couplings allowed by them. Using Eqs.~\eqref{eq:upYukawa_woVLQ}, \eqref{eq:composite_running}, and \eqref{eq:b}, with $\alpha_c(m_Z)=0.1184$, the value of $y_u$ can be computed for given a $\LSp$ and $N$.
%=============================================================================
\begin{figure}[t!]
      \centering
\includegraphics[width=0.6\textwidth]{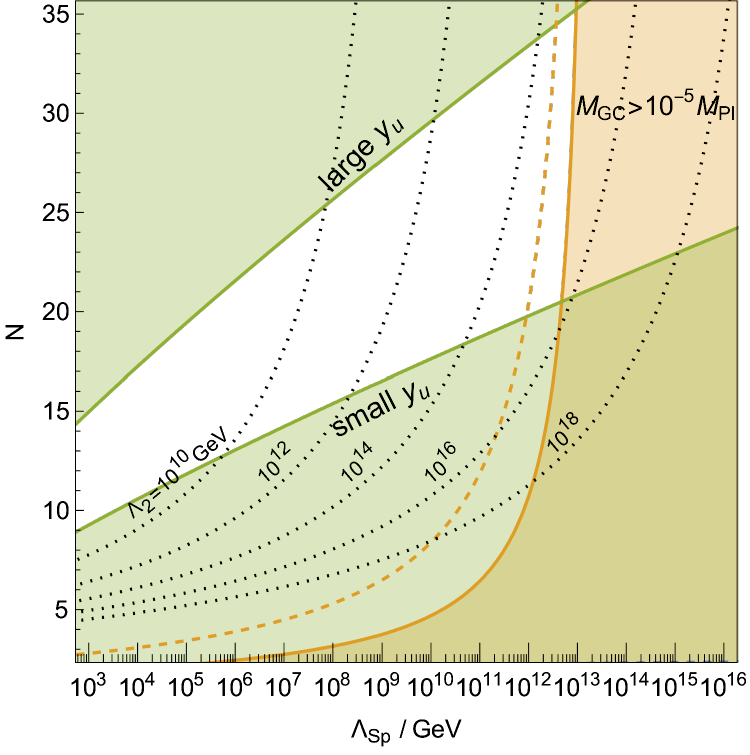}
\caption{
Constraints on the $Sp(2N)$ confinement scale $\LSp$ and $N$ in the minimal one-generation model where the up Yukawa coupling, $y_u$ is generated from $Sp(2N)$ dynamics. In the lower (upper) green region, the up Yukawa is too small (large) when $M_{\rm Sp}=\LSp \,(M_{\rm GC})$.
Between these two regions, shown in white, the observed up Yukawa can be obtained corresponding to $\LSp <M_{\rm Sp} < M_{\rm GC}$.
In the orange region, $M_{\rm GC}>10^{-5} M_{\rm Pl}$ 
when $M_{\rm Sp}=\LSp$,
and dimension-six operators may generate too large a strong CP phase. 
The dashed line shows the boundary when $M_{\rm Sp}=M_{\rm GC}$.
The dotted contours show the Landau pole scale of $SU(2)_L$. 
}\label{fig:one}
\end{figure}
%============================================================================

In figure~\ref{fig:one}, we show the constraints on $\LSp$ and $N$. In the lower (upper) green region, the up Yukawa coupling is too small (large) when $M_{\rm Sp}= \LSp\, (M_{\rm GC})$.  Between these two regions (shown in white), the observed value of the up Yukawa coupling can be obtained for $\LSp <M_{\rm Sp} < M_{\rm GC}$.  
It is clear that a reasonably large value of $N$ is required in the minimal case when there is no up-quark Yukawa coupling at the UV scale. For large $N$, there are many $SU(2)_L$ charged states above $\LSp$, so the Landau pole scale of $SU(2)_L$ may be low.
The dotted contours of figure~\ref{fig:one} show the $SU(2)_L$ Landau pole scale, $\Lambda_2$. The Landau pole is seen to be much above $\LSp$ and therefore the theory can be safely analyzed without knowing the UV completion of $SU(2)_L$.

Furthermore, it should be noted that a sufficiently large grand-color symmetry breaking scale can generate a non-zero strong CP phase of QCD. This arises from the dimension-six term $\supset \phi_{\rm GC}^\dagger\phi_{\rm GC} G{\widetilde G}$ between the grand-color breaking scalar fields,
$\phi_{\rm GC}$, and the $SU(2N+3)$ gauge fields with gauge field strength $G$, which leads to $\theta_{Sp(2N)}\neq \theta_{SU(3)_c}$. Assuming these terms are suppressed by the Planck scale and a loop factor, imposing $\bar{\theta} < 10^{-10}$ requires that $M_{\rm GC} < 10^{-5} M_{\rm Pl}$.
\,\footnote{As in Ref.~\cite{Valenti:2022tsc}, we assume that the dimension-five term $\phi_{\rm GC}G{\widetilde G}$ is suppressed, which can be guaranteed by an additional symmetry such as a $\mathbb{Z}_2$ symmetry under which $\phi_{\rm GC}$ is odd.
If the dimension-five term exists, the upper bound on $M_{\rm GC}$ becomes $10^{-10} M_{\rm Pl}$ and there are still viable parameter regions even with this more stringent constraint.
  
}
In the orange-shaded region (to the right of the orange-dashed line) of figure~\ref{fig:one}, $M_{\rm GC} > 10^{-5} M_{\rm Pl}$ when $M_{\rm Sp}=\LSp \,(M_{\rm GC})$.
This constraint can be relaxed when the grand-color breaking occurs via strong dynamics. In addition, there are also constraints from higher-dimensional CP-violating operators~\cite{Dine:1986bg,Bedi:2022qrd} without grand-color breaking fields such as $GG\widetilde{G}$.
Assuming the operator is suppressed by the Planck scale, the constraint $\bar{\theta} < 10^{-10}$ requires that  $\LSp < 10^{-5} M_{\rm Pl}$. This constraint is weaker than $M_{\rm GC} < 10^{-5} M_{\rm Pl}$, but to avoid the constraint without imposing CP symmetry will require a drastic assumption such as the compositeness of the $SU(2N+3)$ gauge field. Note that even supersymmetry cannot forbid $GG\tilde{G}$.

\subsubsection{Adding vector-like quarks}
\label{sec:One_gen_wVLQ}

The required up Yukawa coupling can be obtained with a smaller value of $N$ by adding a vector-like fermion to the theory. In particular, we introduce a pair of Weyl fermions $\Psi_U$ and $\Psi_{\bar{U}}$ in the fundamental and anti-fundamental representations of $SU(2N+3)$, respectively, where $\Psi_{\bar{U}}$ has the same gauge charge as $\Psi_{\bar{u}}$ in Table~\ref{tab:GC_fermion_charges} while $\Psi_{U}$ has conjugate charges. The vector-like fermions have a mass $m_U$ and a Yukawa interaction 
\begin{align}
    {\cal L} \supset - m_U \Psi_U \Psi_{\bar{U}}- {\lambda_U}  \Psi_q \Psi_{\bar{U}}H  +{\rm h.c.}
    \label{eq:lag}
\end{align}
The parameters $m_U$ and $\lambda_U$ are assumed to be real and positive by taking advantage of the phase rotation of $\Psi_U$ and $\Psi_{\bar{U}}$, respectively. The $\theta$ term associated with the grand-color group can then be removed by a phase rotation of $\Psi_{\bar{u}}$, which is assumed to have a zero Yukawa coupling.\,\footnote{There is also the alternative possibility of imposing a chiral symmetry on $\Psi_{U}$ which will be discussed later.}

The $SU(2)_L$ interaction prefers the condensate $\langle \psi_{q_u}\psi_{q_d}\rangle>0$ (which is made positive by a baryon number rotation), while the $U(1)_Y$ interaction forces $\psi_{\bar{d}}$ to condense with $\psi_{\bar{u}}$ and $\psi_{\bar{U}}$.
This then leads to the following possible condensates
\begin{align}
    \langle \psi_{q_u}\psi_{q_d}\rangle=\frac{\langle \psi_{\bar{d}}\psi_{\bar{U}}\rangle}{\cos\phi}= \frac{\langle \psi_{U}\psi_{\bar{u}}\rangle}{\cos\phi}= \frac{\langle \psi_{\bar{d}}\psi_{\bar{u}}\rangle}{\sin\phi} =-\frac{\langle \psi_{U}\psi_{\bar{U}}\rangle}{\sin\phi}=\CSp\,,
    \label{eq:2+1_condensate}
\end{align}
where $\phi$ corresponds to the VEV of one of the NGB directions (see appendix~\ref{app:CPT}). Note that the condensate signs in \eqref{eq:2+1_condensate} are determined by requiring that the vacuum is connected to $\vev{\psi_{q_u} \psi_{\bar{u}}}=\vev{\psi_{q_d} \psi_{\bar{d}}}= \vev{\psi_U \psi_{\bar{U}}}<0$ by a non-anomalous flavor transformation (see Eq.~\eqref{eq:1genvacpara}). 
The value of $\phi$ is determined by the relative importance of the Dirac 
mass $m_U$ with the Yukawa couplings $\lambda_U$ and 
$y_d$. As shown in appendix~\ref{app:CPT} (see \eqref{eq:tanphi}), we obtain
\begin{equation}
\phi\simeq \arctan\left(\frac{16 \pi^2m_U}{\lambda_U y_d\LSp}\right)\,. 
\label{eq:phi}
\end{equation}
This equation shows that for $m_U\gg \lambda_U y_d \Lambda_{\rm Sp}$, $\phi\simeq \pi/2$, so that $\psi_{\bar{U}}$ forms a condensate mainly with $\psi_U$. This is because the Dirac mass term, which gives rise to a tadpole term for the meson composed of $\psi_U \psi_{\bar{U}}$, favors the condensation of $\psi_U \psi_{\bar{U}}$. On the other hand, for $m_U\ll \lambda_U y_d \Lambda_{\rm Sp}$, we see that $\phi\simeq 0$, implying that the Higgs boson exchange favors the condensation of $\psi_{\bar{U}}$ with $\psi_{\bar{d}}$. This can be intuitively understood by an effective interaction $\propto \lambda_U y_d \,\psi_q \psi_{\bar{d}} \psi_q \psi_{\bar{U}}$ generated by the exchange of the Higgs boson. The condensation of $\psi_q \psi_q$ then generates an effective mass term $\psi_{\bar{U}}\psi_{\bar{d}}$.

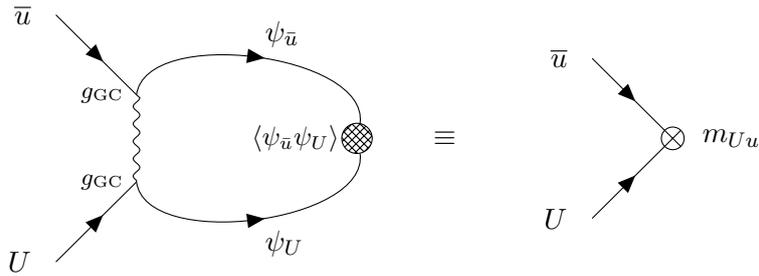
\begin{figure}[t]
    \centering
 \begin{tikzpicture}[scale=0.4] 
  \begin{feynman}  
\vertex   (b) ; 
\vertex [above=1.5em of b] (b1);
\vertex [below=1.5em of b] (b2);
\vertex [below left=of b2](a1) ;
\vertex [above left=of b1](a2) ;
\vertex [Spblob,right  =7em of b] (e)  {};
\vertex [ right  =3em of e] (e2) {$\equiv$}; 
\vertex [  left=.1em of b1](x ) {\footnotesize\( g_{\rm GC}\)} ;
\vertex [  left=.1em of b2](y ) {\footnotesize\(g_{\rm GC}\)} ; 
\vertex [left=2.1em of e] { \small\(   \vev{\psi_{\bar{u}}\psi_{U}}\)};
\vertex [left=0.5em of a2] {\( \overline{u}\)};
\vertex [left=0.5em of a1] {\( U\)};
\diagram* {
(a1)--[fermion](b2)--[photon](b1),(a2)--[fermion](b1), 
(b1)[crossed dot] -- [fermion, half left,looseness=.8 ,  edge label=\( \psi_{\bar{u}} \) ] (e),(b2)-- [fermion, half right, looseness=.8, edge label'=\( \psi_U \) ] (e) ,
};    
\vertex [crossed dot, right=3cm of e2](ax) {};  
\vertex [above left=of ax] (b1x);
\vertex [left=0.5em of b1x] {\({\overline{u}  }\)};
\vertex [below left=of ax] (b2x);
\vertex [left=0.5em of b2x] {\( {U}\)}; 
\vertex [right=2em of ax] {\({m_{Uu}  }\)}; 
\diagram* { 
(b1x)--[fermion ](ax), (b2x)--[fermion](ax),
}; \end{feynman} 
\end{tikzpicture}
      \caption{Feynman diagrams depicting how the mass mixing, $m_{Uu}$, of the up quark, $\bar u$, with the vector-like quark, $U$, is generated from the $\psi_{\bar u}\psi_U$ condensate in the models with an extra vector-like quark with a mass $m_U>|m_{Uu}|$. 
      }
     \label{fig:barBaruU_term}
\end{figure}

The condensates involving $\psi_{\bar{u}}$ spontaneously break the $U(1)$ chiral symmetry of $\psi_{\bar{u}}$, which is then communicated to $\bar{u}$ via the following four-fermion operator, 
\begin{align}
     \frac{g_{\rm GC}^2}{2M_{\rm GC}^2} \psi_{U}^\dag \bar{\sigma}^\mu U \psi_{\bar{u}}^\dag \bar{\sigma}_\mu \bar{u}  + {\rm h.c.} = \frac{g_{\rm GC}^2}{M_{\rm GC}^2} U {\bar u}\psi_{\bar u}^\dag \psi_U^\dag + {\rm h.c.} 
\label{eq:4FermiUbaru}
\end{align}
Using \eqref{eq:2+1_condensate}, the condensation of $\psi_{U}$ and $\psi_{\bar{u}}$ then generates a $U\bar{u}$ mass term
\begin{align}\label{eq:mUu_def}
   \frac{N g_{\rm GC}^2}{16\pi^2} \frac{\Lambda_{\rm Sp}^3}{M_{\rm GC}^2}\cos\phi\,U\bar{u} \equiv - m_{Uu}\, U\bar{u}\,.
\end{align}
The generation of this mass term in the quark picture can be understood by the Feynman diagram shown in figure~\ref{fig:barBaruU_term}.

When $m_{U} > |m_{Uu}|$, we may integrate out $U \bar{U}$ and then via $\bar{u}-U$ mixing, $\bar{u}$ obtains a Yukawa coupling with $q H$ given by
\begin{align}
    y_u^{(1)} = -\frac{m_{Uu}}{m_U}\lambda_U = \frac{N g_{\rm GC}^2}{16\pi^2}\frac{  \Lambda_{\rm Sp}^3}{m_U M_{\rm GC}^2 }\lambda_U\cos\phi.
    \label{eq:up_Yukawa_wVLQ}
\end{align}
Furthermore, as in the minimal model without vector-like quarks, there is an additional contribution to the up Yukawa coupling generated from the four-fermion operator in Eq.~\eqref{eq:one_flavor_model}. The $\psi_{\bar{u}}\psi_{\bar{d}}$ condensate then leads to an up Yukawa coupling proportional to the down Yukawa coupling given by
\begin{equation}
   y_u^{(2)}  =  \frac{N}{4}\frac{g_{GC}^2}{g_2^2}\frac{\LSp^2}{M_{\rm GC}^2}y_d\, \sin\phi \,.
    \label{eq:upYukawa_wVLQyd}
\end{equation}
The total contribution to the up Yukawa coupling is then $y_u = y_u^{(1)} + y_u^{(2)}$ where the relative contributions of the two effects depends on the value of $\phi$. 

\begin{figure}[t]
      \centering
\includegraphics[width=0.6\textwidth]{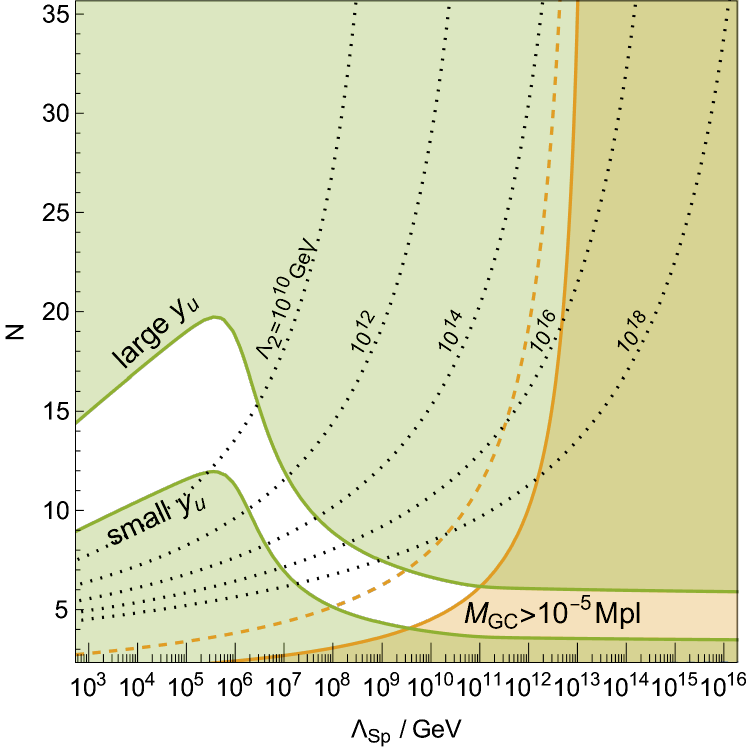}
\caption{
Constraints on the $Sp(2N)$ confinement scale $\LSp$ and $N$ in the one-generation model with a vector-like quark with $m_U = 1$ TeV and $\lambda_U =0.1$, assuming $m_U> |m_{Uu}|$. The up Yukawa coupling is generated either from the mixing with a new colored fermion \eqref{eq:up_Yukawa_wVLQ} or from the down Yukawa coupling \eqref{eq:upYukawa_wVLQyd}.
}
\label{fig:oneVLQ1}
\end{figure}

The allowed parameter space of $\LSp$ and $N$ that generates the required up Yukawa coupling is shown in figure~\ref{fig:oneVLQ1} where the shaded regions and contours are similar to those in figure~\ref{fig:one}. 
The straight boundaries of the green regions ($\LSp \lesssim 10^6$ GeV) are determined by the up Yukawa contribution in Eq.~\eqref{eq:upYukawa_wVLQyd}, while the curved boundaries ($\LSp \gtrsim 10^6$ GeV) are determined by Eq.~\eqref{eq:up_Yukawa_wVLQ}.  We have assumed $m_U = 1$ TeV, $\lambda_U =0.1$, and included the vector-like quark contribution in the running of the gauge coupling.
The assumed value of $m_U$ in figure~\ref{fig:oneVLQ1} is consistent with the lower bound on the vector-like quark mass which is approximately 500 GeV if it decays dominantly into the first two generations~\cite{ATLAS:2015lpr}, whereas it is approximately
1.3 TeV if it decays dominantly into the third-generation quarks~\cite{ATLAS:2022tla}. 
As $m_U$ becomes larger or $\lambda_U$ becomes smaller, the curved boundaries move to the right.
Thus, as shown in figure~\ref{fig:oneVLQ1} in the vector-like quark case, a value of $N$ as low as $4$ can generate the observed up Yukawa coupling, which is much less compared to the case without a vector-like quark. This arises because the mass mixing contribution \eqref{eq:up_Yukawa_wVLQ} is enhanced by a factor of $(g_2^2/64\pi^4)(\lambda_U \LSp/m_U)^2$ compared to the down quark Yukawa coupling contribution \eqref{eq:upYukawa_wVLQyd} and therefore with vector-like quarks, the up Yukawa coupling does not require large values of $N$ for $\LSp \gg m_U$. 

\begin{figure}[h!]
      \centering
\includegraphics[width=0.6\textwidth]{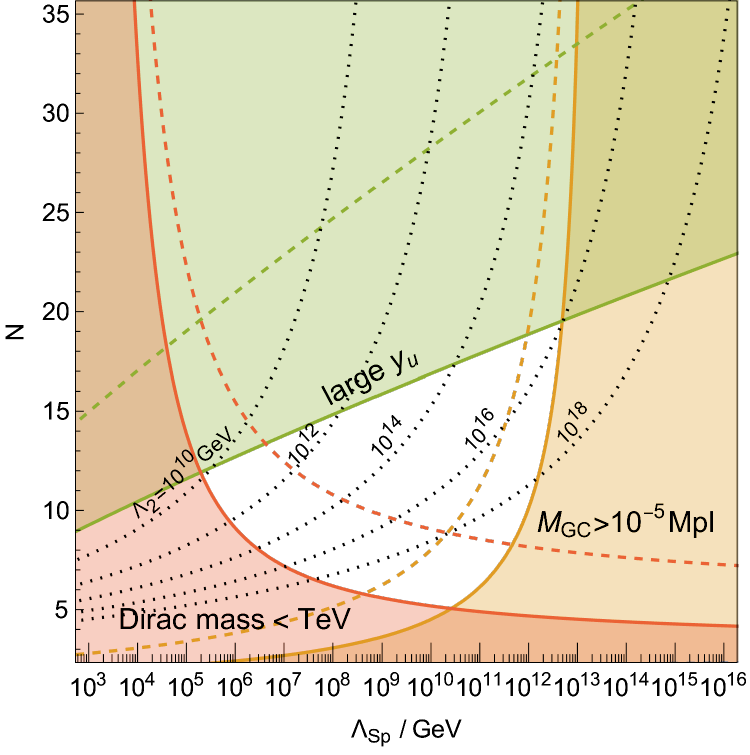}
\caption{
Constraints on the $Sp(2N)$ confinement scale $\LSp$ and $N$ from generating a new colored fermion mass in the one-generation model with vector-like quarks, assuming $m_U<|m_{Uu}|$. In the green region (below the dashed green line), the mass of the new colored fermion is below the TeV scale 
when $M_{\rm Sp}=\LSp$ (below $M_{\rm GC}$.) 
The orange region and the dotted contours are the same as figure~\ref{fig:one}.}
\label{fig:oneVLQ2}
\end{figure}
 
Note that the condensate $\langle\psi_{\bar{d}} \psi_{\bar{U}}\rangle$ also gives a contribution to the down-Yukawa coupling in a similar manner as the up-Yukawa generation \eqref{eq:upYukawa_wVLQyd}, namely 
\begin{equation}
    y_d^{(2)}=\frac{N}{4}\frac{g_{GC}^2}{g_2^2}\frac{\LSp^2}{M_{\rm GC}^2}\lambda_U \cos \phi~.
    \label{eq:yd(2)}
\end{equation}
This contribution is in addition to the tree-level down Yukawa coupling $y_d$, but can be made subdominant when $\lambda_U < \mathcal{O}(1)$. Therefore, the down Yukawa coupling is not excessively generated via condensation. Note that when $y_d=0$ we cannot use \eqref{eq:yd(2)} to generate the down Yukawa coupling because in this limit both \eqref{eq:up_Yukawa_wVLQ} and \eqref{eq:yd(2)} vanish, since due to \eqref{eq:phi}, we obtain $\phi=\pi/2$ .

Finally, when $m_U < |m_{Uu}|$, we may integrate out $U\bar{u}$ and identify $\bar{U}$ with the right-handed up quark and $y_u = \lambda_U$, while $U$ and $\bar{u}$ are extra heavy colored states. The mass $m_{Uu}$ should be greater than the TeV scale to avoid collider constraints on new colored particles. In this case, the constraints on $\LSp$ as a function of $N$ are shown in figure~\ref{fig:oneVLQ2}. The parameter space is excluded in the red region (below the red dashed line) in figure~\ref{fig:oneVLQ2}, because $m_{Uu}$ is below the TeV scale assuming $SU(2N)$ is broken to $Sp(2N)$ just above $\LSp$ (below $M_{\rm GC}$.)
In the green region, the up Yukawa coupling becomes too large from the $y_d$ contribution  \eqref{eq:upYukawa_wVLQyd}.

The special limiting case with $m_U=0$ can be achieved by imposing a chiral symmetry on $\Psi_U$ rather than on $\Psi_{\bar{u}}$. In this case, both $\Psi_{\bar{u}}$ and $\Psi_{\bar{U}}$ have Yukawa couplings, but we can perform a flavor transformation on $\Psi_{\bar{u}}$ and $\Psi_{\bar{U}}$ so that only $\Psi_{\bar{U}}$ has a Yukawa coupling.

Note that the generated up Yukawa or the mass of the new vector-like fermion is not suppressed by the product of the Yukawa couplings $y_d$, $\lambda_U$ or the Dirac mass term $m_U$, owing to the spontaneous breaking of the chiral symmetry. This differs from generating the Yukawa coupling or mass using instanton effects.

\subsubsection{Accidental chiral symmetry}
\label{sec:accidental}

The model with vector-like quarks can be  promoted to a model where the anomalous chiral symmetry is accidental as discussed in section~\ref{sec:toy}. To do so, we add an extra pair of vector-like quarks $\Psi_D$ and $\Psi_{\bar{D}}$ and assume both $\Psi_{\bar{u}}$ and $\Psi_{D}$ are odd under a $\mathbb{Z}_2$ symmetry. The quarks $\Psi_{\bar{D}}$ can have a Yukawa coupling to $\Psi_{q} \widetilde{H}$, but one can take a linear combination of $\Psi_{\bar{d}}$ and $\Psi_{\bar{D}}$, re-branded as $\Psi_{\bar{d}}$, so that only $\Psi_{\bar{d}}$ couples to $\Psi_{q} \widetilde{H}$. 
There is an $SU(2)$ flavor symmetry of $(\psi_{\bar{u}}, \psi_D)$ at the $Sp(2N)$ scale and associated massless directions, but the exchange of heavy $SU(2N)$ gauge bosons breaks the symmetry and gives masses $\sim \LSp^2/M_{\rm Sp}$ along those directions.
As a result, $\psi_{q,\bar{u},\bar{d},U,\bar{U}}$ 
condense in the same way as discussed in section~\ref{sec:One_gen_wVLQ} to generate the up Yukawa coupling, while $\psi_D$ forms a condensate with $\psi_{\bar{D}}$ to give a mass to $D$ and $\bar{D}$.
Alternatively, we may impose an odd $\mathbb{Z}_2$ charge to $\Psi_{\bar D}$ rather than $\Psi_{D}$. In this case, a Dirac mass term between $\Psi_{\bar{d}}$ and $\Psi_{D}$ is allowed by the discrete symmetry. However, unless the Dirac mass term is small, the analysis in section~\ref{sec:One_gen_wVLQ} will need to be modified. 
For $m_U=0$, we may instead impose a $\mathbb{Z}_2$ symmetry under which $\Psi_{U}$ and $\Psi_D$ are odd and other fields, including $\Psi_{\bar{u}}$, are even.

Finally, recall that the $\mathbb{Z}_2$ symmetric  model in section~\ref{sec:toy} has a domain wall problem. Similarly, in the $Sp(2N)$ model with $\mathbb{Z}_2$ symmetry, a domain wall problem arises if the $Sp(2N)$ confinement occurs after inflation. This problem can be avoided by extending the symmetry to $\mathbb{Z}_3$ with the addition of one more pair of vector-like quarks, $D^\prime,\,\bar{D}^\prime$ where $\bar{u},\, D$, and $D^\prime$ are charged under this symmetry.

\subsection{Two-generation model}
\label{sec:Two_gen}

\subsubsection{The minimal setup}

We next analyze the two-generation case with $\Psi_{q_{1,2}},\Psi_{\bar{u}_{1,2}}$, and $\Psi_{\bar{d}_{1,2}}$. However, unlike the one-generation model we will find that this minimal setup does not work and vector-like quarks will eventually need to be added. First, consider the minimal model with a chiral symmetry on $\Psi_{\bar{d}_1}$ and introduce the following Yukawa interactions
\begin{equation}
{\cal L} = -\widetilde{Y}^u_{ia}\Psi_{q_i} \Psi_{\bar{u}_a} H - \widetilde{Y}^d_{i2} \Psi_{q_i} \Psi_{\bar{d}_2} {\widetilde H}+{\rm h.c.},
\label{eq:2genLmin}
\end{equation}
where $i=1,2$, $a=1,2$, and the up-type quark Yukawa coupling matrix $\widetilde{Y}^u$ is taken to be diagonal by flavor transformations of $\Psi_{q_{1,2}}$,$\Psi_{\bar{u}_{1,2}}$, while $\widetilde{Y}^d$ has only $\Psi_{\bar d_2}$ interactions, i.e.,
\begin{equation}
\widetilde{Y}^u = \begin{pmatrix}
    y_u & 0 \\
    0 & y_c
\end{pmatrix},~~
\widetilde{Y}^d = \begin{pmatrix}
    0 & y_1 \\
    0 & y_2
\end{pmatrix}.
\label{eq:HiggsfermionL2}
\end{equation}
We denote the upper (lower) component of the doublet $q_1$ as $q_{u}(q_{d})$ and similarly for the doublet $q_2$ as $q_{c}(q_{s})$, although $q_{d}$ and $q_{s}$ are not purely the left-handed down and strange quarks. 
All of the coupling constants are taken to be real and positive by the phase rotation of quarks and $\theta=0$ by the (anomalous) chiral rotation of $\bar{d}_1$.

The $Sp(2N)$ strong dynamics generates the condensate as parameterized
in Eq.~\eqref{eq:quark condensate}. 
As shown in appendix~\ref{app:two}, the condensate is given by
\begin{align}
    \vev{\psi \psi^T}=\CSp \times
    \begin{pmatrix}
  0 & \cos\phi & 0 &   \sin\phi & 0 & 0& 0& 0 \\
  -\cos\phi & 0 & {-}\sin\phi & 0 & 0 & 0& 0& 0 \\
  0 &   \sin\phi & 0 & - \cos\phi & 0 & 0& 0& 0 \\
 {-} \sin\phi & 0 & \cos\phi & 0 & 0 & 0& 0& 0 \\
 0 & 0 & 0& 0& 0 & -1 & 0 & 0 \\
 0 & 0 & 0 & 0 & 1 & 0 & 0 & 0 \\
0 & 0 & 0& 0& 0 & 0 & 0 & 1 \\
0 & 0 & 0 & 0& 0& 0 & -1 & 0
    \end{pmatrix},~~\phi = {-}\arctan \frac{y_1}{y_2},
    \label{eq:2gencondensate}
\end{align}
\\
in the basis $\psi=(\psi_{q_u},\psi_{q_d},\psi_{q_c},\psi_{q_s},\psi_{\bar{u}_1},\psi_{\bar{d}_1},\psi_{\bar{u}_2},\psi_{\bar{d}_2})^T$.
Note that $\psi_{\bar{u}_1}$ and $\psi_{\bar{u}_2}$ condense only with $\psi_{\bar{d}_1}$ and $\psi_{\bar{d}_2}$, respectively. This can be intuitively understood as follows. The exchange of the Higgs generates effective interactions
\begin{align}
{\cal L}\sim \frac{1}{\LSp^2}
\left(
y_u \psi_{q_u} \psi_{\bar{u}_1} + y_c \psi_{q_c} \psi_{\bar{u}_2}
\right)
\left(
y_1 \psi_{q_d} \psi_{\bar{d}_2} + y_2 \psi_{q_s} \psi_{\bar{d}_2}
\right).
\label{eq:2genL}
\end{align}
For $y_c > y_u$, to minimize the energy given by this interaction, $\psi_{\bar{d}_2}$ should condense exclusively with $\psi_{\bar{u}_2}$, since a non-zero $\psi_{\bar{d}_2}\psi_{\bar{u}_1}$ condensate would reduce the magnitude of the  $\psi_{\bar{d}_2}\psi_{\bar{u}_2}$ condensate and increase the energy.
More explicitly, for the field space parameterized by two parameters $\phi$ and $\alpha$ as  
\begin{align}
    \vev{\psi_{q_u}\psi_{q_d}} &=- \vev{\psi_{q_c}\psi_{q_s}} = \CSp \cos\phi\,,\qquad
    \vev{\psi_{q_u}\psi_{q_s}} =- \vev{\psi_{q_d}\psi_{q_c}} =\CSp \sin\phi\,,\nonumber\\
    -\vev{\psi_{\bar{u}_1} \psi_{\bar{d}_1}} &=\vev{\psi_{\bar{u}_2} \psi_{\bar{d}_2}}= \CSp \cos\alpha\,,\qquad
    -\vev{\psi_{\bar{u}_1} \psi_{\bar{d}_2}} = -\vev{\psi_{\bar{u}_2} \psi_{\bar{d}_1}} = \CSp \sin\alpha\,,
\end{align}
the potential energy obtained from \eqref{eq:2genL} is
\begin{align}
V \sim - \LSp 
\left( y_c \cos\alpha (-y_1 \sin\phi + y_2 \cos\phi) + y_u \sin\alpha (y_1 \cos\phi + y_2 \sin\phi) \right)\,,
\end{align}
which is minimized at $\alpha=0$ and $\phi =- \arctan(y_1/y_2)$.

However, the vanishing $\psi_{\bar{d}_2}\psi_{\bar{u}_1}$ condensate leads to a phenomenological problem. 
Similar to the mechanism in the minimal one-generation model, the quark condensate generates the following Yukawa couplings, 
\begin{align}
    Y_d =& \begin{pmatrix}
        \epsilon y_u \frac{y_2}{\sqrt{y_1^2 + y_2^2}}  & y_1 - \epsilon y_c \frac{y_1}{\sqrt{y_1^2 + y_2^2}}  \\
       - \epsilon y_u \frac{y_1}{\sqrt{y_1^2 + y_2^2}} & y_2 + \epsilon y_c \frac{y_2}{\sqrt{y_1^2 + y_2^2}}
    \end{pmatrix}, \nonumber \\
        Y_u = & \begin{pmatrix}
        y_u  & 0  \\
       0 & y_c + \epsilon \sqrt{y_1^2 + y_2^2}
    \end{pmatrix}\,,
\end{align}
where $\epsilon\sim\LSp^2/M_{\rm GC}^2$ and we have omitted $\cal{O}(\epsilon)$ off-diagonal terms in $Y_u$ since they result in $\mathcal{O}(\epsilon^2)$ corrections to the up-type Yukawa couplings.
The determinant of the Yukawa matrix $Y_d$ is positive and the strong CP phase is zero. However, we cannot reproduce the observed Yukawa couplings. Since $\psi_{\bar{d}_1}$ only condenses with $\psi_{\bar{u}_1}$, the Yukawa coupling of $\bar{d_1}$ is proportional to $y_u$ and is at most $\epsilon y_u$. Given that we must have $\epsilon = y_d/y_u$ to obtain the observed down Yukawa coupling, the charm Yukawa then generates a down-type Yukawa as large as $y_c y_d/y_u \gg y_s$, leading to an unacceptably large strange quark mass. Alternatively, we could impose a chiral symmetry on $\Psi_{\bar{u}_1}$, but the failure persists. In this case, to obtain the observed up Yukawa coupling, we need $\epsilon = y_u/y_d$, but then the strange quark mass is again too large.
The minimal two-generation model is therefore not phenomenologically viable.

\subsubsection{Adding vector-like quarks}
\label{sec:Two_gen_wVLQ}
A successful model can be constructed by adding  vector-like quarks. For example, we can introduce $\Psi_U$ and $\Psi_{\bar{U}}$ with the following interactions
\begin{align}\label{eq:2gen_vlq}
{\cal L} = -\widetilde{Y}^u_{i2}  \Psi_{q_i} \Psi_{\bar{u}_2}H- \widetilde{Y}^d_{ia} \Psi_{q_i} \Psi_{\bar{d}_a} {\widetilde H} - \lambda_{Ui}\Psi_{q_i} \Psi_{\bar{U}} H - m_U \Psi_{U}\Psi_{\bar{U}}+ {\rm h.c.},
\end{align}
where $\Psi_{\bar{u}_1}$ is assumed to have a chiral symmetry and $\Psi_{\bar{u}_2},\Psi_{\bar{d}_{1,2}}$ now have nonzero Yukawa interactions with Yukawa matrices
\begin{equation}
\widetilde{Y}^u = \begin{pmatrix}
    0 & y_{1}^\prime \\
    0 & y_{2}^\prime
\end{pmatrix},\quad
\widetilde{Y}^d  = \begin{pmatrix}
    y_d & 0 \\
    0 & y_s
\end{pmatrix}\,.
\end{equation}
Note that, in principle, there is a $\Psi_{U}\Psi_{\bar{u}_2}$ mass term, but we have redefined the $\Psi_{\bar{U}},\,\Psi_{\bar{u}_2}$ fields since only one linear combination of them couples to $\Psi_{U}$. Furthermore, we use the remaining $\Psi_{U},\, \Psi_{\bar{U}}$ phase rotations to make $m_U,\,\lambda_{U2}$ real. As such,  $\lambda_{U1}$ is the only complex coupling appearing in \eqref{eq:2gen_vlq}.
The vacuum structure of the model for generic couplings and masses in \eqref{eq:2gen_vlq} is complicated, but there are few limiting cases that can be easily analyzed.

%======================================================
\begin{figure}[t!]
      \centering
\includegraphics[width=0.6\textwidth]{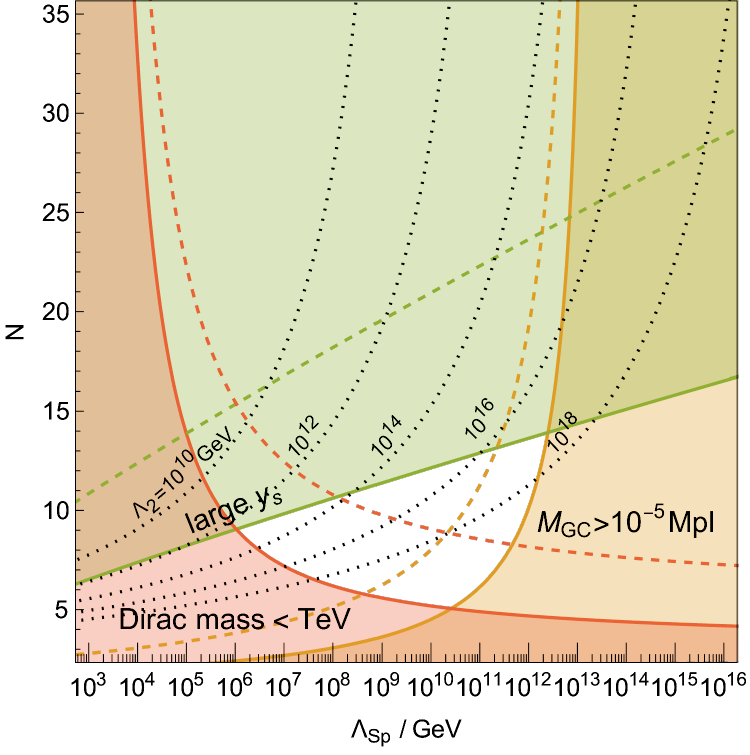}
\caption{
Constraints on the $Sp(2N)$ confinement scale $\LSp$ and $N$ in the two-generation model where a mass of a new colored fermion is generated.
In the green region (below the dashed green line), the mass of the new colored fermion is below the TeV scale 
when  $M_{\rm Sp}=\LSp$ (below $M_{\rm GC}$.)
The orange region and the dotted contours are the same as figure~\ref{fig:one}.}
\label{fig:twoVLQ2}
\end{figure}
%=======================================================
 
When $m_U=0$, $\psi_{U}$ and $\psi_{\bar{u}_1}$ do not have a mass or Yukawa couplings with other quarks, so they condense with each other and generate a mass term $U \bar{u}_1$. As in the one-generation model, $m_U=0$ can be achieved by imposing a chiral symmetry on $\Psi_{U}$, rather than to $\Psi_{\bar{u}_1}$.
The quarks $U$ and $\bar{u}_1$ are new Dirac fermions while $\bar{U}$ is identified with one of the SM right-handed up-type quarks. The observed up Yukawa coupling is then explained by appropriately choosing $\widetilde{Y}^{u,d}$ and $\lambda_{Ui}$.
The constraints on the parameter space is similar to the one-generation model with $m_U < |m_{Uu}|$, except that an extra constraint from generating a too large strange Yukawa from the charm Yukawa should be added. In figure~\ref{fig:twoVLQ2}, we show the constraints on $\LSp$ and $N$.

When $m_U\neq 0$, some of the SM quark Yukawa couplings can be generated in the same manner as in section~\ref{sec:One_gen_wVLQ}.
This is in particular effective when $m_U \ll \lambda_{U1} y_d \LSp$ or $\lambda_{U2} y_s \LSp$, for which the condensation pattern of $\psi_{\bar{U}}$ is mainly determined by the Yukawa couplings rather than by the mass terms. The vector-like fermion $\psi_{\bar{U}}$ mainly condenses with $\psi_{\bar{d}_i}$, while $\psi_U$ condenses with $\psi_{\bar{u}_1}$, generating a mass $m_{U\bar{u}_1} \simeq - N g_{\rm GC}^2 \LSp^3 / (16\pi^2 M_{\rm GC}^2)$ to $U \bar{u}_1$. Assuming $m_U > |m_{U\bar{u}_1}|$, $\bar{u}_1$ obtains a Yukawa coupling $-\lambda_{Ui} m_{U\bar{u}_1}/ m_U$ to $q_i H$ via the mixing with $\bar{U}$.

Instead of $\Psi_{U}$ and $\Psi_{\bar{U}}$, we may add vector-like quarks with different gauge charges and construct a similar model. Furthermore, as in the one-generation model, the anomalous chiral symmetry can be realized as an accidental symmetry at low energies. We will discuss such a model in section~\ref{sec:twogen_accidental}, together with the constraints from long-lived relics and a domain wall problem.

\section{Three-generation models}
\label{sec:3gen}

In this section we discuss the three-generation case in the SM. The natural generalization of the one- and two-generation models, however, has a Higgs-pion mixing problem that results in a large electroweak symmetry breaking scale.\,\footnote{This should also be an issue in the grand-color axion model~\cite{Valenti:2022tsc}.}
Nevertheless, this problem can be avoided as will be subsequently discussed.

\subsection{Higgs-pion mixing}
\label{sec:difficulty}

In order to obtain a large top quark mass, the Higgs must couple to the third-generation quarks with an ${\cal O}(1)$ Yukawa coupling
\begin{align}
{\cal L} \supset -y_3  \Psi_{q_3} \Psi_{\bar{u}_3}H + {\rm h.c.}= -y_3 ( q_3 \bar{u}_3 +  \psi_{q_3} \psi_{\bar{u}_3})H+ {\rm h.c.}\,. 
\label{eq:topYukawaL}
\end{align}
The second term in \eqref{eq:topYukawaL} generates a large mixing between the Higgs and a $\Pi^{H}$ pion\,\footnote{The SM Higgs also mixes with pions that arise from the Yukawa 
interactions with the first two generations (see Eq.~\eqref{eq:one_flavor_model_PiLag}), but this mixing is negligible.} given by
\begin{align} 
    y_3\kappa\frac{\sqrt{N}}{4\pi} \Lambda_{\rm Sp}^2 
   \Pi^{H}  H^\dagger + {\rm h.c.},
\end{align}
where $\kappa$ is an order one constant and the third-generation index on the pion field will be suppressed in the following for simplicity. 
This mixing destabilizes the Higgs and/or pion. Indeed, the mass-squared matrix is given by
\begin{equation}
    V\supset
    \begin{pmatrix}
    H^\dagger&\Pi^{H\dagger}
    \end{pmatrix}
    \begin{pmatrix}
    m_H^2& y_3\kappa\frac{\sqrt{N}}{4\pi}\Lambda_{\rm Sp}^2
  \\
y_3\kappa\frac{ \sqrt{N}}{4\pi}\Lambda_{\rm Sp}^2 &  \frac{\kappa'y_3^2}{16\pi^2}\Lambda_{\rm Sp}^2\end{pmatrix} \begin{pmatrix}
    H\\ \Pi^{H}
    \end{pmatrix}\,,
\label{eq:mass_matrix_compositeH}
\end{equation}
where $m_H$ is the tree level Higgs mass and the $(2,2)$ entry arises from quantum corrections due to the Yukawa coupling, $y_3$. The sign and magnitude of the ${\cal O}(1)$ coefficient $\kappa'$ is model dependent.%
\footnote{The sum of the quantum corrections due to the mass mixing between $H$ and $\Pi^{H}$ and the quartic terms from the Yukawa couplings or the kinetic term gives a positive $\kappa'$, but the corrections due to a trilinear coupling is negative. See figure~\ref{fig:diagrams_c3} for the analogous corrections to the mass of neutral pions, where the size of the corrections depends on the matter content of the model.}
If $\kappa' <0$, then $\Pi^{H}$ obtains a VEV of ${\cal O}(f)$ which breaks electroweak symmetry since $\Pi^H$ is an $SU(2)_L$ doublet.

Even if one assumes $\kappa' >0$ to avoid electroweak symmetry breaking, it is difficult to obtain an ${\cal O}(1)$ top Yukawa coupling. In order to have the electroweak scale much below $\LSp$, the determinant of the mass-squared matrix in \eqref{eq:mass_matrix_compositeH} must nearly vanish, which requires $m_H^2 \sim \kappa^2 N \LSp^2/\kappa'$. The $(1,1)$ entry is then much larger than the other entries, so we may integrate out $H$. The SM-like Higgs is dominantly 
$\Pi^{H}$ that mixes with $H$ by an angle $\sim y_3\kappa' / (4\pi \kappa \sqrt{N})$. The top Yukawa is then $\sim y_3^2 \kappa' / (4\pi \kappa \sqrt{N})$. Assuming $\kappa\sim \kappa' $, an ${\cal O}(1)$ top Yukawa coupling requires that $y_3$ should almost be non-perturbative. The  coupling $y_3$ then develops a Landau pole just above the $Sp(2N)$ confinement scale.

\subsubsection{Possible ways out}
Here we list a few possible ways to avoid the
difficulty with the top Yukawa coupling:
\begin{itemize}
    \item The third generation quarks are $SU(2N+3)$ singlets and are charged under another gauge group $SU(3)_T$. The product group $SU(3)_T\times SU(3)$ is then broken down to a diagonal $SU(3)$ subgroup, which is identified with the QCD color group, $SU(3)_c$. The bottom Yukawa coupling vanishes at tree level and is generated by $SU(3)_T$ instantons, so that CP violation from the $SU(3)_T$ $\theta$ term is absent.
    \item
    The third generation fermions are charged under $SU(2)_T$ which has a large gauge coupling. The first two generations of fermions are charged under $SU(2)_{FS}$. The product group $SU(2)_{FS}\times SU(2)_T$ is broken down to $SU(2)_L$ at a scale below the Sp confinement scale. The large $SU(2)_T$ gauge coupling generates a large positive correction to the $(2,2)$ entry of Eq.~\eqref{eq:mass_matrix_compositeH} which
    ameliorates the necessity of requiring a large Yukawa $y_3$. Furthermore, the top Yukawa at high energy scales can be smaller because of the RGE corrections from $SU(2)_T$.
        \item
    Introduce a term $\psi_{q_3}\psi_{q_i}$ or $\psi_{\bar{u}_3} \psi_{\bar{d}_i}$ with a mass larger than $\LSp$ to decouple $\Pi^{H}$. Indeed, these mass terms preserve the $Sp(2N)\times SU(3)\times U(1)_Y$ symmetry and may arise from $SU(2N+3)\times U(1)$ breaking. As long as there is only one mass term, the phase of the mass term can be removed by the baryon number rotation and does not introduce a strong CP phase. Additional mass terms, if they exist, should be smaller than $10^{-9} \LSp$ in order not to introduce a strong CP phase. Such significant suppression of the mass terms will require additional structures in the model.
    \item
    Introduce supersymmetry, which will suppress the off-diagonal entry of Eq.~\eqref{eq:mass_matrix_compositeH}. For sufficiently large $N$, strong dynamics yields the ADS potential~\cite{Affleck:1983mk} or the deformed moduli constraint~\cite{Seiberg:1994bz} and the chiral symmetry may be spontaneously  broken. 
\end{itemize}

In the next section, we analyze in more detail the first two possibilities that charge the third generation fermions under an extra gauge interaction, leaving a detailed investigation of the supersymmetric possibility for future work.

\subsection{Extra gauge interactions}
\label{sec:extragaugeint}

In this section, we discuss a model with an extra gauge symmetry, $SU(3)_T$ or $SU(2)_T$, under which only the third generation quarks are charged. The first two generations have the same structure and lead to the same constraints as presented in section~\ref{sec:Two_gen_wVLQ}. Note that in section~\ref{sec:Two_gen_wVLQ}
 $\Psi_{\bar{u}_1}$ was massless, and now $\Psi_{\bar{d}_3}$ will also be assumed to be massless in section~\ref{sec:extrasu3}.
 
\subsubsection{Extra $SU(3)$}
\label{sec:extrasu3}
In Ref.~\cite{Agrawal:2017evu}, the QCD color group $SU(3)_c$ is embedded into $SU(3)^3$, and the instanton effects of the three gauge groups generates the mass of lighter fermions in each generation.\,\footnote{The minimal model in Ref.~\cite{Agrawal:2017evu} also requires additional $SU(3)$ factors. However, the computation in~\cite{Agrawal:2017evu} overestimates the instanton effect as pointed out in Ref.~\cite{Csaki:2019vte}.} In our model, we only need to generate the bottom Yukawa coupling by the instanton effect of an extra $SU(3)_T$. This is because $SU(3)_c$ is embedded into $SU(2N+3)\times SU(3)_T$, where $SU(2N+3)$ is eventually broken down to $Sp(2N)\times SU(3)$ and $Sp(2N)$ confines. The remaining groups $SU(3)\times SU(3)_T$ are then broken down to $SU(3)_c$.
\begin{figure}[h]
    \centering
    \begin{tikzpicture}[scale=0.70]
    \begin{feynhand}
    \setlength{\feynhandblobsize}{12mm}
 \vertex  (h2)  [ dot] at (-2.5,0) {};  
 
 \vertex    [above left=.1em of h2]   {$y_t$};  
 
 \vertex  (h4)   at (-5,0) {$H$};  
 
    \vertex (v1) at (0.707107/2,0.707107/2);
    \vertex (v2) at (0.707107/2,-0.707107/2);
   \vertex (v3) at (-0.707107/2,0.707107/2);
    \vertex (v4) at (-0.707107/2,-0.707107/2);

 \vertex [above right=of v1] (h0)  {$\bar{d}_3$}  ; 
        
 \vertex  [below right=of v2](h1)   {$q_3$} ; 
 
 \vertex [above left=1.2em of v3]   {$\bar{u}_3$}  ; 
    \propag[fer] (h2) to [half left ](v3);
     \propag[fer] (h2) to [half right, edge label'=$q_3$](v4);
      \propag[fer] (h0) to  (v1);
     \propag[fer] (h1) to   (v2);
     \propag[sca] (h2) to (h4);
       \vertex[grayblob] (tv) at (0,0) {};
       \node at (0,0) {{\footnotesize $SU(3)_T$}};
    \end{feynhand}
    \end{tikzpicture}
    \caption{The Feynman diagram showing how the bottom quark Yukawa coupling is generated via $SU(3)_T$ instantons.  }
    \label{fig:SU3T_Inst}
\end{figure}
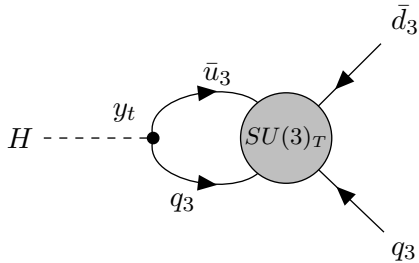

The third-generation quarks $q_3$ and $\bar{u}_3$ have the following Yukawa coupling,
\begin{align}
{\cal L} = - y_t q_3 \bar{u}_3H + {\rm h.c.}\,,
\end{align}
where $y_t$ is taken to be real and positive by a chiral rotation. The $SU(3)_T$ theta parameter is taken to be   $\theta_T=0$ by the rotation of $\bar{d}_3$.
The right-handed bottom quark $\bar{d}_3$ does not have a Yukawa interaction and enjoys a chiral symmetry.
We assume that the breaking $SU(3)_T\times SU(3) \rightarrow SU(3)_c$ occurs at the scale $v_T$ via a Higgs field $\Phi_3$ and the $SU(3)_T$ gauge coupling is semi-perturbative so that the dilute instanton gas approximation is valid.
Then the $SU(3)_T$ instanton can generate a sizable bottom Yukawa coupling~\cite{tHooft:1976snw,Csaki:2019vte},
\begin{align}
    y_b&\simeq 
      2\times 10^{-4}\,y_t \, \left( \frac{2\pi}{\alpha_T(\mu)}\right)^{6}  \int \frac{d\rho}{\rho^5}\,    \frac{4\pi^2}{3}\rho^{4} \,e^{-\frac{2\pi}{\alpha_T(1/\rho)}{-i {\theta}_T}}\,
        e^{-4\pi^2v_T^2\,\rho^2}\,,\nonumber\\
       & \simeq 3\times 10^{-4}\,
       \left(\frac{\Lambda_T}{v_T}\right)^{19/2} \left({\rm ln } \frac{\mu}{\Lambda_T }\right)^6  \,,
       \label{eq:Three_gen_inst_bott_yuk}
\end{align}
where the renormalization scale $\mu$ should be near $v_T$. The would-be confinement scale, $\Lambda_T$ of $SU(3)_T$, when the  $SU(3)_T\times SU(3) $ symmetry is unbroken (i.e., $\Lambda_T < v_T$), is defined to be
\begin{align}
\Lambda_T \simeq v_T \times {\rm exp} \left(- \frac{2\pi}{\alpha_T(v_T) 
\times b_T}\right)\,,\label{eq:Lambda_T_def}
\end{align}
where $b_T=19/2$ is the $\beta$-function coefficient of the $SU(3)_T$ gauge coupling which includes the gauge, fermion, and Higgs contributions.
Note that the gauge coupling of $SU(3)_T$ is a free parameter so that $\alpha_T(v_T)$ can be chosen as large as possible to maximize the instanton effect. This is unlike the minimal model of Ref.~\cite{Agrawal:2017evu}. Despite this, the largest possible bottom Yukawa coupling that can be generated is, for $\mu=v_T$, $y_b\simeq 10^{-7}$ when $\Lambda_T  \simeq 0.5\,v_T$, or equivalently $\alpha_T(v_T)\simeq 1$. 
To estimate the uncertainty of the generated bottom Yukawa coupling in \eqref{eq:Three_gen_inst_bott_yuk}, we can vary $\mu$ by an ${\cal O}(1)$ factor. For $\mu=(1/6-6)v_T$, we find that the maximal value of $y_b$ ranges from $10^{-15}-10^{-2}$.
Given this large uncertainty, it is possible that the observed bottom Yukawa could be explained in the minimal model, but instead we propose an extension of the model where the observed bottom Yukawa coupling can be more concretely obtained.

\begin{figure}[t!]
    \centering
\includegraphics[width=.85\textwidth]{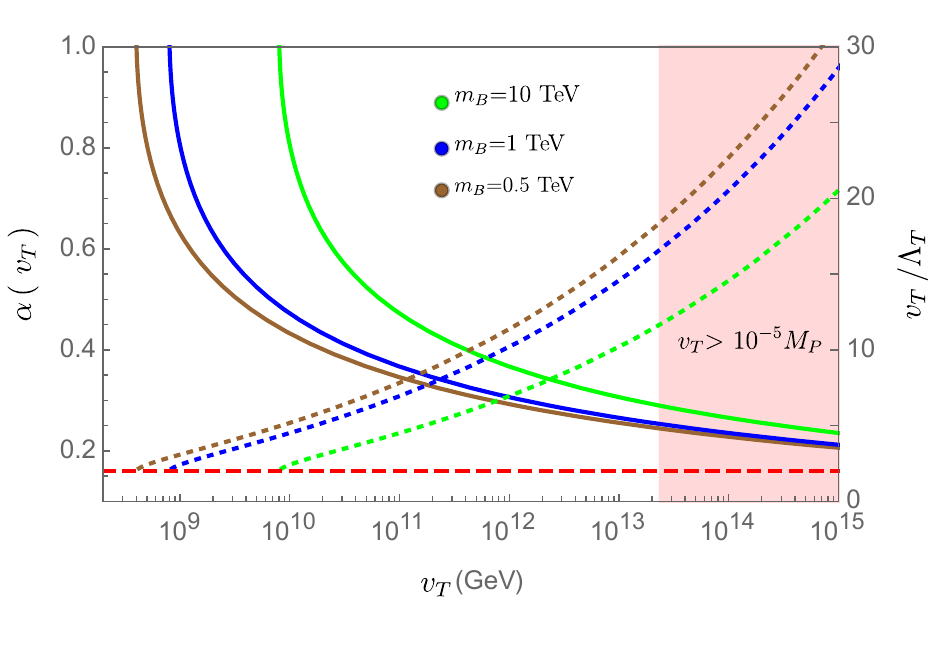}  
\caption{
  The values of the gauge coupling $\alpha(v_T)$ (solid lines) and the ratio $v_T/\Lambda_T$ (dashed lines) as a function of the symmetry breaking scale $v_T$ needed to obtain the required bottom Yukawa coupling in the $SU(3)_T$ gauge extension with a vector-like quark, assuming $\lambda_B=1 $.
  The red region corresponds to a possible constraint from CP violating operators at the Planck scale which can shift the $\bar{\theta}_T$ angle from its QCD aligned value. The horizontal, dashed red line corresponds to the minimum possible value of $v_T/\Lambda_T=e^{2\pi/(53/6)}\simeq 2$.
  }
\label{fig:bottom yukawa}
\end{figure}

To fix the problem of the minimal setup and enhance the bottom Yukawa coupling, we introduce vector-like quarks $B$ and $\bar{B}$, where $\bar{B}$ has the same gauge charges as $\bar{d_3}$, with the following couplings and masses,
\begin{align}
{\cal L} = - m_B B\bar{B} - y_t  q_3 \bar{u}_3 H - \lambda_B q_3 \bar{B} \widetilde{H}  +{\rm h.c.}.
\end{align}
The phases in the Yukawa coupling and mass terms can be removed by rotations of the fermions ${\bar B}, q_3 $, and $  {\bar u}_3$, so that $m_B, y_t$ and $\lambda_B$ are real parameters. The $SU(3)_T$ instanton generates a Dirac mass for $\bar{d}_3$ and $B$,
\begin{align}
{\cal L} = m_{Bd} B \bar{d}_3+{\rm h.c.}\,,
\end{align}
where
\begin{align}
m_{Bd}&\simeq  
2 \times 10^{-4}\,\frac{y_t\lambda_B}{24\pi^2}  \,\, e^{-\frac{2\pi}{\alpha_T(v_T)} }\left( \frac{2\pi}{\alpha_T(\mu)}\right)^{6}
     \int \frac{d\rho}{\rho^5} \,  \frac{4\pi^2}{3}\rho^{3}  \,\left(v_T\rho\right)^{53/6}\,
        e^{-4\pi^2v_T^2\,\rho^2}\,,\nonumber\\
     &   \simeq 10^{-4}\,\lambda_B\, v_T \left(\frac{\Lambda_T}{v_T}\right)^{\frac{53}{6}} \left({\rm ln } \frac{\mu}{\Lambda_T }\right)^6~.
\end{align}
The would be confinement scale, $\Lambda_T$, is defined as before in \eqref{eq:Lambda_T_def}, with $b_T=53/6$ due to the gauge, fermion and Higgs contributions. 
When $m_{Bd} < m_B$, we may integrate out $B\bar{B}$ and the bottom Yukawa in the low energy effective theory is $y_b = \lambda_B m_{Bd}/m_B$.
In figure~\ref{fig:bottom yukawa}, we show the required value of $\alpha_T(v_T)$ (or equivalently $v_T/\Lambda_T$) as a function of $v_T$ for $\lambda_B=1$ and a few choices of $m_B$. In the red-shaded region, CP-violating higher dimensional operators such as $G_TG_T\widetilde{G}_T$, where $G_T$ is the $SU(2)_T$ gauge field strength, can introduce too large a strong CP phase~\cite{Bedi:2022qrd}, assuming the cutoff scale around the Planck scale.%
\footnote{
We assume that any combinatoric factors arising from the color and Lorentz indices are absorbed into the definition of the cutoff scale, $M_{\rm UV}$, of these operators.}
One can see that the observed bottom Yukawa coupling can be obtained for $\alpha_T(v_T)\lesssim 1$ and $v_T\gg \Lambda_T$, for which the instanton computation is reliable. Note that the same mechanism with vector-like quarks can be applied to the model in Ref.~\cite{Agrawal:2017evu} to obtain a sufficiently large bottom, strange, and up Yukawa couplings from the top, charm, and down Yukawa couplings, respectively. 
To obtain the CKM mixing, $q_3$ must couple to the second and first generation right-handed quarks. This can be achieved by the following coupling
\begin{align}
\label{eq:SU(3) CKM}
 {\cal L} = y^B_a B \bar{d}_a\Phi_3 +{\rm h.c.},
\end{align}
where $a=1,2$ is the flavor index and $\Phi_3$ is the $SU(3)_T\times SU(3)$ breaking Higgs field. The phases in $y_a^B$ cannot be removed and a particular linear combination of them is responsible for the CKM phase. After integrating out the vector-like fermions, we obtain generational mixing,
\begin{align}
\label{eq:SU(3) CKM IR}
{\cal L} = -\lambda_B y^B_a \frac{ v_T}{m_B}q_3 \bar{d}_a \widetilde{H}+{\rm h.c.}\,.
\end{align}
For $m_B \sim$ TeV, $v_T\sim 10^{13}$ GeV and $\lambda_B y^B_a\sim 10^{-12}$, the CKM mixing of the third generation with the first two generations can then be explained. 
\subsubsection{Extra $SU(2)$}
\label{sec:SU(2)}

We embed $SU(2)_L$ into $SU(2)_{FS}\times SU(2)_T$ where the first two generations of left-handed fermions are charged under $SU(2)_{FS}$ while the third generation left-handed fermions are charged under $SU(2)_T$.  The product group $SU(2)_{FS}\times SU(2)_T$ is then broken down to $SU(2)_L$ by the VEV of a pseudo-real bifundamental Higgs $\Phi_2$. Furthermore, there is an $SU(2)_{FS}$ doublet Higgs $H_{FS}$ and $SU(2)_T$ doublet Higgs $H_T$ whose VEVs give masses to the first two generations of fermions and the third generation, respectively. The Lagrangian is given by
\begin{align}
{\cal L}= - \widetilde{Y}^u_{3a} \Psi_{q_3} \Psi_{\bar{u}_a} H_T - \widetilde{Y}^d_{3a} \Psi_{q_3} \Psi_{\bar{d}_a} \widetilde{H}_{T} - \widetilde{Y}^u_{ia} \Psi_{q_i} \Psi_{\bar{u}_a} H_{FS} -  \widetilde{Y}^d_{ia} \Psi_{q_i} \Psi_{\bar{d}_a} \widetilde{H}_{FS} + {\rm h.c.},\, \nonumber \\
\label{eq:3genLSU(2)}
\end{align}
where $i=1,2$ and $a=1,2,3$. The vector-like fermion can have Yukawa interactions with $H_T$ and/or $H_{FS}$ depending on their gauge charges.  The two Higgses also mix with each other via the interaction
\begin{align}
{\cal L} = A \Phi_2 H_{FS} H_T^\dag + {\rm h.c.}
\end{align}
The coupling $A$ may be taken real by the phase rotation of $H_T$, so that the VEVs of $H_T$ and $H_{FS}$ are real.
Besides the CKM phase, there remain complex phases in the Yukawa couplings in Eq.~\eqref{eq:3genLSU(2)} which can induce corrections to the strong CP phase. These will be considered in section~\ref{sec:CPcorrections}.

We assume that $SU(2)_{FS}\times SU(2)_T$ symmetry breaking occurs below the $Sp(2N)$ confinement scale and that the $SU(2)_T$ gauge coupling $g_{2,T}$ is much larger than the $SU(2)_{FS}$ coupling $g_{2,FS}$. Then the $(2,2)$ entry of the mass-squared matrix in Eq.~\eqref{eq:mass_matrix_compositeH} receives a large positive correction $\sim g_{2,T}^2 \LSp^2/(16\pi^2)$.

When $g_{2,T}^2 > 4\pi \sqrt{N} y_3 $, the small electroweak scale is achieved by requiring $m_H^2$ to be smaller than the $(2,2)$ entry, so that the Higgs field which couples to the third generation below the confinement scale is dominantly $H_T$, and the top Yukawa is $y_3$. This case does not require large $y_3$, but requires large $g_{2,T}$. An $SU(2)$ gauge theory with the number of flavors between 6 and 11 is considered to be in the conformal window~\cite{Sannino:2009aw,Lee:2020ihn}. In our case, this implies that
$12 < 2N+4\leq 22$, since we have $2N+3$ grand-color copies of $\Psi_{q_3}$ as well as the third generation SM lepton doublet. Thus, $SU(2)_T$ can flow into a conformal fixed point above the $Sp(2N)$ confinement scale for $4<N\leq 9$. Near the lower edge of the window, the fixed point value of $g_{2,T}$ is large.  Below the confinement scale, since the number of $SU(2)_T$ charged fields decreases, the $SU(2)_T$ gauge coupling increases, causing $SU(2)_T$ to eventually confine. This means the $SU(2)_{FS}\times SU(2)_T$ symmetry breaking should occur before $SU(2)_T$ confines. Therefore, since $g_{2,T}$ at $\LSp$ is required to be large, the $SU(2)_{FS}\times SU(2)_T$ symmetry breaking scale should be just below $\LSp$.

When $g_{2,T}^2 < 4\pi \sqrt{N} y_3 $, the Higgs field that couples to the third generation below the confinement scale is dominantly $\Pi^{H}$ and the top Yukawa coupling is approximately $g_{2,T}^2 / (4\pi \sqrt{N})$. The required value of $g_{2,T}^2$ is the same as the $g_{2,T}^2 > 4\pi \sqrt{N} y_3$ case with the bound saturated.

\subsection{Corrections to the strong CP phase}
\label{sec:CPcorrections}

We next discuss possible corrections to the strong CP phase that occur in the three-generation models. The quantum correction below the $Sp(2N)$ confinement scale due to the CKM phase arises at the seven-loop level and therefore is negligible~\cite{Ellis:1978hq}. However, there can be threshold corrections near the $Sp(2N)$ confinement scale, $\LSp$. These corrections are model dependent and nontrivially depend on $m_U$. To simplify the analysis
we consider a case where the SM Yukawa coupling of the quarks $q$, $\bar{u}$, $\bar{d}$ is non-zero above the confinement scale and a new massless $(m_U=0)$ Dirac fermion $U$ and $\bar{U}$  obtains a mass by the $Sp(2N)$ dynamics. A more general analysis for all $m_U$ is given in appendix \ref{app:flavor_inv} where we explicitly show that the corrections to the strong CP phase remain small by constructing the possible flavor invariant combinations that can appear in the corrections.  
Note that $\bar{U}$ and $\bar{u}_1$ in this subsection correspond to $\bar{u}_1$ and $\bar{U}$ in section~\ref{sec:Two_gen_wVLQ} with $m_U=0$, respectively. 
Let us for now ignore the issue of a large electroweak symmetry breaking scale and consider a three-generation model without any extra gauge interactions. A non-zero strong CP phase can arise from the VEVs of the $Sp(2N)$ pions, which are determined by the Yukawa couplings. The pion interactions from the Yukawa couplings are given by
\begin{align}
     - \frac{\sqrt{N}}{4\pi}\LSp^2 Y^u_{i a} \widehat{\Pi}_{q_i \bar{u}_a}H  -  \frac{\sqrt{N}}{4\pi} \LSp^2Y^d_{i a} \widehat{\Pi}_{q_i \bar{d}_a}\widetilde{H} + {\rm h.c.}, 
\end{align}
where $\widehat{\Pi}_{XY}$ is the pion containing the $Sp(2N)$ fermions $\psi_X$ and $\psi_Y$. The $\widehat{\Pi}_{q\bar{d}}$ pion is related to $\widehat{\Pi}_{q\bar{u}}$ by the quark condensation,
\begin{align}
   \widehat{\Pi}_{q_i \bar{d}_b} = \frac{\vev{\psi_{q_i} \psi_{q_j} }}{\CSp} \frac{\vev{\psi_{\bar{d}_b} \psi_{\bar{u}_a} }}{\CSp}\widehat{\Pi}^\dagger_{q_j \bar{u}_a}.
\end{align}
Note that $\psi_U$ and $\psi_{\bar{U}}$ do not have Yukawa couplings and hence do not form condensates with $\psi_{\bar{u},\bar{d}}$.
The Higgs-pion loop then gives rise to a potential
\begin{align}
\label{eq:V3gen}
V \simeq  \frac{N \LSp^4}{(16\pi^2)^2} Y^u_{ia}Y^d_{jb} \frac{\vev{\psi_{q_i} \psi_{q_j} }}{\CSp} \frac{\vev{\psi_{\bar{d}_b} \psi_{\bar{u}_a} }}{\CSp} + {\rm h.c.}
= - \frac{N \LSp^4}{(16\pi^2)^2} {\rm Tr}\left[Y^u \Sigma_{\bar{u}\bar{d}} (Y^d)^T \Sigma_q\right]e^{i \theta_{\eta}} + {\rm h.c.}\,,
\end{align}
where $\Sigma_{\bar{u}\bar{d}}$ and $\Sigma_q$ are the non-linear sigma fields corresponding to the flavor symmetries $SU(3)_{\bar{u}}\times SU(3)_{\bar{d}}/SU(3)$ and $SU(3)_q/SO(3)$, respectively, and $SU(3)_{\bar{u},\bar{d},q}$ are the flavor symmetry groups
of the SM quarks $\bar{u}$, $\bar{d}$, and $q$.
The condensation of $\psi_U\psi_{\bar{U}}$ enters via
the pion field $\theta_\eta$ that
corresponds to the $U(1)$ symmetry $\psi_{q_i}(1)$, $\psi_{\bar{u}_a}(1)$, $\psi_{\bar{d}_a}(1)$, $\psi_U(-6)$, $\psi_{\bar{U}}(-6)$.
The potential \eqref{eq:V3gen} determines the alignment of the neutral pions.
If $\theta_\eta$ obtains a non-zero VEV, the mass term of $U\bar{U}$ 
generated by $\vev{\psi_U \psi_{\bar{U}}}$ will obtain a complex phase. 
 This potential was numerically minimized in Ref.~\cite{Valenti:2022tsc}, which found that the VEV of $\theta_\eta$ remains nearly zero
and is suppressed by small Yukawa couplings and the CKM mixing angles. 
Therefore, in our setup the correction to the strong CP phase will also be
smaller than the experimental upper bound
$\bar\theta\lesssim 10^{-10}$.

We next discuss the extensions of the model that include an extra $SU(3)$ or $SU(2)$ gauge group.
In the $SU(3)$ extension, only the first and second generations have $Sp(2N)$ charged fermions. Thus, we can remove the phases of the Yukawa couplings of the first two generations, and the $Sp(2N)$ dynamics does not generate any new phase. There are non-zero phases in Eq.~\eqref{eq:SU(3) CKM IR}, but they do not generate the strong CP phase at leading-order. This is because the down Yukawa coupling matrix obtained from \eqref{eq:2gen_vlq} and \eqref{eq:SU(3) CKM IR} is given by
\begin{align}
    Y^d = \begin{pmatrix}
    \widetilde{Y}^d_{11} & \widetilde{Y}^d_{12} & 0 \\
    \widetilde{Y}^d_{21} & \widetilde{Y}^d_{22} & 0 \\
    \lambda_B y_1^B\frac{v_T}{m_B} & \lambda_B y_2^B\frac{v_T}{m_B} & y_b
    \end{pmatrix}\,,\label{eq:SU3T_Yuk}
\end{align}
where $y_b$ is generated from $SU(3)_T$ instantons.
The determinant of $Y^d$ does not depend on the complex parameters $y_{1,2}^B$ and is real.
There can be higher-order threshold corrections when $B,\bar{B}$ are integrated
out, but the corrections are suppressed by the smallness of the Yukawa couplings and CKM mixings, similar to the SM corrections.

In the $SU(2)$ extension, assuming $m_{H_{\rm FS}}\gg \LSp$, we can integrate out $H_{\rm FS}$ to obtain the effective theory with Lagrangian
\begin{align}
{\cal L}= - \widetilde{Y}^u_{3a} \Psi_{q_3} \Psi_{\bar{u}_a} H_T - \widetilde{Y}^d_{3a} \Psi_{q_3} \Psi_{\bar{d}_a} \widetilde{H}_{T} - \frac{A \Phi_2}{m_{H_{FS}}^2} \widetilde{Y}^u_{ia} \Psi_{q_i} \Psi_{\bar{u}_a} H_{T} -\frac{A \Phi_2}{m_{H_{FS}}^2}  \widetilde{Y}^d_{ia} \Psi_{q_i} \Psi_{\bar{d}_a} \widetilde{H}_{T} + {\rm h.c.}\,, \nonumber \\
\end{align}
where $i=1,2$ and $a=1,2,3$.
The contribution from the corrections where $\Phi_2$ is treated as a  background field with a VEV is equivalent to the case without any $SU(2)$ extension by identifying $\widetilde{Y}_{3a}^{u,d}\rightarrow Y_{3a}^{u,d}$ and $A \vev{\Phi_2}\widetilde{Y}_{ia}^{u,d}/m^2_{H_{FS}} \rightarrow Y_{ia}^{u,d}$, and does not generate too large a strong CP phase.
However, the corrections with a dynamical $\Phi_2$ field can be different.
In fact, when the $\Phi_2$ legs are closed into a loop instead of taking $\Phi_2$ VEVs, we obtain 
{$A^2\widetilde{Y}^u_{ia} \widetilde{Y}^d_{jb}/16\pi^2 m_{H_{FS}}^4 \sim Y^u_{ia} Y^d_{jb}/ \LSp^2$, }which leads to a factor of $\LSp^2/(16\pi^2 \vev{\Phi_2}^2)$ in comparison with the VEV contribution. Since $\vev{\Phi_2} \lesssim\LSp$,
the loop contribution is smaller than the VEV contribution. 

Although the total pion potential is different from the case without an $SU(2)$ extension, the strong CP phase should remain small. This can be seen by taking the basis where
$\Psi_{q_{1,2}}$ couples only to $\Psi_{\bar{u}_{1,2}}$ and $\Psi_{\bar{d}_{1,2}}$ with real Yukawa couplings and
$\widetilde{Y}^{u,d}_{33}$ are real. In this basis, the determinant of the SM Yukawa couplings is real and the strong CP phase is proportional to $\vev{\theta_\eta}$. Since $\Phi_2$ only couples to $\Psi_{q_{1,2}}$, the extra corrections from the $\Phi_2$ loop only modifies the product of the real Yukawa couplings in Eq.~\eqref{eq:V3gen}.
Furthermore, the $\Phi_2$ loop correction is subdominant in comparison with the VEV contribution, and therefore the VEV of $\theta_\eta$ should remain small.

\subsection{Phenomenology}
\label{sec:pheno}
In this section we discuss the phenomenological implications of our scenario.
Recall that since the third generation fermions are treated differently from the first two generations,
we can still employ the mechanisms discussed in section~\ref{sec:Two_gen_wVLQ} to generate the up quark mass (or the mass of new vector-like fermions). In particular, the spectra of pions and vector-like quarks will have interesting experimental consequences.  

\subsubsection{Pion spectrum}

In section~\ref{sec:Two_gen_wVLQ}, vector-like quarks were introduced with a mass term $m_U$. The $Sp(2N)$ dynamics  generates a mixing between the vector-like quark and $\bar{u}_1$, given by $m_{Uu}$. The $Sp(2N)$ dynamics spontaneously breaks the flavor symmetry, giving rise to a pion spectrum.
For both $m_U=0$ and $m_U\neq 0$, the pion corresponding to the spontaneous breaking of the baryon symmetry remains massless and does not couple to the photon or gluon. We discuss the possibility of this pion being the dark matter in section~\ref{sec:conclusion}.

The next-to-lightest pions may also have phenomenological implications. We first discuss the case with $m_U=0$, for which they may be below the electroweak scale. There are three next-to-lightest pions, which are associated with the symmetries restored in the limit $y_{u,d}\rightarrow 0$, have masses $m_{\Pi_{\rm NL}}\sim\sqrt{y_u y_d}\,\LSp/(4\pi)\sim 10^{-6}\LSp$ (see Eq.~\eqref{eq:pion mass} with $\lambda_U \rightarrow y_u$), but two of them do not have anomalous couplings to SM gauge bosons.
One of them, corresponding to the $U(1)$ symmetry $\Psi_{\bar{U}}(1)$, $\Psi_{\bar{d}_1}(1)$, $\Psi_{U}(-1)$, and $\Psi_{\bar{u}_1}(-1)$, appears in the Yukawa and mass terms of $\bar{u}_1$, $\bar{d}_1$, $U$, and $\bar{U}$ via the four-fermion operators in Eq.~\eqref{eq:GC_int}. Performing a phase rotation of $U\bar{u}_1$ then generates anomalous couplings to photons and gluons.\footnote{It may appear that since the $U(1)$ symmetry, which is a part of an anomaly-free flavor symmetry, does not have an $Sp(2N)$ or $SU(3)_c$ anomaly, the corresponding pion does not couple to gluons. However, the shift symmetry is explicitly broken by the Yukawa interactions and consequently, the pion interactions cannot be determined purely by a symmetry argument. 
}
Note that the contributions from the phase rotation of $\bar{U}$ and $\bar{d}_1$ are negligible, since the up and down Yukawa couplings are dominated by the Yukawa couplings that already exists before the $Sp(2N)$ confinement. 
For $\LSp \sim 10^{6-7}$ GeV, this pion has a mass $\sim 1-10$~GeV and the couplings with photons and gluons are suppressed by $f  \simeq\sqrt{N}\LSp/(4\pi) \sim 10^{6-7}$ GeV. Such a particle can be discovered by axion-like particle searches at DUNE~\cite{Kelly:2020dda}.

It is important to note that this pion can also be considered as a ``heavy QCD axion" in the following sense.
Since $\Psi_U,\Psi_{\bar{u}_1}$ are massless, the theory has an anomalous PQ symmetry, which can remove the strong CP phase.
The PQ symmetry is then spontaneously broken by the $Sp(2N)$ dynamics yielding a NGB that corresponds to the phase direction of $\psi_U\psi_{\bar{u}_1}$ and couples to the $SU(3)_c$ gluon. Because of the explicit PQ breaking by the $Sp(2N)$ anomaly, the NGB appears to obtain a mass solely by the $Sp(2N)$ dynamics.
However, a linear combination of the PQ and
quark chiral symmetry does not have an $Sp(2N)$ anomaly, but is instead explicitly broken by the Yukawa interactions, giving rise to a pseudo-NGB with mass $\sim \sqrt{y_u y_d}\,\LSp/(4\pi) $ rather than $\LSp$.
This state, which may be referred to as a ``heavy QCD axion", is a mixture of the phase direction of $\psi_U\psi_{\bar{u}_1}$ and the $ Sp(2N)$ $\eta'$ meson.

Our setup can be compared with the conventional light QCD axion models that have a similar feature where the light axion state is also a mixture of the NGB resulting from the spontaneous PQ breaking and the $SU(3)_c$ $\eta', \eta$, and $\pi^0$ mesons. A slight difference, however, is that in light QCD axion models, the spontaneous PQ symmetry breaking occurs by dynamics not related to the explicit PQ breaking dynamics, while in our setup, the spontaneous and explicit PQ breaking dynamics are unified and occur simultaneously at the same scale.

In~\cite{Valenti:2022tsc}, instead of introducing massless fermions $\Psi_U,\Psi_{\bar{u}_1}$, there is a NGB that couples to the $SU(2N+3)$ gauge field, where the associated PQ symmetry is spontaneously broken at a scale $f_a$, not related to the $SU(2N+3)$ strong dynamics. The heavy QCD axion obtains a mass $\sim \sqrt{y_u y_d}\LSp/(4\pi) \times f/f_a$. In the limit $f_a = f$, this QCD axion has a similar property to our pion.

For $m_U\neq 0$, the next-to-lightest pions receive masses from the $m_U$ mass term that are at least as large as ${\cal O} (\sqrt{m_U \LSp})$.
As $m_U$ is increased to be $\gtrsim \LSp$, the lightest pion associated with the $U(1)$ symmetry $\Psi_{\bar{U}}(1)$, $\Psi_{\bar{d}_1}(1)$, $\Psi_{U}(-1)$, and $\Psi_{\bar{u}_1}(-1)$ becomes the $Sp(2N)$ $\eta'$ since the state behaving as a ``heavy QCD axion" decouples. This is a formal limit, but mimics the massless up quark solution in QCD, where the $SU(3)_c$ $\eta'$ would be the lightest neutral meson in the formal limit of decoupling the down quark. Thus, even with $m_U\sim$ TeV, it might be possible to discover the $\eta'$-like pions 
in this ``massless up-quark limit".

For large $m_U$, the next-to-lightest pions are associated with the breaking $SU(2)_u\times SU(2)_d \rightarrow SU(2)$, where $SU(2)_u$ and  $SU(2)_d$ act on $\Psi_{\bar{u}_{1,2}}$ and $\Psi_{\bar{d}_{1,2}}$, respectively, and the breaking of a generation non-universal $U(1)$ baryon symmetry under which $\Psi_{q_1}(1)$, $\Psi_{q_2}(-1)$, $\Psi_{\bar{u}_1}(-1)$, $\Psi_{\bar{d}_1}(-1)$, $\Psi_{\bar{u}_2}(1)$, and $\Psi_{\bar{d}_2}(1)$.  
The masses of the former are approximately $\sqrt{y_c y_s}\LSp/4\pi \sim 10^{-4} \LSp$. One of them, associated with a $U(1)$ subgroup $\Psi_{\bar{u}_1}(1)$, $\Psi_{\bar{d}_1}(1)$, $\Psi_{\bar{u}_2}(-1)$, $\Psi_{\bar{d}_2}(-1)$, couples to the gluon and photon. Note that dynamical scales of $\LSp \lesssim 10^4$ GeV are excluded by beam-dump experiments (see~\cite{Kelly:2020dda} for a summary) while $\LSp \sim 10^4-10^5$ GeV can be probed by the LHC~\cite{Hook:2019qoh}. 
The mass of the latter is further suppressed by $\sqrt{\sin\theta_c}$, which is only an ${\cal O}(1)$ factor.

Two of these next-to-lightest pions have flavor-violating couplings to the first two generation quarks, which are constrained by the rare decay of $K$ and $D$ mesons if the next-to-lightest pions are lighter than them. Since $D$ mesons are heavier, we discuss the constraint from the decay of $D$ into the next-to-lightest pions, which is kinematically allowed if $\LSp \lesssim 10^4$ GeV.
The flavor-violating coupling can arise from the following two contributions. One is due to the four-fermion interaction $q_2\bar{u}_1 \psi_{q_2} \psi_{\bar{u}_1}$ in Eq.~\eqref{eq:GC_int} which induces a Yukawa coupling between
$q_2$, $\bar{u}_1$, and $\widehat{\Pi}_{ {q}_2 \bar{u}_1}$. 
 The scalar 
$\widehat{\Pi}_{ {q}_2 \bar{u}_1}$ couples to $H$ and $\widehat{\Pi}_{\bar{u}_1\bar{d}_2}$
via $y_s$, and after electroweak symmetry breaking, mixes with $\widehat{\Pi}_{\bar{u}_1\bar{d}_2}$. 
A second contribution arises from the coupling $\bar{u}_2^\dag \bar{\sigma}^\mu \bar{u}_1 \psi_{\bar{u}_1}^\dag \bar{\sigma}_\mu \psi_{\bar{u}_2}$, where $\psi_{\bar{u}_1}^\dag \bar{\sigma}_\mu \psi_{\bar{u}_2}$ can be identified as 
$f \partial_\mu \widehat{\Pi}_{\bar{u}_1\bar{d}_2}$. 
We find that the former contribution dominates and the Yukawa coupling between $q_2$, $\bar{u}_1$, and $\widehat{\Pi}_{\bar{u}_1\bar{d}_2}$ is
\begin{align}
\frac{g_{\rm GC}^2}{M_{\rm GC}^2}(4\pi)^2 m_c f.
\end{align}
The rare $D$ meson constraints derived in~\cite{MartinCamalich:2020dfe} requires $(4\pi)^2 g_{\rm GC}^2 f / M_{\rm GC}^2 < 10^{-8}$ GeV$^{-1}$.
With this lower bound on $M_{\rm GC}$, we cannot obtain a sufficiently large quark mass for $\LSp < 10^4$ GeV;
$m_{Uu}$ in Eq.~\eqref{eq:mUu_def} is below 1 TeV, and the up Yukawa coupling in Eqs.~\eqref{eq:up_Yukawa_wVLQ} is smaller than the observed value. We conclude that $\LSp \gtrsim 10^4$ GeV is required, so that the next-to-lightest pions are heavier than $D$ meson.

\subsubsection{Long-lived relics}

When the vector-like quarks introduced in
section~\ref{sec:Two_gen_wVLQ} are massless ($m_U=0$), the theory possesses an accidental $U(1)$ symmetry with charges $\Psi_U(1)$ $\Psi_{\bar{u}_1}(-1)$, giving rise to long-lived particles. The lightest $U(1)$ charged particles are the quarks $U$ and $\bar{u}_1$ with a mass $m_{Uu}$ or a pion, $\widehat{\Pi}_{U\bar{U}}$ made from $\psi_U$ and $\psi_{\bar{U}}$, with a mass $m_{\Pi_{\rm NL}}\sim \sqrt{y_u y_d}\LSp/(4\pi)\sim 10^{-6} \LSp$.

For $m_{Uu}< 10^{-6} \LSp$, the quarks $U$ and $\bar{u}_1$ are stable.
They will be abundantly produced in the early universe by the $SU(3)_c$ gauge interaction and will form bound states with the up quark. The bound states have strong interactions with nucleons, and such particles might be excluded by direct detection experiments~\cite{Digman:2019wdm}.%
\footnote{\label{fn:EMchargedSt}
Furthermore, these states can be bound  with protons to become electromagnetic charged states, which are subject to even stronger constraints~\cite{Dunsky:2018mqs}. 
}
The vector-like quarks $\bar{u}_1$ and $U$ may decay via a higher-dimensional operator
$\Psi_U \Psi_{\bar{U}} (\Psi_U \Psi_{\bar{u}_1})^\dag/M_{\rm UV}^2$ or  $\Psi_U \Psi_{\bar{U}} \Psi_U \Psi_{\bar{u}_1}/M_{\rm UV}^2$. This operator induces a tadpole term of $\widehat{\Pi}_{U\bar{U}}$, which mixes $\bar{U}$ with $\bar{u}_1$ via the four-fermion operator in Eq.~\eqref{eq:GC_int}. The mixing angle is
\begin{align}
\theta_{\bar{U}\bar{u}_1}\sim \frac{N}{(4\pi)^2} \frac{\LSp^4}{M_{\rm UV}^2 m_{\Pi_{\rm NL}}^2} \sim 10^{-4} \left(\frac{\LSp}{10^9~{\rm GeV}}\right)^2 \left(\frac{10^{16}~{\rm GeV}}{M_{\rm UV}}\right)^2  \frac{N}{4}.
\label{eq:thUu1}
\end{align}
The decay rate of $U$ and $\bar{u}_1$ is $0.1 \,\theta_{\bar{U}\bar{u}_1}^2y_u^2 m_{Uu}$. The decay occurs before BBN if 
\begin{align}
\theta_{\bar{U}\bar{u}_1} > 10^{-8} \left(\frac{\rm TeV}{m_{Uu}}\right)^{1/2},
\end{align}
which requires
\begin{align}
\label{eq:MUVup_decay}
    M_{\rm UV} < 10^{18}~{\rm GeV} \left(\frac{\LSp}{10^9~{\rm GeV}}\right) \left(\frac{m_{Uu}}{\rm TeV}\right)^{1/4} \left(\frac{N}{4}\right)^{1/2}.
\end{align}
Hence, for $m_{Uu}>$ TeV corresponding to $\LSp> 10^9$ GeV,
the Planck-suppressed operator leads to rapid enough decay of $U$ and $\bar{u}_1$. Note that the mixing angle \eqref{eq:thUu1} is enhanced because of the smallness of $m_{\Pi_{\rm NL}}$. Therefore, operators involving the second generation quarks such as $\Psi_U \Psi_{\bar{u}_2} (\Psi_U \Psi_{\bar{u}_1})^\dag/M_{\rm UV}^2$ 
lead to smaller decay rates because of the larger mass of the pion made of $\psi_U$ and $\psi_{\bar{u}_2}$ is proportional to $\sqrt{y_c y_s}$. 

For $m_{Uu}> 10^{-6} \LSp$, $\widehat{\Pi}_{U\bar{U}}$ is stable. Unless the reheating temperature of the universe is much below $f$, the pion will be abundantly produced from the thermal bath and over-close the universe. It can decay via the operator $\Psi_U \Psi_{\bar{U}} (\Psi_U \Psi_{\bar{u}_1})^\dag/M_{\rm UV}^2$ or $\Psi_U \Psi_{\bar{U}} \Psi_U \Psi_{\bar{u}_1}/M_{\rm UV}^2$, which mixes $U$ with $\bar{U}$, and the $\widehat{\Pi}_{U\bar{U}}\mathchar`-U\bar{U}$ interaction arising from the four-fermion operator in Eq.~\eqref{eq:GC_int}. The resulting interaction is  
\begin{align}
{\cal L} \sim \frac{\sqrt{N}}{4\pi} \frac{\LSp^3}{M_{\rm UV}^2 m_{\Pi_{\rm NL}}^2} \partial_\mu\widehat{\Pi}_{U\bar{U}} \bar{U}^\dag \bar{\sigma}^\mu\bar{U} + {\rm h.c.},
\end{align}
and $\widehat{\Pi}_{U\bar{U}}$ can decay into a pair of up quarks.
Requiring that the decay occurs before BBN, we obtain
\begin{align}
    M_{\rm UV} < 10^{14}~{\rm GeV} \left(\frac{\LSp}{10^7~{\rm GeV}} \right)^{3/4}\left(\frac{N}{4}\right)^{1/4}.
\end{align}
Thus, a sufficiently rapid decay of $\widehat{\Pi}_{U\bar{U}}$ occurs when $M_{\rm UV}$ is much below the Planck scale.

\subsection{Accidental chiral symmetry}
\label{sec:twogen_accidental}

The third generation model in section~\ref{sec:extragaugeint} assumes that there are two massless SM quarks ${\bar u}_1$ and ${\bar d}_3$, associated with two anomalous chiral symmetries. These anomalous symmetries may arise accidentally due to an exact symmetry that can be promoted to a gauge symmetry.

In the two generation model of section~\ref{sec:Two_gen_wVLQ} we introduce a $\mathbb{Z}_2$ symmetry which can accidentally realize the chiral symmetry responsible for a massless quark.
We first discuss the quality of this symmetry assuming $m_U=0$, for which there exists light pions and $\bar{\theta}$ can be more easily shifted by higher-dimensional operators.
We impose a $\mathbb{Z}_2$ symmetry on $\Psi_U$, which can be exact
if an extra vector-like quark pair $\Psi_D$ and $\Psi_{\bar{D}}$ is added with an odd $\mathbb{Z}_2$ charge assigned to one of the quarks. 
These charge assignments allow
a higher-dimensional operator $(\Psi_U \Psi_{\bar{u}_1})^2/M_{\rm UV}^2$ whose coefficient is generically complex. This operator generates a tadpole term for the heavy QCD axion which then gives a complex mass to $U$ and $\bar{u}_1$, generating a non-zero $\bar{\theta}$:
\begin{align}
\label{eq:deltatheta}
\bar{\theta} \sim \frac{N \LSp^4}{M_{\rm UV}^2 m_{\Pi_{\rm NL}}^2}.
\end{align}
Requiring $\bar{\theta}<10^{-10}$, we obtain
\begin{align}
\label{eq:Mlow}
    M_{\rm UV} \gtrsim 10^{19}~{\rm GeV} \left(\frac{\LSp}{10^9~{\rm GeV}}\right) \left(\frac{N}{4}\right)^{1/2}.
\end{align}
For $M_{\rm UV}$ near the Planck scale, $\LSp< 10^8$ GeV is required.

A large scale $M_{\rm UV}$ in Eq.~\eqref{eq:Mlow}, however, is in tension with the requirement on the suppression scale from long-lived relics in Eq.~\eqref{eq:MUVup_decay}. This requires either a low reheating temperature of the universe so that the long-lived relics are not abundantly produced, or more symmetry should be introduced to control higher-dimensional operators. For example, there can be a flavor symmetry under which $\Psi_{\bar{U}}$ and $\Psi_{\bar{u}_1}$ have the same charge, 
so that the operator $\Psi_U \Psi_{\bar{U}} (\Psi_U \Psi_{\bar{u}_1})^\dag$ is allowed while $(\Psi_U \Psi_{\bar{u}_1})^2$ is not.
Alternatively, instead of $\mathbb{Z}_2$, we may impose a $\mathbb{Z}_3$ symmetry on $\Psi_U$, which forbids $(\Psi_U \Psi_{\bar{u}_1})^2$. The $\mathbb{Z}_3$ symmetry can be exact if one more extra massless vector-like quark pair is introduced. The $\mathbb{Z}_3$ symmetry then also solves the domain wall problem as discussed in section~\ref{sec:toy} and \ref{sec:Two_gen_wVLQ}.

For $m_U\neq 0$, we may construct a $\mathbb{Z}_2$ symmetric model without stable particles at the renormalizable level.
For example, we introduce two vector-like quark pairs $\Psi_U,\Psi_{\bar{U}}$ and $\Psi_D,\Psi_{\bar{D}}$ with Dirac mass terms $m_{U}\Psi_U\Psi_{\bar{U}} + m_{D\bar{d}_i} \Psi_D\Psi_{\bar{d}_i} $, and impose an anomaly-free $\mathbb{Z}_2$ symmetry on $\Psi_{\bar{u}_1}$ and $\Psi_{\bar{D}}$. Note that we impose an odd $\mathbb{Z}_2$ charge on $\Psi_{\bar{D}}$ rather than on $\Psi_{D}$, so that the Dirac mass terms of $\Psi_D,\Psi_{\bar{d}_i}$ are allowed.
In this setup, there is no unbroken symmetry that can prevent the decay of new particles and therefore all new particles are unstable.
The up and down Yukawa couplings are generated in the same way as in section~\ref{sec:Two_gen_wVLQ}, as long as the masses $m_{D\bar{d}_i}$ are sufficiently small, otherwise a reanalysis of the vacuum structure is required. There exists a domain wall problem, as discussed in Secs.~\ref{sec:toy} and \ref{sec:One_gen_wVLQ}, which means that the reheating temperature should be below $\LSp$, or the model should be extended with a $\mathbb{Z}_3$ symmetry.

Finally, for the third generation sector of the model in section~\ref{sec:extragaugeint},
it is possible to understand the anomalous chiral symmetry of $\bar{d}_3$ as an accidental symmetry, by
imposing a $U(1)$ symmetry under which $\bar{d}_3$ has charge $k$. Furthermore, we introduce $k$ pairs of vector-like fermions $D$ and $\bar{D}$ with $U(1)$ charges $0$ and $-1$, respectively. This $U(1)$ symmetry does not have an $SU(3)_T$ anomaly and is an exact symmetry. It is also possible to promote this $U(1)$ global symmetry to a gauge symmetry by adding extra $SU(3)_T$ neutral and $U(1)$ charged fermions to cancel the $U(1)^3$ gauge anomaly. The $U(1)$ symmetry is then assumed to be broken by the VEV of an operator ${\cal O}$, with $U(1)$ charge $+1$, to give masses to $D$ and $\bar{D}$ via the coupling ${\cal O} D \bar{D}$. The coupling or mass terms containing $\bar{d}_3$ instead require couplings with ${\cal O}^{k\dagger}$, so for sufficiently large $k$, the chiral symmetry of $\bar{d}_3$ is maintained to be of good quality. For example, when ${\cal O}$ is a fundamental scalar field and ${\langle{\cal O}\rangle} \sim 10^{14}$ GeV, the operator $({\cal O}^{\dagger}/M_{\rm Pl})^k q_3 \bar{d}_3 \widetilde{H}$ can be sufficiently suppressed by requiring $k>2$. In this way, the chiral symmetry of $\bar{d}_3$ can be understood as an accidental symmetry in the low energy theory.

\section{Other types of models}
\label{sec:Higgs}

So far we have considered the simplest case where the chiral symmetry arises from a massless quark that obtains a mass by grand-color strong dynamics via the exchange of heavy gauge bosons. In this section, we discuss other classes of models that contain extra Higgses charged under the chiral symmetry or fermions in higher representations of $SU(2N+3)$.

\subsection{Extra Higgses}

We first consider models that contain extra Higgs fields.
In particular, we impose a chiral symmetry on an extra Higgs doublet as well as some of the quarks as in the Weinberg-Wilczek model~\cite{Weinberg:1977ma,Wilczek:1977pj}. The quarks charged under the chiral symmetry do not couple to the SM Higgs and only couple to the extra Higgs doublet. The $Sp(2N)$ dynamics spontaneously (and explicitly) breaks the chiral symmetry and generates a mixing between the Higgs doublets. After electroweak symmetry is broken by the SM Higgs, a VEV is induced for the extra Higgs which generates masses for the chirally-charged quarks. 
In this setup, the extra Higgs doublet mediates the chiral symmetry breaking from $Sp(2N)$ to $SU(3)_c$, so $M_{\rm GC}$ does not have to be close to $\LSp$ and hence $N$ does not have to be large.
Instead, the extra Higgs should be sufficiently light since the VEV of the extra Higgs is inversely proportional to its mass squared.
The light extra Higgses will introduce another Higgs mass fine-tuning problem which should be eventually addressed, e.g., by supersymmetry or anthropic requirements on the Yukawa couplings.

\begin{figure}[h]
    \centering
    \begin{tikzpicture}[scale=0.70]
    \begin{feynhand}
    \setlength{\feynhandblobsize}{12mm}

 \vertex  (h1)  [ dot] at (3,0) {}; 
 \vertex  (h2)  [ dot] at (-3,0) {};  
 
 \vertex  (h3)   at (5,0){$H_b$}; 
 \vertex  (h4)   at (-5,0) {$H$};  
 
    \vertex (v1) at (0.707107/2,0.707107/2);
     \vertex (v2) at (0.707107/2,-0.707107/2);
       \vertex (v3) at (-0.707107/2,0.707107/2);
        \vertex (v4) at (-0.707107/2,-0.707107/2);
    \propag[fer] (h2) to [half left, edge label=$\bar{u}_3$](v3);
     \propag[fer] (h2) to [half right, edge label'=$q_3$](v4);
      \propag[fer] (h1) to [half right, edge label'=$\bar{d}_3$](v1);
     \propag[fer] (h1) to [half left, edge label=$q_3$](v2);
     \propag[sca] (h1) to (h3); \propag[sca] (h2) to (h4);
       \vertex[grayblob] (tv) at (0,0) {};
       \node at (0,0) {{\footnotesize $SU(3)_T$ }};
    \end{feynhand}
    \end{tikzpicture}
    \caption{The Feynman diagram showing how the mass mixing between the two Higgs fields, $H_b$ and $H$, is generated via $SU(3)_T$ instantons.
    }
    \label{fig:Higgs_mixing}
\end{figure}
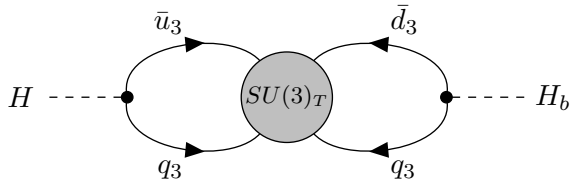

The model still suffers from having too large an electroweak scale by Higgs-pion mixing, but the $SU(3)$ or $SU(2)$ extension can fix this problem.
For the $SU(3)$ extension, instead of forbidding the bottom Yukawa coupling, we can in total introduce three Higgs doublets $H$, $H_b$, and $H_u$ and impose two chiral symmetries,
\begin{align}
U(1)_b:~& \bar{d}_3(1),~ H_b(-1),\nonumber \\
U(1)_u:~& \Psi_{\bar{u}_{1}}(1),~H_u(-1),
\end{align}
to remove the strong CP phases of $SU(2N+3)$ and $SU(3)_T$. The Yukawa interactions are
\begin{align}
    {\cal L} = - y_t q_3 \bar{u}_3 H - \tilde{y}_b q_3 \bar{d}_3 H_b -  \widetilde{Y}^u_{ia} \Psi_{q_i} \Psi_{\bar{u}_a} H - \widetilde{Y}^d_{ia} \Psi_{q_i} \Psi_{\bar{d}_a} \widetilde{H} -\tilde{y}^u_i \Psi_{q_i}\Psi_{\bar{u}_1} H_u + {\rm h.c.}\,,
\end{align}
where $\widetilde{Y}^u_{i1}=0$, $i=1,2$ and $a=1,2,3$.
The $SU(3)_T$ instanton effect generates the mixing $H H_b$ (see figure~\ref{fig:Higgs_mixing}), while the $Sp(2N)$ dynamics generates a mixing $H_u \widetilde{H}$. After $H$ obtains a VEV, these mass mixing terms induce VEVs for $H_b$ and $H_u$. The CKM mixing arises from introducing vector-like fermions $B,\bar{B}$ charged under $SU(3)_T$ that couples to $q_3 \widetilde{H}$ and mix with $\bar{d}_{1,2}$ by $SU(3)^2$ breaking into $SU(3)_c$.

Other variants of the model are  possible. For example, we can eliminate $H_b$ and generate the bottom Yukawa by $SU(3)_T$ instantons as in section~\ref{sec:extrasu3}. Alternatively, instead of imposing a chiral symmetry on $\Psi_{\bar{u}_1}$, we can impose a chiral symmetry on $\Psi_{\bar{d}_1}$ and introduce $H_d$ that couples to $\Psi_{\bar{d}_1}$.

For the $SU(2)$ extension, instead of a single $H_{FS}$, we introduce $H_{FS}$ and $H_{FS,u}$ and impose a chiral symmetry with the following charge assignment,
\begin{align}
   \Psi_{\bar{u}_{1}}(1), H_{FS,u}(-1).
\end{align}
The $Sp(2N)$ dynamics then generates the mixing $H_{FS}^\dag H_{FS,u}$. A VEV of $H_{FS}$ induces a VEV of $H_{FS,u}$ to generate the up quark mass.

We may also construct a model analogous to the model in~\cite{Hook:2014cda} by introducing a pair of vector-like quarks $\Psi_Q,\Psi_{\bar{Q}}$ and a complex scalar field $S$ with a $U(1)$ chiral symmetry
\begin{align}
    \Psi_Q(1),~\Psi_{\bar{Q}}(0),~S(-1).
\end{align}
These fields couple to each other via a Yukawa coupling
\begin{align}
    {\cal L}= - \lambda_S \Psi_Q \Psi_{\bar{Q}}S + {\rm h.c.} = - \lambda_S \psi_Q \psi_{\bar{Q}}S -\lambda_S Q \bar{Q}S + {\rm h.c.}\,,
\end{align}
where $\Psi_Q=(Q, \psi_Q)$ (and similarly for $\Psi_{\bar Q}$) are the grand-color multiplets.
With the $U(1)$ chiral rotation, the strong CP phase can be removed and is unphysical.
The $Sp(2N)$ dynamics generates a $\psi_Q \psi_{\bar{Q}}$ condensate, which then induces a VEV of $S$ and hence a non-zero mass for $Q$ and $\bar{Q}$.

With extra Higgses, it is also possible to utilize instanton effects to generate sufficiently large quark masses.
For example, let us impose a chiral symmetry on the right-handed up quark and introduce an extra Higgs that is charged under the chiral symmetry. The up quark couples to the extra Higgs while the other quarks couple to the SM Higgs. The gauge group is $SU(N+3)$, which breaks down to $SU(N)\times SU(3)$, and $SU(N)$ is further Higgsed down. The $SU(N)$ instanton effect generates the mixing between the Higgses proportional to the product of the Yukawa couplings. For a sufficiently large $SU(N)$ symmetry breaking scale, a large $SU(N)$ gauge coupling at that scale, and a light extra Higgs, the observed up Yukawa can be obtained. These type of scenarios have the advantage that the difficulty pointed out in section~\ref{sec:difficulty} is absent.

\subsection{Higher fermion representations}

If all massless $SU(3)_c$ charged particles are also charged under $G_{c'}$, as in the model~\cite{Hook:2014cda} with bi-fundamental fermions, the chiral symmetry of these fermions can be directly broken by the $G_{c'}$ dynamics. In our $SU(2N+3)$ grand-color model, this can be achieved by the rank 3 anti-symmetric tensor representation of $SU(2N+3)$ and its anti-representation, which is similar to the rank 3 anti-symmetric tensor representation of $SU(6)$ considered in~\cite{Gaillard:2018xgk,Gherghetta:2020ofz}.  
In this type of model, there is no need for large $N$ or extra Higgses.

However, because the anti-symmetric rank 3 tensor representation contributes significantly to the $\beta$-function of the gauge coupling, it remains to be shown that $G_{c'}$ exhibits chiral symmetry breaking, rather than flowing to a conformal fixed point. The contribution of the anti-symmetric three tensor representation to the $\beta$-function is $\Delta b = -(2N)(2N+1)/3$. Adding the contribution from the $Sp(2N)$ gauge boson and the three generations, $Sp(2N)$ is asymptotically free if $N=1$ or $2$ (assuming the extra $SU(3)_T$ model to remove the third generation fermions does not change the range of $N$). The value $N=1$ leads to $\LSp< \Lambda_{\rm QCD}$ and does not work. Instead, $N=2$ can have $\LSp> \Lambda_{\rm QCD}$, but given that the theory is near the boundary between asymptotically free and non-free, it seems plausible that $Sp(2N)$ does not exhibit chiral symmetry breaking and rather flows into a conformal fixed point.

\section{Dark matter}
\label{sec:darkmatter}
In this section we discuss the possibility that the lightest pion arising from the $Sp(2N)$ confinement can provide the missing dark matter component of the Universe.
An interesting feature of $Sp(2N)$ strong dynamics is that there are no stable $Sp(2N)$ baryons and most of the $Sp(2N)$ pions are unstable, except for one;
in all models, the SM baryon symmetry is spontaneously broken by the $Sp(2N)$ dynamics. The baryon symmetry can be explicitly broken by introducing a small coupling between the $SU(2N+3)$ quark and the grand-color breaking field. This generates a tiny mass for the corresponding NGB, which can then be dark matter.

The dark matter can be produced in the early universe from the thermal bath.
The dark matter couples to hypercharge and $SU(2)_L$ gauge bosons through the anomaly of the baryon symmetry. 
When the reheating temperature $T_{\rm RH}$ is above the weak scale, the dark matter is dominantly produced at $T=T_{\rm RH}$ with a rate $\sim g_2^2(\alpha_2/4\pi)^2 T^3 / (8\pi f^2)$, where $ g_2$ is the $SU(2)_L$ gauge coupling. If $T_{\rm RH}$ is below the weak scale, the production still dominantly occurs at $T=T_{\rm RH}$ but via scattering with a photon and an off-shell $Z$ boson exchange. The rate is suppressed by a factor of $(T_{\rm RH}/m_Z)^4$, where $m_Z$ is the $Z$ boson mass. The resultant dark matter abundance is
\begin{align}
\frac{\rho_{\rm DM}}{s} \simeq 0.4~{\rm eV} \left( \frac{10^{10}~{\rm GeV}}{f} \right)^2 \left(\frac{m_{\rm DM}}{1~{\rm MeV}}\right) \left(\frac{T_{\rm RH}}{10^3~{\rm GeV}}\right){\rm min}\left(1, \left(\frac{T_{\rm RH}}{m_Z}\right)^4\right)\,,
\end{align}
where $m_{\rm DM}$ is the dark matter mass. Note that 
dark matter is sufficiently cold for $m_{\rm DM}\gtrsim 25$ keV~\cite{Yeche:2017upn}.

The dark matter can also be produced via the misalignment mechanism~\cite{Preskill:1982cy, Dine:1982ah,Abbott:1982af}, which gives
\begin{align}
\frac{\rho_{\rm DM}}{s} \simeq 0.4~{\rm eV}  \left( \frac{f}{10^{12}~{\rm GeV}} \right)^2 \left(\frac{T_{\rm RH}}{10^4~{\rm GeV}}\right) \theta_i^2 \,{\rm min}\left(1, \frac{\sqrt{m_{\rm DM} M_{\rm Pl}}}{T_{\rm RH}}\right),
\end{align}
where $\theta_i$ is the initial misalignment angle.
The first/second case in ``min" corresponds to the beginning of oscillations before/after the completion of reheating.

%======================================================
\begin{figure}[t!]
      \centering
\includegraphics[width=0.6\textwidth]{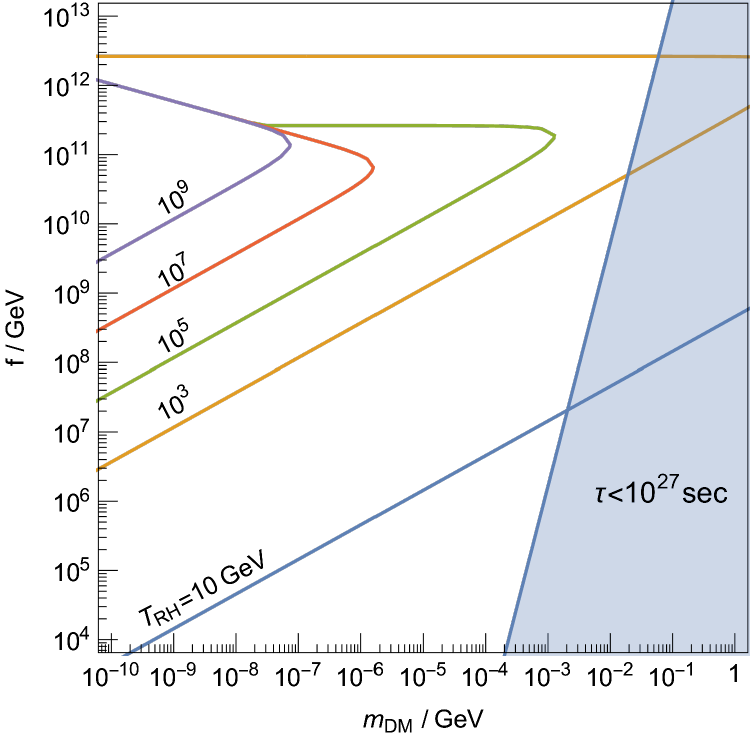}
\caption{
Constraints on the dark matter mass $m_{\rm DM}$ and the decay constant $f$. The blue-shaded region is excluded by gamma-ray and X-ray observations. The contours show the required reheating temperature $T_{\rm RH}$ to explain the observed dark matter abundance by the misalignment mechanism (horizontal or negatively sloped parts of the contours) or production from the thermal bath (positively sloped parts.)}
\label{fig:mDMf}
\end{figure}
%=======================================================

The parameter space is constrained by indirect detection experiments.
For the dark matter mass below the weak scale, the dark matter can decay into a photon and a pair of SM fermions via off-shell $Z$ boson exchange
with a rate
\begin{align}
\Gamma \simeq \frac{g_2^2}{128\pi^3}\left(\frac{e g_2}{16\pi^2}\right)^2 \frac{m_{\rm DM}^7}{f^2m_Z^4},
\end{align}
where $e $ is the electromagnetic gauge coupling.
For $m_{\rm DM}=0.1-10^4$ MeV, which is the relevant mass range for setting limits, 
indirect-detection constraints from X-ray and gamma ray observatories require $\Gamma^{-1}\gtrsim 10^{27}$~sec~\cite{Essig:2013goa}.

In Fig.~\ref{fig:mDMf}, we show the bound on $(m_{\rm DM},f)$. For a given reheating temperature, $(m_{\rm DM},f)$ on each contour can explain the observed dark matter abundance.
On the horizontal and negatively sloped segments of the contours, the misalignment mechanism determines the dark matter abundance, with the oscillation of dark matter begins before and after the completion of reheating, respectively. On the positively sloped segments, scattering from the thermal bath determines the abundance. Therefore, on the positively sloped  segments of the contours, only $m_{\rm DM}> 25$ keV provides sufficiently cold dark matter. The blue-shaded region is excluded by indirect-detection experiments.

\section{Discussion and Summary}
\label{sec:conclusion}

In this paper, we have proposed a new massless quark solution to the strong CP problem by embedding the QCD group,
$SU(3)_c$, into a larger, simple (grand-color) gauge group. The chiral symmetry of the massless quark makes the $\theta$ term unphysical and the grand color gauge interaction preserves the CP symmetry.
The chiral symmetry can be realized as an accidental symmetry at low energies, arising from 
an exact discrete symmetry at UV scales.
The grand color gauge group is then spontaneously broken down to $SU(3)_c \times G_{c'}$. The $G_{c'}$ strong dynamics
exhibits chiral symmetry breaking, which is transferred to the $SU(3)_c$ charged fermions by dimension-six, four-fermion operators generated by the exchange of heavy gauge bosons, and the massless quark obtains a non-zero mass. Furthermore, we showed that the strong CP phase of $SU(3)_c$ remains below the experimental upper bound after including quantum corrections from the Yukawa interactions.

In the simplest class of models, the electroweak vacuum becomes unstable due to a mixing between the Higgs and a NGB leading to a tachyonic direction. This problem can be avoided in several ways such as by introducing a mass term from grand-color breaking, introducing supersymmetry, or extending the gauge group.
 The first solution requires a non-trivial flavor structure of the mass term, while the supersymmetric case will be analyzed in future work. Instead, we focused on extending the gauge group, and considered two possibilities.
In the first model, the third-generation fermions are charged under an $SU(2)_T$ gauge symmetry with a large gauge coupling where quantum corrections proportional to the gauge coupling stabilizes the tachyonic direction. The first and second generation fermions are charged under a different $SU(2)_{FS}$ gauge symmetry with a weak coupling and the two $SU(2)$'s are broken down to $SU(2)_L$.
In the second model, the third generation fermions are charged under $SU(3)_T$ and 
not under $SU(2N+3)$. This means there are no $Sp(2N)$ charged partners of the third generation quarks and the tachyonic direction is absent. The $SU(3)_T$ and $SU(3)\subset SU(2N+3)$ groups are then broken down to $SU(3)_c$. The possible strong CP phase from the extra $SU(3)$ is removed because of a massless bottom Yukawa coupling at UV scales. The bottom Yukawa coupling is generated by $SU(3)_T$ instantons. We also considered a class of models with extra Higgses that are charged under the chiral symmetry. 
The chiral symmetry breaking in the $Sp(2N)$ dynamics is communicated to $SU(3)_c$-charged fermions by the Higgses, so large $N$ is not necessary. Another possibility is a model where all $SU(3)_c$ massless fermions are also charged under $Sp(2N)$, so that the chiral symmetry breaking in $Sp(2N)$ is directly communicated to $SU(3)_c$. It remains, however, to be shown that the $Sp(2N)$ dynamics breaks chiral symmetry, rather than the theory flowing to a conformal fixed point.

Even though our massless quark solution has no light axion there may still be phenomenological signals. In some models, one of the NGBs has a mass $\sim 10^{-6}\LSp$ that couples to gluons and photons. For $\LSp\sim 10^6$ GeV, the mass is $\sim 1$ GeV and may be discovered at DUNE. In fact, this pseudo-NGB behaves as a ``heavy QCD axion" that arises from the mixing between one of the NGBs and the Sp(2N) $\eta'$. This is similar to what occurs in usual QCD axion models, except that the pseudo-NGB (or heavy axion) is composite and the spontaneous breaking of the PQ symmetry occurs via the same dynamics that also has an $Sp(2N)$ anomaly. 
Furthermore, the vector-like quarks were crucial to obtain viable scenarios. Since they are charged under QCD with possible masses near the TeV scale, they could potentially be directly produced at colliders and provide a hint for our mechanism. 
There are also cosmological consequences of our model. Since $N$ is large, we expect that the confinement of $Sp(2N)$ is associated with a first-order phase transition~\cite{Holland:2003kg}, which may produce primordial gravitational waves. While interesting, a more detailed study of the phase transition is beyond the scope of this work. In addition, the lightest NGB which arises from the spontaneous breaking of baryon number can be a decaying dark matter candidate.

We comment on possible future directions. It will be interesting to embed the theory into supersymmetric theories. Supersymmetry can solve the problem of Higgs-pion mixing and may also (partially) explain the small mass scales in the theory such as the electroweak scale, grand-color symmetry breaking scale, and the extra Higgs masses. Supersymmetric extensions, however, may have extra CP phases in the masses or couplings of superpartners, and thus it remains to be carefully checked whether the corrections to the strong CP phase from these possible phases are sufficiently suppressed. It will also be interesting to use the $Sp(2N)$ dynamics to solve the problems in (beyond) the Standard Model in addition to the strong CP problem. 
At the very least, it provides a new massless up-quark type solution to the strong CP problem, providing an intriguing alternative to a light axion.

\section*{Acknowledgments}
We thank Raymond Co for collaborating in the early stages of this work and Luca Vecchi for useful discussions.
The work of R.B. and T.G. is supported in part by the Department of Energy under Grant No.~DE-SC0011842 at the University of Minnesota. K.H. is supported by the Department of Energy under Grant No.~DE-SC0025242 at the University of Chicago, a Grant-in-Aid for Scientific Research from the Ministry of Education, Culture, Sports, Science, and Technology (MEXT), Japan (20H01895), and by World Premier International Research Center Initiative (WPI), MEXT, Japan (Kavli IPMU).
T.G. also acknowledges the Aspen Center for Physics, which is supported by National Science Foundation grant PHY-2210452, where part of this work was performed.

\appendix
\section{Grand color symmetry breaking}
\label{sec:GC_breaking}

In this appendix, we present details of the grand color symmetry breaking. There are two ways to break the grand color group either as a two-step breaking considered in Eq.~\eqref{eq:GC_breaking_pattern} or just  breaking the grand color group in a single step. We show that the latter is not possible.

We first discuss the two-step symmetry breaking in Eq.~\eqref{eq:GC_breaking_pattern}. The first symmetry breaking, $SU(2N+3)\rightarrow SU(2N)\times SU(3)_c \times U(1)$, can be achieved by the VEV of scalar field transforming in the adjoint representation of $SU(2N+3)$, which is proportional to
\begin{align}
\begin{pmatrix}
3 \mathbb{I}_{2N} & \\
 & -2N \mathbb{I}_{3}
\end{pmatrix}\,,
\end{align}
where $\mathbb{I}_{2N}\, (\mathbb{I}_{3})$ is a $2N\times 2N \, (3\times 3) $ identity matrix.
It is shown in~\cite{Ruegg:1980gf} that this VEV can be the absolute minimum of the scalar potential. The second symmetry breaking in Eq.~\eqref{eq:GC_breaking_pattern}, $SU(2N)\times U(1)\times U(1)_{Y'} \rightarrow Sp(2N)\times U(1)_Y$, is achieved by a scalar field, $A_{ij}$, transforming as a rank 2 anti-symmetric tensor of $SU(2N+3)$ with a non-zero $U(1)_{Y'}$ charge. Under $SU(2N)\times SU(3)$, $A_{ij}$ decomposes into a rank 2 anti-symmetric tensor $B_{ij}$ of $SU(2N)$, a $\bar{3}$ of $SU(3)$, and a bi-fundamental of $SU(2N)\times SU(3)$. By coupling the adjoint representation of $SU(2N+3)$ to $A_{ij}$, we can generate a negative mass squared for $B_{ij}$ while keeping the mass squared of all other terms positive, so that only $B_{ij}$ obtains a non-zero VEV among the components of $A_{ij}$.

Next, we show that the VEV of $B_{ij}$ that breaks $SU(2N)$ into $Sp(2N)$ is the absolute minimum of the potential~\cite{Li:1973mq,Kim:1980ec}. The scalar potential at the renormalizable level is 
\begin{align}
\label{eq:VB}
V = - m_{B}^2 {\rm Tr}\left(B B^\dag\right) + \lambda_{B1}  \left({\rm Tr}\left(B B^\dag\right)\right)^2 + \lambda_{B2} {\rm Tr}\left(B B^\dag B B^\dag\right).
\end{align}
Note that the first two terms depend on the norm of $B_{ij}$, so the extremization of the potential is determined by the last term with the norm fixed. The
$B_{ij}$ VEV can be generically parameterized as
\begin{align}
\vev{B} =
\begin{pmatrix}
i \sigma_2 b_1 & & \\
 & \ddots  & \\
  & & i \sigma_2 b_N
\end{pmatrix},~~ b_i \geq 0\,.
\end{align}
Fixing the norm of $B$ to be $v_B$ via $b_N^2=v_B^2- \sum_{i=1}^{N-1} b_i^2$,
we extremize the following function of $b_i$ ($i=1,\dots,N-1$),
\begin{align}
 {\cal F}(b_i) = \sum_{i=1}^{N-1} b_i^4 + \left(v_B^2- \sum_{i=1}^{N-1} b_i^2 \right)^2,
\end{align}
which corresponds to the last term of Eq.~\eqref{eq:VB}.
The derivative of ${\cal F}$ with respect to $b_i$ is
\begin{align}
\frac{\partial {\cal F}}{\partial b_i} =
4 b_i \left( b_i^2 -b_N^2\right)~,
\end{align}
so the potential is extremized at $b_i=0$ or $b_i^2=b_N^2$ for $i=1,\dots, N-1$.
This means that any non-zero $b_i$ should take the same value.
Without loss of generality, we take $n\leq N$ nonzero $b_i$, with the remainder zero i.e.
\begin{align}
b_1 =\cdots =b_{n}\equiv b,\qquad b_{n+1} = \cdots = b_{N}=0,
\qquad b^2 = \frac{v_B^2}{n}\,.
\end{align}
The last term in Eq.~\eqref{eq:VB} is then proportional to
\begin{align}
\lambda_{B2} n b^4 = \lambda_{B2} \frac{v_B^4}{n}. 
\end{align}
When $\lambda_{B2} >0$, the absolute minimum of the potential occurs for the largest possible value of $n$,
i.e. $n=N$. Thus, $SU(2N)$ is spontaneously broken to $Sp(2N)$.

We next discuss the alternative possibility of breaking the grand color group with a one-step symmetry breaking~\cite{Jetzer:1983ij}
\begin{equation}
    SU(2N+3)\times U(1)_{Y'}\to Sp(2N)\times SU(3)_c\times  U(1)_Y.
\end{equation}
This can be achieved by the following VEV of $A_{ij}$,
\begin{align}
\label{eq:AVEV}
\vev{A} \propto
\begin{pmatrix}
i \sigma_2 & & & \\
 & \ddots  & & \\
  & & i \sigma_2 & \\
  &&& 0_{\bf 3}
\end{pmatrix}.
\end{align}
where $0_{\bf 3}$ is a zero $3\times 3$ matrix.
However, the VEV in Eq.~\eqref{eq:AVEV} is a saddle point rather than a local minimum and is not stable. To see this, we can extend the previous analysis on the breaking, $SU(2N)\rightarrow Sp(2N)$ to $SU(2N+3)$. Since $2N+3$ is odd, the absolute minimum occurs by taking the $(2N+3)$th diagonal entry of $A$ to be zero and the remaining $2N+2$ entries nonzero. This gives the parametrization
\begin{align}
\vev{A} =
\begin{pmatrix}
i \sigma_2 a_1 & & & & \\
 & \ddots  & & & \\
  & & i \sigma_2 a_N & & \\
  &&& i \sigma_2 a & \\
&&&&0
\end{pmatrix}.
\end{align}
The potential of $A$ is
\begin{align}
\label{eq:VA}
V &= - m_A^2 {\rm Tr}\left(A A^\dag\right) + \lambda_{A1}  \left({\rm Tr}\left(A A^\dag\right)\right)^2 + \lambda_{A2} {\rm Tr}\left(A A^\dag A A^\dag\right) \nonumber \\
&= -2m_A^2 \left(\sum_{i=1}^N a_i^2 +a^2\right) +4 \lambda_{A1} \left(\sum_{i=1}^N a_i^2 +a^2\right)^2 + 2\lambda_{A2} \left(\sum_{i=1}^N a_i^4 +a^4\right). 
\end{align}
We consider the extremum with Eq.~\eqref{eq:AVEV},  where $a_i^2=m_A^2/(4 \lambda_{A1} N+2\lambda_{A2})\equiv v_A^2$ for $i=1,\dots,N$ and $a^2=0$. At this extremum, the Hessian of the potential  is
\begin{align}
 32\lambda_{A1}
 \begin{pmatrix}
 1 & \cdots & 1 & 0 \\
 \vdots & \ddots & \vdots & 0 \\
 1 & \cdots & 1 & 0 \\
 0&0&0&0 
 \end{pmatrix} v_A^2
 +8\lambda_{A2}\begin{pmatrix}
 2 \mathbb{I}_N&0 \\
 0 & -1
 \end{pmatrix} v_A^2\,,
\end{align}
with the $N+1$ eigenvalues 
\begin{align}
    16(2 N \lambda_{A1}+ \lambda_{A2} )v_A^2,~16\lambda_{A2} v_A^2,~\cdots, 16\lambda_{A2} v_A^2,~ - 8 \lambda_{A2} v_A^2.
\end{align}
Clearly, for $\lambda_{A2} >0$ there is one negative eigenvalue, while for $\lambda_{A2} <0$ there are $N-1$ negative eigenvalues. Thus, in both cases, the VEV in Eq.~\eqref{eq:AVEV} is not stable. For $|\lambda_{A2}|\ll 1$, quantum corrections due to gauge interactions will determine the stability of the VEV. 
However, these corrections should favor the VEV with maximal residual gauge symmetry, i.e., $a_1\neq 0$,~$a_2 = a_3 = \cdots a_N=a=0$. We conclude that the one-step gauge symmetry breaking is not possible.

\section{Proof that \texorpdfstring{$\bar{\theta}=0$}{} using the quark picture}
\label{sec:proof}

In this appendix, we argue that the strong CP phase of QCD is zero rather than $\pi$ using the quark picture. We choose the $Sp(2N)$ sign convention so that 
\begin{align}
\Psi_f \Psi_{\bar{f}}\equiv
\Psi_f^A \Psi_{\bar{f},A} = f^a \bar{f}_a + J_{ij} \psi_f^i \psi_{\bar{f}}^j \equiv f \bar{f} + \psi_f \psi_{\bar{f}},
\end{align}
where $A$, $a$, $i$ are $SU(2N+3)$, $SU(3)$, and $Sp(2N)$ indices, respectively and $J_{ij}$ is an anti-symmetric $Sp(2N)$ invariant tensor.

\subsection{One flavor toy model}
\label{sec:proof1}

We consider the one flavor toy model in section~\ref{sec:toy} and show that $\vev{\psi_{u} \psi_{\bar{u}}}<0$, which is the key point. Since we are interested in the vacuum of the theory, we use the Euclidean path-integral. 
Introducing a mass $m$ for the vector-like fermions $\psi_{u},\psi_{\bar{u}}$, the partition function is defined as
\begin{align}
Z(m) = \int D\psi_u D\psi_{\bar{u}}DA~e^{- S_E -\int d^4x_E\,\left(m \psi_u \psi_{\bar{u}} + {\rm h.c.}\right) }\,,
\end{align}
where $x_E$ are Euclidean spacetime coordinates and $S_E$ is the Euclidean action of the gauge bosons $A$ and quarks $\psi_{u},\psi_{\bar{u}}$ not including the mass terms. With $\theta$ set to zero above the $SU(N)$ confinement scale, the Euclidean action $S_E$ is real. 
Then
\begin{align}
\vev{\psi_u \psi_{\bar{u}}} = - V^{-1}\left.\frac{1}{Z}\frac{\partial Z}{\partial m}\right|_{m=0},
\label{eq:Z2pt}
\end{align}
where $V$ is the spacetime Euclidean volume.%
\footnote{Note that since the chiral symmetry is explicitly broken by the strong dynamics, 
the order of the limits $V\rightarrow\infty$ and $m\rightarrow 0$ is not crucial.}
Performing the path integral over fermions, we obtain
\begin{align}
Z(m) = \int DA~{\rm det}(\slashed{D} + m) = \int DA \prod_{\lambda_n >0}m^{n_0(A)}(\lambda_n^2(A) + m^2),
\end{align}
where $n_0$ is the number of fermion zero modes and $i\lambda_n$ are the non-zero eigenvalues of the Dirac operator $\slashed{D}$.
The only non-zero contribution to $\partial Z/\partial m|_{m=0}$ arises from gauge field configurations with $n_0=1$, and in this case $\partial Z/\partial m|_{m=0} >0$. In addition, $Z(0)$ is also positive, which then using \eqref{eq:Z2pt} gives the result $\vev{\psi_u \psi_{\bar{u}}} <0$.
\subsection{One-generation models}

\label{sec:proof2}
We first consider the minimal one-generation model in section~\ref{sec:onegenmin} where the key point required that $\vev{\psi_{q_u}\psi_{\bar{u}}}<0$.
To show this, we assume $\theta =0$, and use the fact that the $Sp(2N)$ vacuum is
connected with $\vev{\psi_{q_u} \psi_{\bar{u}}} = \vev{\psi_{q_d} \psi_{\bar{d}}} = - \CSp$ by a non-anomalous flavor transformation, chosen to be an $SO(2)$ rotation: $(\psi_{q_d},\psi_{\bar{u}})\to (\psi_{\bar{u}}, -\psi_{q_d})$.
In the electroweak symmetric limit, the vacuum is $\vev{\psi_{q_u} \psi_{q_d}} = - \vev{\psi_{\bar{u}} \psi_{\bar{d}}} = \CSp$.
With a non-zero VEV of $H$, a nonzero condensate 
$\vev{\psi_{q_d} \psi_{\bar{d}}}<0$ is induced, since $y_d >0$.
Given that this vacuum is connected to the $Sp(2N)$ vacuum via $\vev{\psi_{q_u} \psi_{\bar{u}}} = \vev{\psi_{q_d} \psi_{\bar{d}}}$ by the $SO(2)$ rotation, the desired negative value of $\vev{\psi_{q_u} \psi_{\bar{u}}}$ is obtained.

Next we consider the one-generation model with vector-like quarks in section~\ref{sec:One_gen_wVLQ}. 
The case with $m_U < m_{Uu}$, where the SM up quark is mostly $\bar{U}$, reduces to the analysis of the minimal one-generation model,
so let us analyze the case with 
$m_U >   m_{Uu}$. In this case, the SM right-handed up quark mainly comes from $\bar{u}$, so choosing $\theta=0$ means that the vacuum of $Sp(2N)$ is connected with $\vev{\psi_{q_u} \psi_{\bar{u}}} = \vev{\psi_{q_d} \psi_{\bar{d}}} = \vev{\psi_U\psi_{\bar{U}}}= -\CSp$ by a non-anomalous flavor transformation.\footnote{Note that the simple exchange of $\bar{U}$ with $\bar{u}$ involves an anomalous chiral rotation by $\pi$ so one should be careful about which fermion is identified with a SM quark and the meaning of $\theta$ for a given choice.}
Such a vacuum is parameterized by
\begin{align}
 \vev{\psi_{q_u}\psi_{q_d}} = \CSp,~~
    \begin{pmatrix}
    \vev{\psi_{\bar{d}} \psi_{\bar{u}}} &
    \vev{\psi_{\bar{d}} \psi_{\bar{U}}}
    \\
    \vev{\psi_{U} \psi_{\bar{u}}} &
    \vev{\psi_{U} \psi_{\bar{U}}}
    \end{pmatrix} = \CSp \begin{pmatrix}
    \sin\phi & \cos\phi \\
    \cos\phi &- \sin\phi
\end{pmatrix}.
\label{eq:1genvacpara}
\end{align}

The exchange of the Higgs neutral component at the confinement scale, $\LSp$ generates an effective operator
\begin{align}
 {\cal L} \sim \frac{\lambda_U y_d}{\Lambda_{\rm Sp}^2}\left(\psi_{q_u}\psi_{\bar{U}} \right) \left(\psi_{q_d} \psi_{\bar{d}}\right).
 \label{eq:effop_1gen}
\end{align}
Since the effective operator \eqref{eq:effop_1gen} depends on the strong dynamics, we cannot precisely determine the magnitude of the coefficient of the operator, but we can still determine the sign.
The condensation $\vev{\psi_{q_u}\psi_{q_d}}=  \CSp$ gives
\begin{align}
{\cal L}\sim \lambda_U y_d\Lambda_{\rm Sp} \psi_{\bar{d}} \psi_{\bar{U}} \equiv - V_{\rm eff}\,.
\end{align}
To minimize the energy (or $V_{\rm eff}$), we require $\vev{\psi_{\bar{d}} \psi_{\bar{U}}}>0$, so that using \eqref{eq:1genvacpara} we obtain $\vev{\psi_U \psi_{\bar{u}}} >0$. For this sign, the mass generated by the dimension-six operator \eqref{eq:4FermiUbaru} is negative, i.e. $m_{Uu}<0$. After integrating out $U\bar{U}$, the up Yukawa coupling obtained from \eqref{eq:up_Yukawa_wVLQ} is given by $- \lambda_U m_{Uu}/ m_U$ and is therefore positive.

\subsection{Two-generation models}
 {In the minimal two generation case, we can again rely on the arguments presented in section~\ref{sec:proof2} for each generation to conclude that $\bar{\theta}=0$. Starting from the vacuum $\vev{\psi_{q_u} \psi_{\bar{u}}},\, \vev{\psi_{q_d} \psi_{\bar{d}}},\,\vev{\psi_{q_c} \psi_{\bar{c}}},\, \vev{\psi_{q_s} \psi_{\bar{s}}} <0$, we can perform a baryon number and an $SO(2)$ rotation to obtain
$\vev{\psi_{q_u} \psi_{q_d}} = - \vev{\psi_{\bar{u}} \psi_{\bar{d}}} = \CSp$. Similarly, we can perform an $SO(2)$ rotation of $(\psi_{q_s},\psi_{\bar{c}})$ to obtain $\vev{\psi_{q_c} \psi_{q_s}}= - \vev{\psi_{\bar{c}} \psi_{\bar{s}}}$. Note that the relative sign of $\vev{\psi_{q_u} \psi_{q_d}}$ and $\vev{\psi_{q_c} \psi_{q_s}}$ does not need to be fixed a priori, since the two vacua are connected by a (non-anomalous) flavor transformation.}

The two-generation model with vector-like quarks in section~\ref{sec:Two_gen_wVLQ} reduces to the case with one generation and a vector-like quark, so the proof in appendix~\ref{sec:proof2} is applicable.

\section{Four-fermion operators due to gauge boson exchange}
\label{app:4fermiops}
In this appendix we derive the interactions mediated by the massive grand-color gauge bosons.

\subsection{\texorpdfstring{$SU(2N+3)\to SU(2N)\times SU(3)_c$}{}}
\label{sec:GC_4-fermi}
The first stage in the grand color symmetry breaking pattern \eqref{eq:GC_breaking_pattern} contains the breaking $SU(2N+3)\to SU(2N)\times SU(3)_c$.
The branching rule of $SU(2N+3)$ to $SU(2N)\times SU(3)_c$ is
\begin{equation}
    (2N+3)\otimes \overline{(2N+3)}\to (adj_{2N},1)\oplus (1,1)\oplus (1,adj_{3})\oplus (1,1)\oplus (2N, \bar{3})\oplus (\overline{2N},3)~.\label{eq:GC_breaking_adj}
\end{equation}
Therefore, there are additional massive vector bosons with mass $M_{\rm GC}$ transforming as $(2N, \bar{3})$ and $(\overline{2N}, {3})$ which can have phenomenological consequences at low energies. These gauge bosons are the $X_{\mu,\,am}$, and $X_{\mu,\,ma}$ subset of the $SU(2N+3)$ gauge fields $A_{\mu,\,ij}$, where $i,j,a,m$ are gauge indices -- $i,j=1,2\dots 2N+3$, $a=1,2,3$ and $m=4,\dots 2N+3$, and $\mu$ is a Lorentz index. The $X_\mu$ matrices refer to the off-diagonal subset of $A_\mu=A_\mu^\alpha t^\alpha$, 
where $\alpha=1,2\dots (2N+3)^2-1$ is an $adj_{2N+3}$ index and can be written as
\begin{equation}
X_{\mu,\,ij}\equiv
    \begin{pmatrix} & X_{\mu,\,am}\\X_{\mu,\,ma} &\end{pmatrix}~.
\end{equation} 
It is convenient to define $X_\mu=X^{\alpha^\prime}_\mu t^{\alpha^\prime}$, where $t^{\alpha^\prime}$ are the off-diagonal generators\footnote{These can always be chosen as matrices with $\sigma_1/2,\,\sigma_2/2$ in the $a,\,m$ subspace -- e.g. $t^{\alpha^\prime}$ defined by $t^{\alpha^\prime}_{am}=-i/2,~t^{\alpha^\prime}_{ma}=i/2$ and $t^{\alpha^\prime}_{ij}=0~\forall ~i,j\neq a,m$.} of $SU(2N+3)$, corresponding to the $(2N, \bar{3})\oplus (\overline{2N},3)$ representation in \eqref{eq:GC_breaking_adj}. The matter content of the $SU(2N+3)$ grand color theory, given in Table~\ref{tab:GC_fermion_charges}, consists of the left-handed Weyl fermions $\Psi_q,\Psi_{\bar u}$ and $\Psi_{\bar d}$.
The interactions between these fermions and the massive gauge bosons $X_\mu$ are
\begin{equation}
    \mathcal{L}\supset i\Psi^\dagger_{q\,a}\Bar{\sigma}^{\mu}(-ig_{GC}X_{\mu\,am} ) \Psi_{q\,m}+i{\Psi}_{\bar{u}\, m}^\dagger\Bar{\sigma}^{\mu}(+ig_{\rm GC} X_{\mu\, am}){\Psi}_{\bar{u}\, a}+ M^2_{\rm GC}X_{\mu\,am}X_{ma}^{\mu}+{\rm h.c. }\,,
    \label{eq:GC_boson_Lag}
\end{equation}
where the normalization for the gauge boson mass term has been fixed using 
$\text{Tr} ~t^{\alpha^\prime}t^{\beta^\prime}=\frac{1}{2}\delta^{\alpha^\prime\beta^\prime}$ so that 
 \begin{equation}
     \mathcal{L}\supset \frac{1}{2}M_{\rm GC}^2 X^{\alpha^\prime}_{\mu}X^{\alpha^\prime\,\mu}=M_{\rm GC}^2
     \text{Tr} X_{\mu} X^{\mu}=M^2_{\rm GC}(X_{\mu\,am}X_{ma}^{\mu}+X_{\mu\,ma}X_{am}^{\mu})~.
 \end{equation} 
 In \eqref{eq:GC_boson_Lag}, we have also used the property that $\left(t_{\overline{2N}}^{\alpha^\prime}\right)_{ij}=-\left(t_{2N}^{\alpha^\prime}\right)_{ji}$ to  write the $\Psi_{\bar{u}\,a}$ interaction in terms of $X^{\mu}_{am}$,
 where $t_{2N}$ 
 are the generators for the fundamental representation of $SU(2N)$. 

The equation of motion obtained from \eqref{eq:GC_boson_Lag} is then given by
\begin{align}
    -2M_{\rm GC}^2X^{\mu}_{ma}&=g_{GC}\left(\Psi^\dagger_{q\,a}\Bar{\sigma}^{\mu}\Psi_{q\,m}-\Psi^\dagger_{\Bar{u}\,m}\Bar{\sigma}^{\mu}\Psi_{\Bar{u}\,a} \right)\,.
    \label{eq:eom}
\end{align}
Integrating out the massive gauge bosons in \eqref{eq:GC_boson_Lag}, using \eqref{eq:eom}, then gives
\begin{align}
    \mathcal{L}&\supset \frac{g_{GC}^2}{2M_{\rm GC}^2}\left((\psi^\dagger_q\Bar{\sigma}^{\mu}q)({\psi}_{\bar{u}}^\dagger\Bar{\sigma}_{\mu} \Bar{u})+(q^\dagger\Bar{\sigma}^{\mu}\psi_q)(\bar{u}^\dagger\Bar{\sigma}_{\mu} \psi_{\Bar{u}})\right)\,,\nonumber\\
    &=\frac{g_{GC}^2}{M_{\rm GC}^2}\left((\psi^\dagger_q{\psi}_{\bar{u}}^\dagger)(\Bar{u}q)+(q^\dagger\bar{u}^\dagger)( \psi_q \psi_{\Bar{u}})\right)\,,
    \label{eq:4fermiL}
\end{align}
where the grand-color fermion multiplets have been split into their $SU(2N)\times SU(3)_c$ components as shown in Table~\ref{tab:GC_fermion_charges}. A Fierz rearrangement has been performed to obtain the second line in \eqref{eq:4fermiL}.
This is the same four-fermion operator given in \eqref{eq:one_flavor_model}.
 { Note that there are no four-fermion operators such as  ${\bar{d}}^\dag{\bar{u}}^\dag\psi_{\bar{d}}\psi_{\bar{u}}$ or $q^\dag q^\dag\psi_q\psi_q$. Therefore, the condensates in Eqs.~\eqref{eq:2condensate} and \eqref{eq:2+1_condensate} do not generate mass for the SM fermions due to the grand color gauge bosons, unlike the one flavor toy model in section~\ref{sec:toy}.
 }

\subsection{\texorpdfstring{$SU(2N)\to Sp(2N)$}{}}

The second stage of the grand color symmetry breaking \eqref{eq:GC_breaking_pattern} involves $SU(2N)\to Sp(2N)$. Due to this breaking at the scale $M_{\rm Sp}$, we obtain four-fermion operators corresponding to the exchange of massive $SU(2N)$ gauge bosons between two pairs of fermions -- two in the $2N$ representation and two in $\overline{2N}$ representation. Therefore, we expect the four-fermion operators to appear with both signs. 
Since, $\psi^I\psi^J\sim e^{i\Pi^{IK}/f}\Sigma_0^{KJ}$, these four-fermion operators have the form $\psi^4/M_{\rm Sp}^2$ and contribute to the pion masses only to subleading order ${\cal O}(\LSp^2/M_{\rm Sp}^2)$, compared to the Higgs-mediated Yukawa contributions that will be discussed in appendix \ref{app:CPT}.

\section{Vacuum alignment by Yukawa couplings and masses}
\label{app:CPT}

In this appendix, we perform the vacuum stability analysis for the $SU(2F) \rightarrow Sp(2F)$ symmetry breaking associated with the confinement of $Sp(2N)$ for the one and two-generation models discussed in sections \ref{subsec:one_gen_model} and \ref{sec:Two_gen}. This involves appropriately parameterizing the broken symmetry directions of the vacuum in terms of pions, and studying their potential. In particular, we will compute the pion VEVs and check whether these vacua are tachyonic or not, hence establishing their stability.

Above the confinement scale, $\LSp$ there is an $SU(2F)$ flavor symmetry for the $2F$ fermions, $\psi$, charged under $Sp(2N)$. Under an $SU(2F)$ transformation, the fermions $\psi$ and the non-linear sigma field $\Sigma$ (as well as $\Sigma_0$) defined in \eqref{eq:fermion_condensate} transform as~\cite{Kogut:2000ek}
\begin{align}
    \psi&\to \mathcal{U}\psi,\nonumber\\
   \frac{1}{\CSp}\psi\psi^T = \Sigma&\to \mathcal{U}\Sigma \,\mathcal{U}^T\,
\end{align}
where $\mathcal{U}$ is a $2F\times 2 F$ unitary matrix.
The ``effective" fermion masses can then be written as
\begin{align}
    {\cal L} =  -\frac{1}{2} \psi^T  M \psi + {\rm h.c.} =  \frac{1}{2}{\rm Tr}\left(\psi \psi^T M\right) + {\rm h.c.} ,
    \label{eq:Mdefn}
\end{align}
where the ``effective" mass matrix $M$ includes Dirac masses, Yukawa couplings and the Higgs field $H$. Once the $Sp$ sector confines, the flavor symmetry breaks down to $Sp(2F)$~\cite{Vafa:1983tf, Kosower:1984aw}, and $M$ can be considered as an $SU(2F)$ spurion transforming as
\begin{align}
M\to \mathcal{U}^*M\,\mathcal{U}^\dagger.
\end{align}
This leads to the low energy Lagrangian
\begin{align}\label{eq:MSigma_spurion}
\mathcal{L}&\supset \frac{\CSp}{2}\text{Tr}(\Sigma\,M +M^\dagger\Sigma^\dagger)+c_1\frac{\CSp}{\LSp}\text{Tr}(\Sigma\,M)\text{Tr}(M^\dagger\Sigma^\dagger)\nonumber\\
&\quad+c_2\frac{\CSp}{\LSp}\{[\text{Tr}(\Sigma\,M)]^2+\text{h.c.}\}+c_3\frac{\CSp}{\LSp}\{\text{Tr}(\Sigma\,M\Sigma\,M)+\text{h.c.}\} \,,
\end{align} 
where we have only included terms linear or quadratic in $M$, absorbed an order one coefficient in the first term into the definition of $\CSp$ and $c_{1,2,3}$ are constants.

 \subsection{One-generation model}

We first analyze the one-generation model with a vector-like quark presented in section \ref{sec:One_gen_wVLQ} where 
\begin{align}
\psi=
\begin{pmatrix}    \psi_{q_u}&\psi_{q_d}&\psi_U&\psi_{\bar{u}}&\psi_{\bar{d}}&\psi_{\bar{U}}
\end{pmatrix}^T~ .
\end{align}
As discussed in appendix \ref{sec:proof2}, $\theta=0$ corresponds to $\vev{\psi_{q_u} \psi_{\bar{u}}}=\vev{\psi_{q_d} \psi_{\bar{d}}}= \vev{\psi_U \psi_{\bar{U}}}<0$. The relevant condensates related to this by a non-anomalous flavor transformation are given by $\vev{\psi_{q_{u}} \psi_{q_d}},\,\vev{\psi_{U} \psi_{\bar{u}}},\,\vev{\psi_{\bar{d}} \psi_{\bar{U}}}>0$. In terms of $\Sigma_0^{IJ}\simeq   \langle \psi_I\psi_J\rangle / \CSp$, this can be written as
\begin{align}
\Sigma_{0}=\begin{pmatrix}
    \label{eq:vac_one}
    i\sigma_2&0&0\\0&  i \sigma_2& 0 \\0& 0 &  i \sigma_2
\end{pmatrix}\,.
\end{align}
Given that the number of flavors $F=3$, there are 14  broken symmetry generators, corresponding to the flavor symmetry breaking $SU(6)\rightarrow Sp(6)$ by the $Sp(2N)$ strong dynamics, which satisfy $\widetilde{T}^\alpha\Sigma_0=\Sigma_0\widetilde{T}^{\alpha\,T}$. These generators $\widetilde{T}^\alpha$ are given by

\begin{scriptsize}
\begin{align} 
&\widetilde{T}^{1}=\frac{1}{2\sqrt{2}}\begin{pmatrix}
     \mathbb{I}_2&0&0\\0&- \mathbb{I}_2&0\\0&0&0 
\end{pmatrix},\quad 
&\widetilde{T}^{2}=\frac{1}{2\sqrt{6}}\begin{pmatrix}
     \mathbb{I}_2&0&0\\0& \mathbb{I}_2&0\\ 0&0&-2 \mathbb{I}_2
\end{pmatrix},  \\
&\widetilde{T}^{3}=\frac{1}{2\sqrt{2}}\begin{pmatrix}
    0&0&i\sigma_1\\0&0&0\\-i\sigma_1&0&0
\end{pmatrix},\quad 
&\widetilde{T}^{4}=\frac{1}{2\sqrt{2}}\begin{pmatrix}
    0&0&i\sigma_2\\0&0&0\\-i\sigma_2&0&0 
\end{pmatrix},\quad 
&\widetilde{T}^{5}=\frac{1}{2\sqrt{2}}\begin{pmatrix}
    0&0&i\sigma_3\\0&0&0\\-i\sigma_3&0&0 
\end{pmatrix},\quad 
&\widetilde{T}^{6}=\frac{1}{2\sqrt{2}}\begin{pmatrix}
    0&0&\mathbb{I}_2\\0&0&0\\ \mathbb{I}_2&0&0 
\end{pmatrix}, \nonumber\\
&\widetilde{T}^{7}=\frac{1}{2\sqrt{2}}\begin{pmatrix}
    0&0&0\\0&0&i\sigma_1\\0&-i\sigma_1&0
\end{pmatrix},\quad 
&\widetilde{T}^{8}=\frac{1}{2\sqrt{2}}\begin{pmatrix}
    0&0&0\\0&0&i\sigma_2\\0&-i\sigma_2&0
\end{pmatrix},\quad 
&\widetilde{T}^{9}=\frac{1}{2\sqrt{2}}\begin{pmatrix}
    0&0&0\\0&0&i\sigma_3\\0&-i\sigma_3&0
\end{pmatrix},\quad 
&\widetilde{T}^{10}=\frac{1}{2\sqrt{2}}\begin{pmatrix}
    0&0&0\\0&0&\mathbb{I}_2\\0&\mathbb{I}_2&0
\end{pmatrix}, \nonumber\\&
\widetilde{T}^{11}=\frac{1}{2\sqrt{2}}\begin{pmatrix}
    0&i\sigma_1&0\\ -i\sigma_1&0&0\\0&0&0
\end{pmatrix},\quad 
&\widetilde{T}^{12}=\frac{1}{2\sqrt{2}}\begin{pmatrix}
    0&i\sigma_2&0\\ -i\sigma_2&0&0\\0&0&0
\end{pmatrix},\quad 
&\widetilde{T}^{13}=\frac{1}{2\sqrt{2}}\begin{pmatrix}
    0& i\sigma_3&0\\ -i\sigma_3&0&0\\0&0&0
\end{pmatrix},\quad 
&\widetilde{T}^{14}=\frac{1}{2\sqrt{2}}\begin{pmatrix}
    0&\mathbb{I}_2&0\\ \mathbb{I}_2&0&0\\0&0&0
\end{pmatrix}\,. \nonumber
\end{align}   
 \end{scriptsize}
We next redefine the generators as 
\begin{align}
    &{T}^{1}=\widetilde{T}^{1},\,  &{T}^{2}=\widetilde{T}^{2},\nonumber\\
     &{T}^{3}=\frac{\widetilde{T}^{3}+i\widetilde{T}^{4}}{\sqrt{2}},\quad 
     &{T}^{4}=\frac{\widetilde{T}^{5}+i\widetilde{T}^{6}}{\sqrt{2}},\nonumber\\
      &{T}^{5}=\frac{\widetilde{T}^{7}+i\widetilde{T}^{8}}{\sqrt{2}},\quad  &{T}^{6}=\frac{\widetilde{T}^{9}+i\widetilde{T}^{10}}{\sqrt{2}},\nonumber\\
        &{T}^{7}=\frac{ \widetilde{T}^{11}+i\widetilde{T}^{12}}{\sqrt{2}},  
&{T}^{8}=\frac{\widetilde{T}^{13}+i\widetilde{T}^{14}}{\sqrt{2}}\,,  
\label{eq:T1T8new}
\end{align}
which will be convenient for identifying the pion constituents.
The non-linear sigma field is written in the form
\begin{align}
\Sigma(x)=\exp\left[
\frac{i}{f}
\Pi\cdot T
\right] \Sigma_0\,, \quad\label{eq:Sigma_Pi_def}
\Pi\cdot T \equiv \sum_{\alpha=1}^2\Pi^\alpha(x)T^\alpha +
\sum_{\alpha=3}^8 \left(\Pi^\alpha(x)T^\alpha + 
\Pi^{\alpha\,\dagger}(x)T^{\alpha\dag}\right)\,,
\end{align} 
where $\Pi^{1,2}$ are real scalar fields and $\Pi^{3\dots 8}$ are complex scalar fields,
so that the pion fields can be identified with quark-antiquark and diquark states. Using the basis \eqref{eq:T1T8new} for the generators in $SU(6)/Sp(6)$, we can expand $\Sigma$ in \eqref{eq:Sigma_Pi_def}
to leading-order beyond $\Sigma_0$
\begin{align}
   \begin{scriptsize} i\Pi\cdot T \cdot \Sigma_0=\frac{1}{2}\begin{pmatrix}
        0&\frac{i}{\sqrt{6}}(\sqrt{3}\Pi^1+\Pi^2)&\Pi^{7}&-\Pi^{8}&\Pi^{3}&-\Pi^{4}\\
        - \frac{i}{\sqrt{6}}(\sqrt{3}\Pi^1+\Pi^2)&0&-\Pi^{8\,\dagger}&-\Pi^{7\,\dagger}&-\Pi^{4\,\dagger}&-\Pi^{3\,\dagger}\\
        -\Pi^{7}&\Pi^{8\,\dagger}& 0&\frac{i}{\sqrt{6}}(-\sqrt{3}\Pi^1+\Pi^2)&\Pi^{5}&-\Pi^{6}\\
        \Pi^{8}&\Pi^{7\,\dagger}&\frac{i}{\sqrt{6}}(\sqrt{3}\Pi^1-\Pi^2)&0&-\Pi^{6\,\dagger}&-\Pi^{5\,\dagger}\\
        -\Pi^{3}&\Pi^{4\,\dagger}&-\Pi^{5}&\Pi^{6\,\dagger}&0&-i\sqrt{\frac{2}{3}}\Pi^{2}\\
        \Pi^{4}&\Pi^{3\,\dagger}&\Pi^{6}&\Pi^{5\,\dagger}&i\sqrt{\frac{2}{3}}\Pi^{2}&0\\
    \end{pmatrix}\,.\end{scriptsize}
    \label{eq:PionMatrixSp6}
\end{align}  
Using \eqref{eq:PionMatrixSp6} and \eqref{eq:fermion_condensate}, we can read off the quark content of the different pion fields to linear order in the pion fields:
\begin{align}
\frac{f}{\CSp}(\psi_{q_u}\psi_{q_d})= \frac{i}{2\sqrt{2}}\Pi^1 + \frac{i}{2\sqrt{6}}\Pi^2,~~&
\frac{f}{\CSp}(\psi_{U}\psi_{\bar{u}})= -\frac{i}{2\sqrt{2}}\Pi^1 + \frac{i}{2\sqrt{6}}\Pi^2,\nonumber \\
\frac{f}{\CSp}(\psi_{\bar{d}}\psi_{\bar{U}})= - i \sqrt{\frac{2}{3}}\Pi^2,~~&
\frac{f}{\CSp} (\psi_{q_u}\psi_{\bar{d}}) = - \frac{f}{\CSp}(\psi_{q_d}\psi_{\bar{U}})^\dag = \frac{1}{2}\Pi^3,\nonumber \\
\frac{f}{\CSp}(\psi_{q_u}\psi_{\bar{U}}) =  \frac{f}{\CSp}(\psi_{q_d}\psi_{\bar{d}})^\dag = -\frac{1}{2}\Pi^4,~~&
\frac{f}{\CSp}(\psi_{U}\psi_{\bar{d}}) = - \frac{f}{\CSp}(\psi_{\bar{u}}\psi_{\bar{U}})^\dag = \frac{1}{2}\Pi^5\,, \nonumber \\
\frac{f}{\CSp}(\psi_{U}\psi_{\bar{U}}) =  \frac{f}{\CSp}(\psi_{\bar{u}}\psi_{\bar{d}})^\dag = -\frac{1}{2}\Pi^6,~~&
\frac{f}{\CSp}(\psi_{q_u}\psi_{U}) = - \frac{f}{\CSp}(\psi_{q_d}\psi_{\bar{u}})^\dag = \frac{1}{2}\Pi^7\,,\nonumber \\
\frac{f}{\CSp}(\psi_{q_u}\psi_{\bar{u}}) =  \frac{f}{\CSp}(\psi_{q_d}\psi_{U})^\dag = -\frac{1}{2}\Pi^8\,.&
\label{eq:2genquarkcontent}
\end{align}
From the quark content in \eqref{eq:2genquarkcontent} one can see that $\Pi^1$, $\Pi^2$ and $\Pi^6$ are electroweak-neutral pions. The electroweak-charged pions receive their dominant mass contribution via gauge boson loops, which results in a positive mass squared for all of them. Therefore, to study the vacuum stability, it suffices to check the signs of the quadratic $\Pi^1$, $\Pi^2$ and $\Pi^6$ terms that arise from \eqref{eq:MSigma_spurion}. The matrix $M$ in \eqref{eq:MSigma_spurion}
can be obtained by identifying the terms
\begin{align}
{\cal L}   \supset  -y_d  \psi_q \psi_{\bar d} {\widetilde H}  - m_U \psi_U \psi_{\bar{U}}- {\lambda_U}   \psi_q \psi_{\bar{U}} H +{\rm h.c.}\,,
\end{align} 
in \eqref{eq:Mdefn} and comparing these with \eqref{eq:MSigma_spurion}, implying
\begin{align} 
M =\begin{pmatrix}
        0&0&0&0& y_d \varphi^-&\lambda_U\,\varphi^0\\
         0&0&0&0&y_d\,(\varphi^0)^\dagger& -\lambda_U \varphi^+\\
         0&0&0&0&0&m_U\\
         0&0&0&0&0&0\\
      {-} y_d \varphi^-&-y_d\,(\varphi^0)^\dagger&0&0&0&0\\ 
         -\lambda_U\,\varphi^0& \lambda_U \varphi^+&-m_U&0&0&0
    \end{pmatrix}~,\label{eq:MassYukawaSpurion}
\end{align} 
where $\varphi^0$ and $\varphi^+$ are the neutral and charged components of $H=(\varphi^+~~\varphi^0)^T$, 
respectively, with $\varphi^-=(\varphi^+)^\dagger$. Note that we work in the convention that $\varepsilon^{12}=+1$ such that $\mathcal{L}\supset -\psi_q\psi_{\bar{U}}H=-\varepsilon^{ij}\psi_{q\,i}H_j\psi_{\bar{U}}=(\varphi^+\psi_{q_d}-\varphi^0\psi_{q_u})\psi_{\bar U}$.

The linear and quadratic potential of the electroweak-neutral pions obtained from Eq.~\eqref{eq:MSigma_spurion} is 
\begin{align}
\mathcal{L}\supset \frac{ m_U}{2}\CSp \left( \frac{1}{f} \Pi^6 - \frac{i }{4 \sqrt{6}f^2} \Pi^6 (\sqrt{3}\Pi^1 + \Pi^2) \right) + {\rm h.c.} \nonumber \\
+ c_3 y_d \lambda_U \frac{\LSp^2}{16\pi^2}  \left( |\Pi^6|^2 +\frac{1}{2} \left(\Pi^1 - \frac{1}{\sqrt{3}} \Pi^2 \right)^2  \right),
\label{eq:neutralPi_V}
\end{align}
where we have ignored ${\cal O}(m_U^2)$ terms, which are subdominant since we have assumed $m_U\ll \LSp$.
The vacuum alignment of the electroweak neutral pions is determined by $c_3$, and this coefficient can be obtained from the quantum corrections due to the pion-Higgs interactions. The terms proportional to $c_{3}$
originally included $|H|^2$ but since we are only interested in the pion potential in the electroweak symmetric limit, $|H|^2$ has been replaced with $\LSp^2/(16\pi^2)$, corresponding to the quantum corrections from the Higgs loop. 

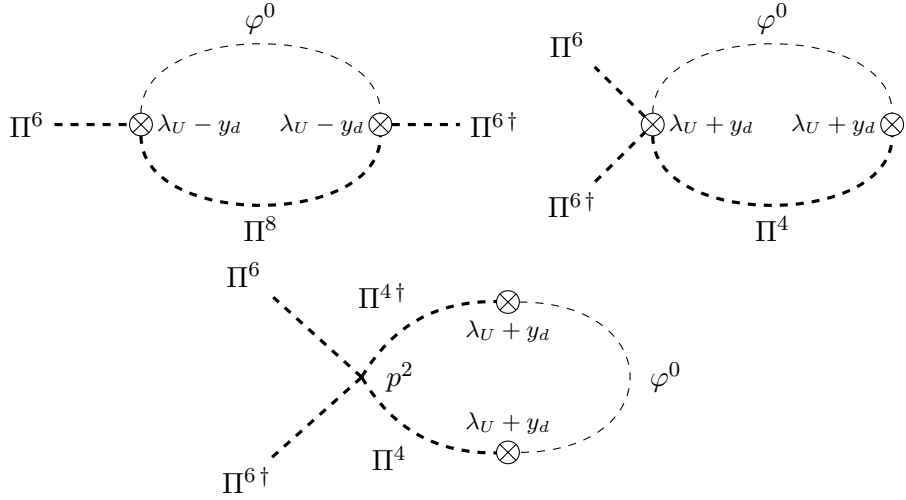
\begin{figure}[!t]
    \centering
\begin{tikzpicture}[scale=0.6] 
\begin{feynman}
\vertex (a) {\(\Pi^{6}\)}; 
\vertex [crossed dot,right=of a] (b) {}; 
\vertex [below right=of b] (c)  {\(\)};
\vertex [right=of b] (e1);\vertex [crossed dot,right=of e1] (e){};
\vertex [right=of e] (f){\(\Pi^{6\, \dagger}\)}; 
\vertex [below left=of e](d) {\(\)};
\vertex [right=02em of b] {\footnotesize\(\lambda_U-y_d\)};
\vertex [ left=2em of e] {\footnotesize\(\lambda_U-y_d\)};  
\diagram* {
(a) -- [scalar, very thick] (b) -- [scalar, very thick,
half right, looseness=1 , edge label'=\( \Pi^8\)] (e) -- [scalar, very thick] (f) ,
(b) -- [scalar, half left,  looseness=1 ,edge label=\(\varphi^0\)] (e), 
,
};
\end{feynman}
\end{tikzpicture}
 \begin{tikzpicture}[scale=0.4]
 \begin{feynman}
 
\vertex [crossed dot]  (b) {}; 
\vertex [below right=of b] (c)  ;
\vertex [right=of b] (e1);\vertex  [crossed dot, right=of e1 ] (e){ };
 
\vertex [below left=of e](d)  ;
  
\vertex [  right=02em of b] {\footnotesize\(\lambda_U+y_d\)};
\vertex [left=2em of e] {\footnotesize\(\lambda_U+y_d\)};
    \vertex [below left=of b](a1){\( \Pi^{6\, \dagger}\)};\vertex [above left=of b](a2){\({\Pi^{6\ } }\)};
\diagram* {
(a1)--[scalar, very thick](b)--[scalar, very thick](a2),
  (b) -- [scalar, very thick, half right, looseness=1,edge label'=\(\Pi^{4}\)] (e), 
(b) -- [scalar, half left, looseness=1, edge label=\(\varphi^0\)] (e), 
};
\end{feynman}
\end{tikzpicture}
 \hspace{1cm}
 \begin{tikzpicture}[scale=0.4]
 \begin{feynman}
 
\vertex   (b) ; 
\vertex [right=5em of b] (e1);\vertex [right=4em of e1] (e);
 
\vertex [crossed dot,above  =2.2em of e1] (f)  {};
\vertex [crossed dot,below  =2.2em of e1] (c)  {};
\vertex [below left=of e](d)  ;
\vertex [above=1em of c] {\footnotesize\(\lambda_U+y_d\)};
\vertex [below=1em of f] {\footnotesize\(\lambda_U+y_d\)};
\vertex [right=0.5em of e] {\(\varphi^0\)};
\vertex [right=0.5em of b] {\(p^2\)};

    \vertex [below left=of b](a1){\( \Pi^{6\, \dagger}\)};\vertex [above left=of b](a2){\({\Pi^{6\ } }\)};
\diagram* {
(a1)--[scalar, very thick](b)--[scalar, very thick](a2),
  (b) -- [scalar, very thick, bend left, looseness=1 , edge label=\( \Pi^{4\, \dagger}\)] (f), 
(b) -- [scalar,very thick, bend right, looseness=1 , edge label'=\(\Pi^{4} \)] (c), 
(f)[crossed dot] -- [scalar,  half left, looseness=2.5 ] (c) ,
}; \end{feynman}
\end{tikzpicture}
\caption{One loop Feynman diagrams used to compute the quantum corrections to the mass of the neutral pion $\Pi^6$. Similar diagrams 
with the replacements $\varphi^0\rightarrow \varphi^+$, $\Pi^8\to\Pi^{7\,\dagger}$ and $\Pi^4\to \Pi^{3\,\dagger}$ give the same mass contribution to $\Pi^6$.}
 \label{fig:diagrams_c3} 
\end{figure}

To compute the coefficient $c_3$ in the chiral Lagrangian, we can analyze the mass corrections of the pions resulting from the kinetic term and the term linear in $M$ in \eqref{eq:MSigma_spurion}, and compare them to the terms in \eqref{eq:MSigma_spurion} that are quadratic in $M$. The quantum correction to the $\Pi^6$ mass can be computed using the following relevant interaction terms

\begin{align}
\label{eq:Lint}   {\cal L} \supset & \frac{\CSp}{8 f^2} \left(-\lambda_U + y_d \right) \Pi^8 \Pi^6 \varphi^0 - \frac{\CSp}{48 f^3} |\Pi^6|^2  \left(\lambda_U + y_d \right)   \Pi^4\varphi^0  + \frac{\CSp}{2f} \left(\lambda_U + y_d \right) \Pi^4 \varphi^0 + {\rm h.c.} \nonumber \\
   & -\frac{1}{48f^2}  \Pi^6 \Pi^{6\, \dagger} \partial^{\mu}\Pi^4\partial_{\mu}\Pi^{4\, \dagger}\,.
\end{align}
The quantum correction from $\varphi^+$ is the same as that from $ \varphi^0$.
The quantum corrections to the mass squared of $\Pi_6$ can be calculated from the diagrams in figure~\ref{fig:diagrams_c3} which gives
\begin{align}
  m_{\Pi_6}^2&= \Biggl( \left(\frac{ -\lambda_U+y_d}{8} \right)^2 +2\left(-\frac{\lambda_U+y_d}{48 } \right)  \left(\frac{ \lambda_U+y_d}{2} \right) 
      - \left(\frac{ -1}{48 } \right)\left(\frac{ \lambda_U+y_d}{2} \right)^2\Biggl) \frac{\Lambda_{Sp}^2}{16\pi^2} \ln \frac{\Lambda_{Sp}^2}{m_{\Pi_4}^2} 
     \nonumber \\
  & =   - \frac{\lambda_Uy_d}{8  } \frac{\LSp^2}{16\pi^2} \ln \frac{\Lambda_{Sp}}{m_{\Pi_4}} 
   ~.\label{eq:Pi6MassSq}
\end{align}Including similar contributions from $\varphi^+$ gives an additional factor of $2$. Thus, comparing \eqref{eq:Pi6MassSq} with the mass squared obtained from \eqref{eq:neutralPi_V} gives the identification
\begin{align}
c_3 = - \frac{1}{4  }\ln{\frac{\LSp}{m_{\Pi_4}}}.
\label{eq:c3}
\end{align}
Since $c_3<0$, the mass squared of $\Pi^6$ is positive. For simplicity, 
the $\mathcal{O}(1)$ $\log$ factor in \eqref{eq:c3} has been ignored in 
section~\ref{sec:One_gen_wVLQ}.
 
The first term in Eq.~\eqref{eq:neutralPi_V} destabilizes $\Pi^6=0$, while the term proportional to $c_3 \lambda_U y_d$ stabilizes it. The vacuum alignment is determined by the balance between these two terms. However, the potential \eqref{eq:neutralPi_V} is obtained from expanding around the vacuum \eqref{eq:vac_one} assuming a zero $\Pi^6$ VEV. To allow for the possibility of a nonzero $\Pi^6$ VEV requires computing the full potential around Eq~\eqref{eq:vac_one}.
The full potential containing all the electroweak neutral pions is then given by
\begin{align}\label{eq:FullPot_Onegen}
   {\cal L}\supset \frac{1}{2}  \CSp \Biggl(&\frac{4 {m_U} }{ {\Pi^{\rm A1}}}\sin \left(\frac{\Pi^{\rm A1}}{2f}\right) \left(\text{Im}(\Pi^6) \sin \left(\frac{\Pi^{\rm B1}}{4\sqrt{6}f}\right)+\text{Re}(\Pi^6) \cos \left(\frac{\Pi^{\rm B1}}{4\sqrt{6}f}\right)\right)\nonumber\\
   &+\frac{c_{3}}{2\sqrt{2}\pi^2} \lambda_U y_d \LSp\Bigg( {\left(\Pi^1-\sqrt{3} \Pi^2\right) \frac{1}{\Pi^{\rm A1}}\sin\left(\frac{\Pi^{\rm A1}}{2f}\right) \sin \left(\frac{\Pi^{\rm B1}}{4\sqrt{6}f}\right)}\nonumber\\
   &\qquad-2 \sqrt{2}
   \cos \left(\frac{\Pi^{\rm A1}}{2f}\right) \cos\left(\frac{\Pi^{\rm B1}}{4\sqrt{6}f}\right)\Bigg)\Biggl)\,,
\end{align}
where we have defined 
$\Pi^{\rm A1}\equiv\sqrt{\Pi^{6\,\dagger}\Pi^6+\frac{1}{8}(\Pi^1-\sqrt{3}\Pi^2)^2}$ and $\Pi^{\rm B1}\equiv \sqrt{3}\Pi^1+\Pi^2$.
Since only ${\rm Re}\,(\Pi^6)$ obtains a VEV, we can set all other pion fields to zero in \eqref{eq:FullPot_Onegen} and work with the much simpler potential
\begin{align}
\mathcal{L}\supset -V(\phi) =  2 m_U\CSp \sin \phi - \frac{c_3}{2\pi^2} \CSp \LSp   \lambda_U y_d \cos\phi,\qquad \phi \equiv {\rm Re}\frac{\Pi^6}{2f}\,.
\label{eq:1genpot}
\end{align}
The minimum of the potential \eqref{eq:1genpot} is simply
\begin{align}
\tan\phi = \frac{4\pi^2 m_U}{|c_3| \lambda_U y_d \LSp}\simeq \frac{16\pi^2 m_U}{\lambda_U y_d \LSp}\,,
\label{eq:tanphi}
\end{align}
where in the second relation we have used \eqref{eq:c3} (ignoring the log).
The relation between $\phi$ and the quark condensation is given by Eq.\,\eqref{eq:2+1_condensate} and \eqref{eq:tanphi} is used in \eqref{eq:phi}.

In the limit, $m_U \ll y_d \lambda_U \LSp$, the squared masses of the neutral pions are
\begin{align}
\label{eq:pion mass}
m_{\Pi}^2 = 
0, ~- c_3\frac{  y_d \lambda_U}{16\pi^2}\LSp^2,~- c_3\frac{ y_d \lambda_U}{16\pi^2 }\LSp^2,~ -c_3  \frac{ y_d \lambda_U}{12\pi^2}\LSp^2\,,
\end{align}
where $c_3<0$ is given in \eqref{eq:c3}.
The nonzero mass-squared values are all positive and the massless neutral pion is associated with  the spontaneous breaking of the baryon symmetry. 

For $m_U \gg y_d \lambda_U \LSp$, three neutral pions have masses squared of ${\cal O}(m_U \LSp)$ and one is massless. In this limit, $\phi\to \pi/2$ and the vacuum reduces to the one used in section \ref{sec:onegenmin}:  $\langle\psi_{q_u}\psi_{q_d}\rangle=-\langle\psi_{\bar{u}}\psi_{\bar{d}}\rangle > 0$.
We can also confirm the results presented in appendix~\ref{sec:proof} in terms of fermions in the pion picture by including $\eta'$ and the corresponding generator $=\mathbb{I}/\sqrt{12}$ in \eqref{eq:Sigma_Pi_def}. As expected, we find that the $\eta'$ mixes with the other pions -- $\Pi^1,\,\Pi^2,$ and Im\,$\Pi^6$, but the vacuum remains stable at $\vev{\eta'}=0$.

Note that there are no neutral pion mass terms proportional to $\lambda_U^2$ or $y_d^2$. This can be understood by collective symmetry breaking. Neglecting the Yukawa couplings, the model has an $SU(2)$ flavor symmetry corresponding to transformations between $\psi_{\bar{U}}$ and $\psi_{\bar{u}}$, and another $SU(2)$ corresponding to 
transformations between $\psi_{{U}}$ and $\psi_{\bar{d}}$. In the IR, this $SU(2)\times SU(2)$ flavor symmetry is broken down to $SU(2)$ upon the quark condensation. Now, if we turn on only one of the Yukawa couplings, the symmetry breaking pattern is $SU(2)\to \varnothing$. Therefore, none of the neutral Nambu-Goldstone bosons receive a mass contribution proportional to $\lambda_U^2$ or $y_d^2$. 
This is evident in \eqref{eq:Pi6MassSq} where we explicitly confirm the absence of a quantum correction to the mass of $\Pi^6$ proportional to $\lambda_U^2$ or $y_d^2$ from the interactions in Eq.~\eqref{eq:Lint}.

\subsection{Two-generation model}
\label{app:two}

We next analyze the two-generation model without a vector-like quark where
\begin{align}
\psi^T = \left(
\psi_{q_u}
\psi_{q_d}
\psi_{q_c}
\psi_{q_s}
\psi_{\bar{u}_1}
\psi_{\bar{d}_1}
\psi_{\bar{u}_2}
\psi_{\bar{d}_2}
\right)\,.
\label{eq:psi2gen}
\end{align}
As discussed in appendix \ref{sec:proof2}, we need $\vev{\psi_{q_u} \psi_{q_d}} >0$ and $\vev{\psi_{\bar{u}} \psi_{\bar{d}}} <0$. Thus, we start from the field space
\begin{align}
\Sigma_{0}= \begin{pmatrix}
    i\sigma_2&0&0&0\\0&-i\sigma_2&0&0\\0&0&-i\sigma_2&0\\0&0&0&i\sigma_2
\end{pmatrix}~.\label{eq:Sigma0_2gen}
\end{align}
Given the number of flavors $F=4$, the symmetry breaking is $SU(8)\rightarrow Sp(8)$ which leads to 27 broken generators, with 6 of them corresponding to neutral pions.
The neutral pions correspond to the following symmetry-breaking patterns. In the limit of vanishing Yukawa couplings, the theory has an $SU(2)_q\times SU(2)_{\bar{u}\bar{c}}\times SU(2)_{\bar{d}\bar{s}} \times U(1)_B\times U(1)_{\bar{u}\bar{c}-\bar{d}\bar{s}}$ flavor symmetry. The $SU(2)_q$ symmetry is broken down to $U(1)$, the $SU(2)_{\bar{u}\bar{c}}\times SU(2)_{\bar{d}\bar{s}}$ symmetry breaks into $SU(2)$, and $U(1)_B$ is completely broken, giving six pions.
The associated six generators are 
\begin{align}
\widetilde{T}^{1}=\frac{1}{4}\begin{pmatrix}
     \mathbb{I}_2&0&0&0\\0& \mathbb{I}_2&0&0\\ 0&0&- \mathbb{I}_2 &0\\ 0&0&0& -\mathbb{I}_2
\end{pmatrix}, \quad
& \widetilde{T}^{2}=\frac{1}{2\sqrt{2}}\begin{pmatrix}
    \mathbb{I}_2&0 & 0 &0 \\0&- \mathbb{I}_2&0&0\\  0&0&0 &0 \\
    0&0&0 &0
\end{pmatrix},
\nonumber\\
\widetilde{T}^{3}=\frac{1}{2\sqrt{2}}\begin{pmatrix}
 0&0&0&0\\
    0&0&0&0\\0&0&\mathbb{I}_2&0\\0&0&0& -\mathbb{I}_2& 
\end{pmatrix}, \quad
&\widetilde{T}^{4}=\frac{1}{2\sqrt{2}}\begin{pmatrix}
    0&i\mathbb{I}_2 & 0 &0 \\-i\mathbb{I}_2&0&0&0\\  0&0&0 &0 \\
    0&0&0 &0
\end{pmatrix},\nonumber \\
\widetilde{T}^{5}=\frac{1}{2\sqrt{2}}\begin{pmatrix}
     0&0&0&0\\
     0&0&0&0\\0&0&0&\sigma_3\\0&0&\sigma_3&0 
\end{pmatrix}, \quad &
\widetilde{T}^{6}=\frac{1}{2\sqrt{2}}\begin{pmatrix}
0&0&0&0\\
    0&0&0&0\\0&0&0&i\mathbb{I}_2 \\0&0&-i\mathbb{I}_2 &0
\end{pmatrix}\,.\quad 
\end{align}    
These generators are then  redefined as
\begin{align}
{T}^{i}=\widetilde{T}^{i} ~(i=1\dots4),\qquad  {T}^{5}=\frac{i \widetilde{T}^{5}+\widetilde{T}^{6}}{\sqrt{2}}, 
\end{align}
in order to determine the pion constituents. The non-linear sigma field associated with this redefinition is then
\begin{align}
\Sigma(x)= \exp\left[
\frac{i}{f}
\Pi\cdot T
\right] \Sigma_0, \qquad
\Pi\cdot T \equiv  
\sum_{\alpha=1}^4\Pi^\alpha(x)T^\alpha +
\Pi^5(x)T^5 +
 \Pi^{5\dagger}(x)T^{5\dag}\,,
 \label{eq:SigT1T5}
\end{align} 
where $\Pi^{1\dots4}$ are real scalar fields and $\Pi^{5}$ is a complex scalar field.

The leading non-trivial term in the expansion of $\Sigma$ in \eqref{eq:SigT1T5} is
\begin{align}
&i\Pi\cdot T \cdot \Sigma_0 
 \\
    = &
    \begin{tiny}
   \frac{i}{2\sqrt{2}}  \begin{pmatrix}
        0&\frac{1}{\sqrt{2}} \Pi^1 + \Pi^2& 0 & -i \Pi^4 & 0 &0 &0 &0\\
       -\frac{1}{\sqrt{2}} \Pi^1 - \Pi^2 & 0 & i \Pi^4 & 0 &0 &0 &0 &0 \\
    0 & -i \Pi^4 & 0 &-\frac{1}{\sqrt{2}} \Pi^1 + \Pi^2 & 0 &0 &0 &0 \\
    i \Pi^4 & 0 & \frac{1}{\sqrt{2}} \Pi^1 - \Pi^2 & 0 &0 &0 &0 &0 \\
    0 &0 &0 & 0 & 0 &  \frac{1}{\sqrt{2}} \Pi^1 - \Pi^3 & 0 &  i \sqrt{2} \Pi^5 \\
    0 &0 &0 &0 &-\frac{1}{\sqrt{2}} \Pi^1 + \Pi^3 & 0 & -i \sqrt{2}\Pi^{5\dagger} & 0 \\
    0 & 0 & 0 & 0 & 0 & i \sqrt{2}  \Pi^{5\dagger} & 0 & -\frac{1}{\sqrt{2}} \Pi^1 - \Pi^3 \\
    0 &0 &0 &0 & -i \sqrt{2}\Pi^5 & 0 &\frac{1}{\sqrt{2}} \Pi^1 + \Pi^3  & 0
\end{pmatrix}~.\end{tiny}
\nonumber
\label{eq:PionMatrixSp8}
\end{align} 
The matrix $M$ in \eqref{eq:Mdefn} then becomes
\begin{align}
 M = \begin{pmatrix}
 0 & 0 & 0 & 0 & y_u \varphi^0 & 0 &0 &  y_1 \varphi^- \\
 0 & 0 & 0 & 0 &  {-}y_u \varphi^+& 0 & 0 & y_1 (\varphi^0)^\dagger \\ 
 0 & 0 & 0 & 0 & 0 & 0 & y_c \varphi^0 &  y_2 \varphi^- \\
 0 & 0 & 0 & 0 & 0 & 0 &  { { -}}y_c \varphi^+& y_2 (\varphi^0)^\dagger \\
- y_u \varphi^0 &   y_u \varphi^+& 0 & 0& 0& 0& 0& 0 \\
0 & 0& 0& 0& 0& 0& 0& 0 \\
0 & 0 & - y_c \varphi^0 &   y_c \varphi^+& 0& 0& 0& 0\\
  {-}y_1 \varphi^- & - y_1 (\varphi^0)^\dagger &  {-} y_2 \varphi^- & - y_2 (\varphi^0)^\dagger & 0& 0& 0& 0  
\end{pmatrix}.
\end{align} 
The linear and quadratic potential of the electroweak-neutral pions obtained from \eqref{eq:MSigma_spurion} is
\begin{align}
  \mathcal{L}\supset
 \frac{c_3}{4\pi^2} \CSp\LSp \Bigg(& \frac{1}{\sqrt{2}f}y_1 y_c \Pi^4 -\frac{1}{f} y_1 y_u {\rm Re}\,\Pi^5 +\frac{y_2 y_c }{8f^2}
\left((\Pi^2+ \Pi^3)^2+(\Pi^4)^2 + 2|\Pi^5|^2 \right) \nonumber \\
& -\frac{y_u}{2\sqrt{2} f^2}\left( y_2  \Pi^4\, {\rm Re}\,\Pi^5 -  y_1 \Pi^2\,{\rm Im}\,\Pi^5 \right) \Bigg)\,,
\end{align}
where $\Pi^4$ and ${\rm Re}\,\Pi^5$ have tadpole terms and can obtain VEVs. The $\Pi^4$ tadpole term is much larger than the ${\rm Re}\,\Pi^5$ tadpole term since $y_c \gg y_u$. The VEV of $\Pi^4$ then generates another ${\rm Re}\,\Pi^5$ tadpole term which exactly cancels the term independent of $\Pi^4$. The cancellation occurs beyond the leading-order which requires the full potential. This is obtained from \eqref{eq:MSigma_spurion} and is given by
\begin{align}
  \mathcal{L}\supset -\frac{c_3}{2\pi^2}  \frac{\CSp\LSp }{\Pi^{\rm A2} \Pi^{\rm B2}}\Biggl(\sin \left(\frac{\Pi^{\rm B2}}{2 \sqrt{2} f}\right) \Bigg(-\sin\left(\frac{\Pi^{\rm A2}}{2 \sqrt{2} f}\right) \bigl(y_2 \left(y_c\Pi^2 \Pi^3-\sqrt{2}y_u
  \Pi^4\,{\rm Re}\,\Pi^5\right)\nonumber\\
  +\sqrt{2}y_1 y_u\Pi^2\,\text{\rm Im}\,\Pi^5\bigl)-y_1y_c\Pi^4\Pi^{\rm A2}\cos \left(\frac{\Pi^{\rm A2}}{2 \sqrt{2} f}\right)\Bigg)\nonumber\\
  + \Pi^{\rm B2} \cos \left(\frac{\Pi^{\rm B2}}{2 \sqrt{2} f}\right) \left(y_2 y_c\Pi^{\rm A2}\cos \left(\frac{\Pi^{\rm A2}}{2 \sqrt{2} f}\right)+\sqrt{2}y_1 y_u  {\rm Re}\,\Pi^5  \sin \left(\frac{\Pi^{\rm A2}}{2 \sqrt{2} f}\right)\Biggl)\right)\,,
\end{align}
where we have defined $\Pi^{\rm A2} \equiv\sqrt{(\Pi ^3)^2+2\Pi ^{5\,\dagger}\Pi ^5}$ and $\Pi^{\rm B2}\equiv\sqrt{(\Pi ^2)^2+(\Pi ^4)^2}$.
Focusing on the potential of the fields that can acquire a VEV, namely $\phi \equiv \Pi^4/(2\sqrt{2}f)$ and $\alpha \equiv {\rm Re}\, \Pi^5/(2f)$, we obtain
\begin{align}
 \mathcal{L}\supset -V(\phi, \alpha) = -\frac{c_3 }{2\pi^2} \CSp \LSp\left(
y_c \cos\alpha \left(y_2 \cos\phi - y_1 \sin\phi\right)
+y_u \sin\alpha \left(y_1 \cos\phi + y_2 \sin\phi
\right)
\right)\,.
\end{align}
For $y_c > y_u$, the minimum is given by
\begin{align}
\vev{\Pi^4} = -2 \sqrt{2} f\arctan\frac{y_1}{y_2},\qquad\vev{\Pi^5} = 0\,,
\end{align}
where $\Pi^2 + \Pi^3$ and the imaginary part of $\Pi^5$ both have positive mass-squared and no tadpole terms, and therefore do not obtain VEVs. The linear combination 
$\Pi^2 - \Pi^3$ is not fixed by the leading-order potential, but we find that it obtains a positive mass squared from a higher-order coupling with $\Pi^4$. With a non-zero $\Pi^4$ VEV, the quark condensate is
\begin{align}
  \vev{ \psi \psi^T} = \CSp \times
    \begin{pmatrix}
  0 & \frac{y_2}{\sqrt{y_1^2 + y_2^2}} & 0 & - \frac{y_1}{\sqrt{y_1^2 + y_2^2}} & 0 & 0& 0& 0 \\
  -\frac{y_2}{\sqrt{y_1^2 + y_2^2}} & 0 & \frac{y_1}{\sqrt{y_1^2 + y_2^2}} & 0 & 0 & 0& 0& 0 \\
  0 & - \frac{y_1}{\sqrt{y_1^2 + y_2^2}} & 0 & - \frac{y_2 }{\sqrt{y_1^2 + y_2^2}} & 0 & 0& 0& 0 \\
  \frac{y_1}{\sqrt{y_1^2 + y_2^2}} & 0 & \frac{y_2}{\sqrt{y_1^2 + y_2^2}} & 0 & 0 & 0& 0& 0 \\
 0 & 0 & 0& 0& 0 & -1 & 0 & 0 \\
 0 & 0 & 0 & 0 & 1 & 0 & 0 & 0 \\
0 & 0 & 0& 0& 0 & 0 & 0 & 1 \\
0 & 0 & 0 & 0& 0& 0 & -1 & 0
    \end{pmatrix}\,,
\end{align}
where $\psi$ is given in \eqref{eq:psi2gen}. This condensate is then rewritten in \eqref{eq:2gencondensate}.

We also comment on the nature of the vacuum when the two-generation Yukawa matrix in \eqref{eq:HiggsfermionL2} is replaced by the SM-like Yukawa structure
\begin{align}
    \widetilde{Y}^d=\begin{pmatrix}
        y_d\cos\theta_c&y_s\sin\theta_c\\-y_d\sin\theta_c&y_s\cos\theta_c
    \end{pmatrix}~.
\end{align}
In this case, 
 \begin{align}
  \mathcal{L}\supset
 \frac{c_3}{4\pi^2} \CSp\LSp \Bigg(& \frac{1}{\sqrt{2}f}\left(y_s y_c+y_dy_u\right)\sin\theta_c \,\Pi^4 -\frac{1}{f} \left(y_d y_c+y_sy_u\right)\sin\theta_c\,{\rm Re}\,\Pi^5\nonumber \\
 &+\frac{1 }{8f^2}\left(y_s y_c+y_dy_u\right)\cos\theta_c 
\left( (\Pi^2+ \Pi^3)^2+(\Pi^4)^2 + 2|\Pi^5|^2 \right) \nonumber \\
& -\frac{1}{2\sqrt{2} f^2}\left(y_d y_c+y_sy_u\right)\left( \cos\theta_c\, \Pi^4\, {\rm Re}\,\Pi^5 -  \sin\theta_c\,\Pi^2\,{\rm Im}\,\Pi^5 \right) \Bigg)\,.\label{eq:NLOPotential_cabibo}
\end{align} 
We see from \eqref{eq:NLOPotential_cabibo} that only $\Pi^4$ and ${\rm Re}\,\Pi^5$ can obtain VEVs due to Yukawa interactions. Including higher order terms in $\phi \equiv \Pi^4/(2\sqrt{2}f)$ and $\alpha \equiv {\rm Re}\, \Pi^5/(2f)$ and setting all other pion VEVs to zero, we obtain
 \begin{align}
 \mathcal{L}\supset   -\frac{c_3 }{2\pi^2} \CSp \LSp\left(
 \left(y_s y_c+y_dy_u\right) \cos\left(\theta_c+\phi \right)\cos\alpha
+\left(y_d y_c+y_sy_u\right)\sin\left(\theta_c+\phi \right) \sin\alpha  
\right)\,.
\label{eq:Potential_cabibo}
\end{align} 
The minimum of this potential is given by 
\begin{align}
\vev{\Pi^4} = -2 \sqrt{2} f\,\theta_c~, 
\end{align} 
with all other pions having a vanishing VEV. The pion masses at this minimum are then
\begin{align} 
m_{\Pi}^2 = &~0, 
~-\frac{c_3}{16\pi^2}(y_d y_u+y_sy_c)\LSp^2,
~-\frac{c_3}{16\pi^2 }(y_s\pm y_d )(y_c\pm y_u)\LSp^2,\nonumber\\
~&- \frac{c_3}{16\pi^2}
\left(y_d y_u+y_sy_c \pm \sqrt{(y_cy_d+y_uy_s)^2\sin^2\theta_c+(y_cy_s+y_uy_d)^2\cos^2\theta_c }\right)\LSp^2\,.
\end{align}
Note that the next-to-lightest state in the limit $y_u,y_d\to 0$ has a mass $\sim \sqrt{y_sy_c} \sin \left(\frac{\theta_c}{2}\right)  \frac{\LSp}{4\pi}$, while for $\theta_c\to 0$, the mass 
would instead be $ \frac{\theta_c^2}{2} \left(y_c^2-y_u^2\right) \left(y_s^2-y_d^2\right) / \left(y_c y_s+y_d y_u\right) \frac{\LSp}{4\pi}$.

We also briefly comment on the case with vector-like quarks $U,\,\bar{U}$. In this case, we start from the vacuum 
\begin{align}
\Sigma_0={\rm diag}\left(i\sigma_2,i\sigma_2,i\sigma_2,i\sigma_2,i\sigma_2\right)~,
\end{align} 
in the basis 
\begin{align}
\psi^T = \left(\psi_{q_u}~
\psi_{q_d}~\psi_{q_c}~\psi_{q_s}~\psi_{U}~
\psi_{\bar{u}_1}~\psi_{\bar{d}_1}~\psi_{\bar{u}_2}~\psi_{\bar{d}_2}~\psi_{\bar{U}}~
\right)\,.
\label{eq:psi2genVLQ}
\end{align} 
The symmetry breaking generates 44 pions, 11 of which are electroweak neutral. In this case, we only show the pions that may obtain a VEV which are given by 
\begin{align}
    i\Pi\cdot T \cdot \Sigma_0 
    = &-   \frac{i}{2\sqrt{2}}  \begin{pmatrix} 0_{4\times 4}&0_{4\times 6}\\
    0_{6\times 4}&A_{6\times 6}\end{pmatrix},\quad A=\begin{pmatrix}
        0&0& 0&0& 0&\Pi^1\\
        0&0& 0&0& \Pi^1&0\\
        0&0& 0&0& 0&\Pi^2\\
        0&0& 0&0& \Pi^2&0\\
        0&-\Pi^1& 0&-\Pi^2& 0&0\\
        -\Pi^1&0& -\Pi^2&0& 0&0
    \end{pmatrix}\,.
\end{align}
The linear terms in the pion potential arising from the Yukawa interactions in \eqref{eq:2gen_vlq} are given by 
\begin{align}
    \mathcal{L}\supset
 \frac{\CSp}{\sqrt{2}f} \left(m_U\Pi^1 + \frac{c_3}{4\pi^2}\LSp( y_d\,{\rm{Re}}\,(\lambda_{U1})- y_2^\prime y_s)  \,\Pi^2 \right) \,.
 \label{eq:ApproxLag_2genVLQ}
\end{align} 
Writing the potential to all orders then gives 
\begin{align}
\vev{\Pi^1} = -2\sqrt{2}\,f\,\alpha \,\sin \phi,\quad\vev{\Pi^2} = -2\sqrt{2}\,f\,\alpha\, \cos \phi\,,
\end{align}
where 
\begin{align}
&\phi= \arctan\frac{4\pi^2m_U}{c_3\LSp ( y_d\,{\rm{Re}}\,(\lambda_{U1})- y_2^\prime y_s)}\,,\\
&\alpha=\arctan \frac{\sqrt{\left(c_3\LSp ( y_d\,{\rm{Re}}\,(\lambda_{U1})- y_2^\prime y_s)\right)^2+16\pi^4m_U^2}}{c_3\LSp (\lambda_{U2}  y_s+ y_1^\prime y_d)}~.
\end{align}
The condensates can then be written as
\begin{align}
    \Sigma_q&=
    \begin{array}{cc}
            \begin{pNiceMatrix}[first-row,first-col]
               & \psi_{q_u} & \psi_{q_c}   \\
               \psi_{q_d}& -1&0 \\
               \psi_{q_s}& 0&-1  
            \end{pNiceMatrix} &   
        \end{array} \nonumber\,,\\
 \Sigma_{ud}&= \begin{array}{cc}
            \begin{pNiceMatrix}[first-row,first-col]
        &\psi_{\bar{d}}&\psi_{\bar{s}}&\psi_U\\
\psi_{\bar{u}}&       - \sin^2\frac{\alpha}{2}\, \sin{2\phi}&\sin{\alpha}\sin\phi&\cos^2\frac{\alpha}{2}+\cos{2\phi}\sin^2{\frac{\alpha}{2}}\\\psi_{\bar{c}}&  \cos^2\frac{\alpha}{2}-\cos{2\phi}\sin^2{\frac{\alpha}{2}}&\sin{\alpha}\cos\phi&-\sin^2\frac{\alpha}{2}\, \sin{2\phi}\\\psi_{U}&  
      - \sin{\alpha}\cos\phi&\cos{\alpha}&-\sin{\alpha}\sin\phi
    \end{pNiceMatrix} &   
        \end{array}\,,\label{eq:2Gen_Cond}
\end{align}
where we have separated the $SU(2)$ doublets and the singlets for convenience.
In particular, note that in the limit $m_U\rightarrow 0$, we obtain 
\begin{align}
\vev{\Pi^1} = 0,\quad\vev{\Pi^2} = -2 \sqrt{2} f\arctan\frac{ y_d\,{\rm Re}(\lambda_{U1})- y_2^\prime y_s}{ y_s\lambda_{U2}+ y_1^\prime y_d}\,.
\end{align}
Also, when $\lambda_{U1}\to y_2^\prime y_s/y_d $, this reduces to the one generation case with a vector-like quark.

\section{Flavor structure of the three-generation models}\label{app:flavor_inv}

In this appendix we study the flavor structure of the three-generation models by computing the flavor invariants which can then be used to estimate the contribution of Yukawa phases to the strong CP phase.
Consider the Yukawa interactions and mass terms in a generic model discussed in section \ref{sec:extragaugeint}
\begin{align}
{\cal L}= - \widetilde{Y}^u_{aj} \psi_{q_a} \psi_{\bar{u}_j} H - \widetilde{Y}^d_{ai} \psi_{q_a} \psi_{\bar{d}_i} \widetilde{H} -  {\widetilde{M} }_{ij} \psi_{\bar{d}_i} \psi_{\bar{u}_j}   \, ,\label{eq:flavorGeneric}
\end{align}
where $a=1\dots F_q,~i=1,\dots F_{\bar{d}}+1,$ and $j=1,\dots F_{\bar{u}}+1$ are flavor indices and we have defined 
\begin{align}
    \psi_{q} & =\begin{pmatrix}
         \psi_{q_1}, & \dots , &\psi_{q_{F_q}} 
    \end{pmatrix}^T\,,\nonumber\\
     \psi_{\bar{u}} & =\begin{pmatrix}
         \psi_{\bar{u}_1}, &  \dots, &\psi_{\bar{u}_{F_{\bar{u}}}},&\psi_{\bar{U}}
    \end{pmatrix}^T\,,\nonumber\\
     \psi_{\bar{d}} & =\begin{pmatrix}
         \psi_{\bar{d}_1}, &  \dots , &\psi_{\bar{d}_{F_{\bar{d}}}},&\psi_U
    \end{pmatrix}^T\,,
\end{align}
which transform under the flavor symmetries\footnote{ Note that in this appendix, we are only interested in the flavor symmetries below the scale $\LSp$, and hence below the grand color breaking scale, $M_{\rm GC}$. In contrast to this, above $M_{\rm GC}$ the flavor symmetries are $SU(F_q)_q \times SU(F_{\bar{u}}+1)_{\bar{u}}\times SU(F_{\bar{d}})_{\bar{d}}$.} $SU(F_q)_q \times SU(F_{\bar{u}}+1)_{\bar{u}}\times SU(F_{\bar{d}}+1)_{\bar{d}}$ as
\begin{align}
    \psi_{q}\to {\cal U}_q \psi_{q},\quad \psi_{\bar{u}}\to {\cal U}_{\bar{u}} \psi_{\bar{u}},\quad \psi_{\bar{d}}\to {\cal U}_{\bar{d}} \psi_{\bar{d}}~,
\end{align} where ${\cal U}_q\in SU(F_q)_q$ and $ {\cal U}_{\bar{u},\bar{d}} \in SU(F_{\bar{u},\bar{d}}+1)_{\bar{u},\bar{d}}$. 
Additionally, there are two more non-anomalous $U(1)$ symmetries, which for $F_{\bar{u}}=F_{\bar{d}}$ are 
\begin{align}
 U(1)_1:& \qquad \psi_{q}(F_{\bar{u},\bar{d}}
 +1),\quad\psi_{\bar{u}}(-F_q), \quad\psi_{\bar{d}}(-F_q)~,\\
 U(1)_2:&\qquad \qquad \psi_{q}(0),~\qquad\psi_{\bar{u}}(1), ~\quad\psi_{\bar{d}}(-1)~.
\end{align}
The Yukawa couplings $\widetilde{Y}^u,\,\widetilde{Y}^d$ and the mass terms $\widetilde{M}$ can be treated as spurions under these symmetries transforming as
\begin{align}
   \widetilde{Y}^u &\to {\cal U}_{q}^* \,   \widetilde{Y}^u\, {\cal U}_{\bar{u}}^\dagger\,,
    \label{eq:yukawa_transfYu}\\
    \widetilde{Y}^d&\to {\cal U}_{q}^* \,   \widetilde{Y}^d\, {\cal U}_{\bar{d}}^\dagger \,,
     \label{eq:yukawa_transfYd}\\
  \widetilde{M}&\to {\cal U}_{\bar{d}}^* \,  \widetilde{M}\, {\cal U}_{\bar{u}}^\dagger\,, 
  \label{eq:yukawa_transfM}
\end{align}
under the non-abelian flavour symmetries, whereas under $ U(1)_{1,2}$, $ \widetilde{M}$ 
has charge $(2F_q,0)$, and $\widetilde{Y}^{u,\,d}$ has charge $(F_q-F_{\bar{u},\bar{d}}-1,\mp 1)$, respectively.

Apart from the Yukawa couplings themselves, the quark condensates, $\Sigma_{ud}$ and $\Sigma_{q}$, described in appendix \ref{app:two} can also contribute to the corrections to the strong CP phase. Under the non-abelian flavor symmetries, these transform as 
 \begin{align}
  \Sigma_{ud}\to {\cal U}_{\bar{u}} \Sigma_{ud}\,{\cal U}_{\bar{d}}^T\,,\\
   \Sigma_{q}\to {\cal U}_{q} \Sigma_{q}\,{\cal U}_{q}^T\,,
\end{align}
and under $U(1)_{1,2}$, their charges are $\Sigma_{ud}(-2F_q,0),$ and $\Sigma_{q}(2F_{\bar{u},\bar{d}}+2,0)$, respectively.
To study CP violation arising from the Yukawa couplings, masses, and condensates, we can construct quantities with a non-zero phase that are invariant under $SU(F_q)_q \times SU(F_{\bar{u}}+1)_{\bar{u}}\times SU(F_{\bar{d}}+1)_{\bar{d}}\times U(1)_1\times U(1)_2$.

We first discuss the phenomenologically viable models discussed in section~\ref{sec:extragaugeint}, specifically the model described in section \ref{sec:extrasu3}. This has the flavor symmetries $SU(2)_q\times SU(3)_{\bar{u}}\times SU(3)_{\bar{d}}$ for the fermions charged under $SU(3)$ (corresponding to $F_q=F_{\bar{u}}=F_{\bar{d}}=2$) and a flavor symmetry $SU(2)_B$ for the quarks $(\bar{d}_3,\, \bar{B})$ charged under $SU(3)_T$. By the rotations of fields, the CP phases in this model can be put in $\lambda_{U1}$ and $y^B_a$. The $\lambda_{U1}$ appear in the $2\times 3$ matrix $\widetilde{Y}^u$ and transforms accordingly as in \eqref{eq:yukawa_transfYu}, 
while for $y^B_a$ we can define $X^B_{ab}\equiv y^B_ay^{B\,*}_b$ which transforms as
\begin{align}
  X^B \to {\cal U}_{\bar{d}}^*X_B{\cal U}_{\bar{d}}^T\,.
\end{align}
Note that the CKM phase for a three-generation model, such as the SM, is given by ${\rm arg \, Tr} \left(X_u^2X_d^2X_uX_d \right)$,  where $X_{u,d}\equiv\widetilde{Y}^{u,d}\widetilde{Y}^{u,d\,\dagger}$ and $\widetilde{Y}^{u,d}$ are $3\times 3$ matrices.
As such, using the effective $3\times 3$ matrix \eqref{eq:SU3T_Yuk}, the CKM phase in the model in section~\ref{sec:extragaugeint} can be identified with ${\rm arg}\left( y_1^By_2^{B\,*}\right)$ for our choice of basis. The leading-order flavor invariants that can have a non-zero phase are given by
\begin{align}
&\text{Tr}( (\widetilde{Y}^{d})^T \Sigma_{q}\widetilde{Y}^{u}\Sigma_{ud})~,\label{eq:cond_inv1}\\ 
&  \label{eq:cond_inv2}\text{Tr}(X_B \, \widetilde{M} \,  \Sigma_{ud})~,\\\label{eq:cond_inv3}
&\text{Tr}((\widetilde{Y}^{d})^T\Sigma_{q}\widetilde{Y}^{u}\widetilde{M}^\dagger)~,\\ \label{eq:cond_inv4}
&\text{Tr}(   \widetilde{Y}^{u}\Sigma_{ud}\widetilde{M} \widetilde{Y}^{u\,\dagger})~,\\ \label{eq:cond_inv5}
&\text{Tr}(X_B (\widetilde{Y}^{d})^T\Sigma_{q}\widetilde{Y}^{u}\Sigma_{ud})~,\\ \label{eq:cond_inv6}
   &  \text{Tr}(X_B \, \widetilde{M} \,  \Sigma_{ud}\widetilde{M} \,  \Sigma_{ud})~.
\end{align}
Using the form of the condensates in \eqref{eq:2Gen_Cond}, the imaginary part of the leading-order invariant \eqref{eq:cond_inv1} is given by
\begin{align}
    \frac{\left| \text{Re\,}(\lambda_{U1}) y_d-y_2^\prime y_s\right| }{\mathcal{I} }   \frac{  \text{Im }( \lambda_{U1}  )}{\lambda_{U2}}\frac{y_d}{y_s}\lesssim 10^{-8}\, \frac{  \text{Im }( \lambda_{U1}  )}{\lambda_{U2}}~,
\label{eq:invar_eval1}\end{align} where we have defined 
\begin{align}
\mathcal{I}= \sqrt{1+\frac{\left(\text{Re\,}(\lambda_{U1}) y_d-y_2^\prime y_s\right)^2}{\lambda_{U2}^2 y_s^2}
+\frac{16 \pi ^4  {m_U}^2}{c_3^2   \lambda_{U2}^2 y_s^2\LSp ^2}}~,
\end{align}
which is always $\gtrsim 1$. Note that, since $y_2^\prime\sim y_c$ and the combination $\text{Re\,}(\lambda_{U1}) y_d-y_2^\prime y_s \sim 10^{-6}$ appears with a factor of $1/16\pi^2$ (see the Lagrangian in \eqref{eq:ApproxLag_2genVLQ}, for example), the quantity in \eqref{eq:invar_eval1} contributes to the strong CP phase by an amount $\sim 10^{-10}$ for $\lambda_{U1}\sim \lambda_{U2}$, close to the current experimental bounds. 
 
Note that \eqref{eq:cond_inv2} and \eqref{eq:cond_inv6} do not involve $\widetilde{Y}^{u,d}$, and the phases in $y^B_{1,2}$ can be rotated away into $\widetilde{Y}^{u,d}$ without changing the traces in \eqref{eq:cond_inv2} and \eqref{eq:cond_inv6}, and hence are purely real. It can be explicitly checked that both \eqref{eq:cond_inv2} and \eqref{eq:cond_inv6} vanish for the model in section~\ref{sec:extrasu3}. Furthermore, the quantity in   \eqref{eq:cond_inv3} vanishes for the model in section~\ref{sec:Two_gen_wVLQ}.
 
The next correction comes from \eqref{eq:cond_inv4} and is given by 
 \begin{align}
    \frac{(1-1/\mathcal{I})\left| \text{Re\,}(\lambda_{U1}) y_d-y_2^\prime y_s\right| y_1^\prime}{1+c_3^2 \LSp ^2 \left(\text{Re\,}(\lambda_{U1}) y_d-y_2^\prime y_s\right)^2/16 \pi ^4  {m_U}^2}  {\text{Im}(\lambda_{U1})} \lesssim10^{-10}\,\text{Im}(\lambda_{U1})~,
 \end{align}
which does not contribute significantly to the corrections to the strong CP phase. 

Finally, we discuss the trace in \eqref{eq:cond_inv5}.
Schematically, its imaginary part is given by 
\begin{align}
   {\rm Im}\left( y_1^By_2^{B\,*}\right)\mathcal{I}_1+ {\rm Im}\left( y_1^By_2^{B\,*}\lambda_{U1}^*\right)\mathcal{I}_2+ {\rm Im}\left(\lambda_{U1} \right)| y_1^B|^2\mathcal{I}_3~,\label{eq:cond_inv5EVal}
\end{align}
where we have introduced $\mathcal{I}_{1,2,3}\ll 1$, which are certain products of Yukawa couplings and sinusoidal functions appearing in \eqref{eq:2Gen_Cond}.
Since $y^B_{1,2}\sim 10^{-8}$, the expression \eqref{eq:cond_inv5EVal} is numerically $\ll 10^{-10}$, and cannot contribute significantly to the strong CP phase. Note that this is the first trace where the CKM phase $\sim {\rm arg}\left( y_1^By_2^{B\,*}\right)$ appears.

We next consider the model with the third generation charged under a different $SU(2)_T$ gauge group, as discussed in section~\ref{sec:SU(2)}, {which has the flavor symmetries $SU(2)_q\times SU(4)_{\bar{u}}\times SU(4)_{\bar{d}}$ (corresponding to $F_q=2,\,F_{\bar{u}}=F_{\bar{d}}=3$).}
 In this case, we have multiple phases appearing in the Yukawa couplings in \eqref{eq:3genLSU(2)}.
We now denote the $H_{FS}$ Yukawa couplings by $\widetilde{Y}^u,\,\widetilde{Y}^d$, which transform as in \eqref{eq:yukawa_transfYu} and \eqref{eq:yukawa_transfYd}. In addition, we define $X^{u,d}_{ab}\equiv \widetilde{Y}^{u,d\, *}_{3a}\widetilde{Y}^{u,d}_{3b}$ which transforms as
 \begin{align}
      X^{u,d} \to {\cal U}_{\bar{u},\bar{d}}X^{u,d}{\cal U}_{\bar{u},\bar{d}}^\dagger\,.
 \end{align}
 Note that $\widetilde{Y}^u$ and $X^u$ also include the Yukawa couplings of the terms containing the quark doublet and the vector-like quark $\bar{U}$. 
 The leading-order flavor invariants are then given by 
 \begin{align} 
   & \label{eq:condSU2_inv1}  \text{Tr} (X^{d\,*} \widetilde{M}\Sigma_{ud})~,\\
 & \label{eq:condSU2_inv2}\text{Tr} (X^{d\,*}(\widetilde{Y}^{d})^T\Sigma_{q}\widetilde{Y}^{u}\widetilde{M}^\dagger)~,\\
   & \label{eq:condSU2_inv3}\text{Tr} (X^{d\,*}(\widetilde{Y}^{d})^T\Sigma_{q}\widetilde{Y}^{u}\Sigma_{ud})~,\\
   & \label{eq:condSU2_inv4} \text{Tr} (X^{u} \Sigma_{ud}\widetilde{M})~,\\
  &  \label{eq:condSU2_inv5} \text{Tr} (X^{u\,*}(\widetilde{Y}^{u})^T\Sigma_{q}\widetilde{Y}^{d}\widetilde{M}^*)~,\\
  & \label{eq:condSU2_inv6}  \text{Tr} (X^{u\,*}(\widetilde{Y}^{u})^T\Sigma_{q}\widetilde{Y}^{d}\Sigma_{ud}^T)~.
 \end{align}
The explicit form of the condensates $\Sigma_{ud}, \,\Sigma_{q}$ was computed for a different model in appendix \ref{app:two}. For the extra $SU(2)$ model discussed in section~\ref{sec:SU(2)}, we additionally have the Yukawa couplings of $\psi_{q_{1,2}}$ 
to the $\psi_{\bar{b}},\,\psi_{\bar{t}}$ quarks. Since the corresponding CKM mixing angles are small compared to those between the first and second generation fermions, the quarks $\psi_{\bar{b}}$ and $\psi_{\bar{t}}$ only condense with each other, i.e., the resulting $\Sigma_{ud}$ can be approximated as a $4\times 4$ matrix, consisting of an upper left block given by \eqref{eq:2Gen_Cond}, approximately $-1$ in the $\psi_{\bar{t}}\,\psi_{\bar{b}}$ entry and almost zero ($\lesssim 10^{-3}$) elsewhere.  
The sign of $\vev{\psi_{\bar{t}} \,\psi_{\bar{b}}}$ is chosen so that $\arg\det \Sigma_{ud}=0$. 
The condensation of $\psi_{q_{1,2}}$ is not expected to change, and hence $\Sigma_q$ is still given by \eqref{eq:2Gen_Cond}.

Using this form for $\Sigma_{ud}$ and $\Sigma_{q}$, the quantities \eqref{eq:condSU2_inv1},\,\eqref{eq:condSU2_inv2},\,\eqref{eq:condSU2_inv3}, and \eqref{eq:condSU2_inv5} vanish for our model. The imaginary part of \eqref{eq:condSU2_inv4} is given by
 \begin{align}
    \left(\frac{m_U}{\LSp}\right)^2\frac{\left| \text{Re\,}(\lambda_{U1}) y_d-y_2^\prime y_s\right| }{ \mathcal{I}\,(\mathcal{I}+1)}   \frac{  \text{Im }( \widetilde{Y}^{u\,*}_{3\bar{U}}\widetilde{Y}^u_{32}  )}{\lambda_{U2}^2 y_s^2} &\lesssim 10^{-11}\frac{ 1}{\lambda_{U2}^2}\left(\frac{m_U}{\LSp}\right)^2~,
\label{eq:condSU2_inv4Eval}
\end{align}
where we have taken $\widetilde{Y}^{u}_{3\bar{U}}$ and $\widetilde{Y}^u_{32} $ to be of order the SM Yukawa couplings of $q_3$ to the first and the second generation fermions, respectively. Furthermore, this quantity appears with a factor of $1/16\pi^2$, as can be seen from the number of Yukawa couplings appearing in \eqref{eq:condSU2_inv4}. We thus require $m_U\lesssim10\,\lambda_{U2}\LSp$, which is easily satisfied for our model. Note that in computing \eqref{eq:condSU2_inv4Eval}, we assumed all the condensates between
$\psi_{\bar{t}}$ and $\psi_{\bar{b}}$ are zero, except those between each other. The correction upon relaxing this assumption is given by
\begin{align}
        \frac{m_U}{\LSp} {  \text{Im }( \widetilde{Y}^{u\,*}_{3\bar{U}}(\Sigma_{ud})_{{3\bar{U}}} )} &\lesssim 10^{-11} \left(\frac{m_U}{\LSp}\right)~,
\label{eq:invar_eval4_SU2}
 \end{align}
where we have used a conservative estimate for $(\Sigma_{ud})_{{3\bar{U}}}\sim 10^{-3}$. Thus, we see that our approximation works well for the estimate \eqref{eq:condSU2_inv4Eval}.
 
Finally, we consider the imaginary part of the trace in \eqref{eq:condSU2_inv6}, which is given by 
\begin{align}\label{eq:condSU2_inv6Eval}
     y_d\,\text{Im}( \widetilde{Y}^{u}_{13}\widetilde{Y}^u_{32}  \widetilde{Y}^u_{33}  )+   \frac{\left| \text{Re\,}(\lambda_{U1}) y_d-y_2^\prime y_s\right| }{ \mathcal{I}}   \frac{  \text{Im}( \widetilde{Y}^{u}_{23}\widetilde{Y}^u_{32}  \widetilde{Y}^u_{33}  )}{\lambda_{U2} } \lesssim  10^{-11} +10^{-12}\frac{ 1}{\lambda_{U2}} ~,
\end{align}
where we have taken the Yukawa couplings $\widetilde{Y}^{u}$ to be their SM values. This quantity is further accompanied by a factor of $(1/16\pi^2)^2$, and hence does not generate a significant correction for $\lambda_{U2}\gtrsim 10^{-3}$. Upon relaxing the assumption that $(\Sigma_{ud})_{3i}=0,\, i\neq 3$, we obtain a correction to \eqref{eq:condSU2_inv6Eval} given by
\begin{align} 
        \sum_{i=1,2}(\Sigma_{ud})_{3i} y_{d,s}\text{Im}\left(\lambda_{Ui}\widetilde{Y}^{u\,*}_{3\bar{U}}+y^\prime_i\widetilde{Y}^{u\,*}_{32}+\widetilde{Y}^{u}_{i3}\widetilde{Y}^{u\,*}_{33}\right)&\sim  10^{-9}+10^{-15}\lambda_{U2}+ 10^{-16}\lambda_{U1}~, 
\label{eq:invar_eval6_SU2}
 \end{align}
where we have taken the Yukawa couplings $\widetilde{Y}^{u}$ to be their SM values and $(\Sigma_{ud})_{{3j}}\sim 10^{-3},\,j=1,2$. With the additional factor of $(1/16\pi^2)^2$, \eqref{eq:invar_eval6_SU2} does not generate any significant correction.   

Finally, we can also check determinant-like invariants in addition to  the trace-like invariants discussed above. Since each determinant expression reduces to a product of separate determinants, we only need to check the determinants of the individual  matrices appearing in the trace-like quantities. In particular, $\arg\det \Sigma_{ud}=\arg\det \Sigma_{q}=0$ for the expressions given in \eqref{eq:2Gen_Cond} (and their generalization for the extra $SU(2)$ model) which also serves as a consistency check on the analysis in appendix \ref{sec:proof} since the determinants are real and positive ($=1$). Furthermore, we can work in the basis in which the down-type Yukawa couplings are diagonal and positive, thus implying $\arg\det \widetilde{Y}^d=0$, and since $\bar{u}_1$ does not couple to any of the quark doublets, $\arg\det \widetilde{Y}^u=0$. As such, the determinant of the product of any of the Yukawa matrices or the condensates is also positive (and real), and therefore the determinant-like invariants do not give any corrections to the strong CP phase. We thus see that the corrections to the strong CP problem appearing at higher-loop order remain small in both of our phenomenologically viable models discussed in section \ref{sec:3gen}.

\bibliographystyle{JHEP}

\bibliography{references}
\newpage
 \end{document}